\documentclass[11pt,reqno]{article}
\usepackage{bm}
\usepackage{amsthm}
\usepackage{amsmath}
\usepackage{amssymb}
\usepackage{pdfsync}
\usepackage[round]{natbib}
\usepackage{color}
\usepackage{authblk}
\usepackage{microtype}
\usepackage{graphicx}
\usepackage{float}
\usepackage{prodint}
\usepackage{paralist}
\usepackage{tabularx}
\usepackage{epstopdf}
\usepackage{rotating}
\usepackage{geometry}
\DeclareMathSymbol{\shortminus}{\mathbin}{AMSa}{"39}
\newcommand{\rnc}{\renewcommand}
\newcommand{\nc}{\newcommand}

\definecolor{color1}{RGB}{122,0,63}

\nc{\mb}{\mathbb}
\nc{\mc}{\mathcal}
\nc{\N}{\mb{N}}
\nc{\R}{\mb{R}}
\nc{\Q}{\mb{Q}}

\nc{\E}{E}
\rnc{\P}{P}
\nc{\var}{ \text{Var} }
\nc{\mbf}{\boldsymbol}
\nc{\I}{I}

\nc{\trans}{^{\top}}
\nc{\assumption}{{Assumption A}}

\newtheorem{lemma}{Lemma}
\newtheorem{assump}{Assumption}
\newtheorem{theorem}{Theorem}

\newtheorem{definition}{Definition}

\newcommand\blfootnote[1]{%
  \begingroup
  \renewcommand\thefootnote{}\footnote{#1}%
  \addtocounter{footnote}{-1}%
  \endgroup
} 

\DeclareSymbolFont{matha}{OML}{txmi}{m}{it}
\DeclareMathSymbol{\varv}{\mathord}{matha}{118}

\begin{document}

\title{\Large \bf
Inference for all variants of the multivariate coefficient of variation in factorial designs
}
\author[1]{Marc Ditzhaus}
\author[2,$^*$]{\L ukasz Smaga}

\affil[1]{Faculty of Mathematics, Otto von Guericke University Magdeburg, Germany}
\affil[2]{Faculty of Mathematics and Computer Science, Adam Mickiewicz University, Poland}

\maketitle

\begin{abstract}
	The multivariate coefficient of variation (MCV) is an attractive and easy-to-interpret effect size for the dispersion in multivariate data. Recently, the first inference methods for the MCV were proposed by \cite{ditzhaus_smaga_2022} for general factorial designs covering $k$-sample settings but also complex higher-way layouts. However, two questions are still pending: (1) The theory on inference methods for MCV  is primarily derived for one special MCV variant while there are several reasonable proposals. (2) When rejecting a global null hypothesis in factorial designs, a more in-depth analysis is typically of high interest to find the specific contrasts of MCV leading to the aforementioned rejection. In this paper, we tackle both by, first, extending the aforementioned nonparametric permutation procedure to the other MCV variants and, second, by proposing a max-type test for post hoc analysis. To improve the small sample performance of the latter, we suggest a novel studentized bootstrap strategy and prove its asymptotic validity. The actual performance of all proposed tests and post hoc procedures are compared in an extensive simulation study and illustrated by a real data analysis.
\blfootnote{${}^*$ Corresponding author. Email address: {ls@amu.edu.pl}} 
\end{abstract}


\section{Introduction}
\label{sec:intro}

The coefficient of variation (CV) is defined as the standard deviation divided by the population mean. By this, it becomes a powerful, unit-free measure of dispersion and is used in diverse areas, e.g. in medicine for reliability and reproducibility of measurements \citep{neumann2021reliability}, for risk evaluation in finance \citep{Ferri1979} or in psychology \citep{Weber2004}. Furthermore, it is used in control charts for monitoring \citep{jalilibal2021monitoring}. However, as pointed out by \cite{yeong2016control} in the context of control charts: ``\textit{There are many situations where multiple characteristics need to be monitored simultaneously.}'' This is certainly true apart from control charts, e.g. when various medical measurements are taken from the same patients. For such scenarios, the CV can be extended to the multivariate setting in various ways \citep{Reyment1960,VanValen1974,voinovNikulin1996,AlbertZhang2010}:
\begin{align}\label{eqn:MCV}
	C^{\text{RR}} = \sqrt{\frac{(\det\mbf{\Sigma})^{1/d}}{\mbf{\mu}^{\top}\mbf{\mu}}},\ C^{\text{VV}} = \sqrt{\frac{\mathrm{tr}\mbf{\Sigma}}{\mbf{\mu}^{\top}\mbf{\mu}}},\ C^{\text{VN}} = \sqrt{\frac{1}{\mbf{\mu}^{\top}\mbf{\Sigma}^{-1}\mbf{\mu}}},\ C^{\text{AZ}} = \sqrt{\frac{\mbf{\mu}^{\top}\mbf{\Sigma}\mbf{\mu}}{(\mbf{\mu}^{\top}\mbf{\mu})^2}},
\end{align}
which all reduces to CV in the univariate case ($d=1$). Here $\mbf{\mu}\neq \mbf{0}$ denotes the mean vector of a $d$-dimensional random variable and $\mbf{\Sigma}$ is the corresponding covariance matrix.  The standardized means as the reciprocal of $C^v$ are of their own interest:
\begin{align}\label{eqn:def_B}
	B^v = \frac{1}{C^v} \quad(v=\text{RR},\,\text{VV},\,\text{VN},\,\text{AZ}).
\end{align}
The differences between the four variants are discussed in great detail by \cite{AlbertZhang2010}. One remarkable difference is that $C^{\text{RR}}$ and $C^{\text{VN}}$ require a regular matrix $\mbf\Sigma$ while the other two variants do not. This regularity assumptions becomes rather restrictive for high-dimensional scenarios, such as microarray data. Moreover, the variant $C^{\text{VV}}$ is the only one which does not take the covariance of the different measurements into account. 

When we turn to the inference problem, the literature regarding the MCV becomes scare while for the univariate CV various two- and $k$-sample testing proposal can be found in the literature, see \cite{AertsHaesbroeck2017} and \cite{PaulySmaga2020}. Recently, \cite{ditzhaus_smaga_2022} addressed the remaining question for more general CV testing procedures, namely in complex factorial designs. The latter are highly relevant  for various fields, e.g. in biomedicine or psychology \citep{gissi:1990,baigent:etal:1998,cassidy:etal:2008,kurz:etal:2015}, where the $k$-sample set-up is often too narrow. Methods for factorial designs allow to discuss main effects (e.g. of the gender, measurement, site) and even interaction effects: '\textit{it is desirable for reports of factorial trials to include estimates of the interaction between the treatments}' \citep{lubsen:pocock:1994}. In addition to the extension towards factorial designs, \cite{ditzhaus_smaga_2022} proposed their nonparametric methods directly for $C^{\text{VN}}$ in the multivariate set-up. This complemented the prior proposal of \cite{AertsHaesbroeck2017}, which, in contrast, relied on (semi-)parametric model assumptions and whose convergence rate was rather slow, see the simulation results of \cite{ditzhaus_smaga_2022}. To the best of our knowledge, these two proposals are the only ones discussing the inference problem for MCV and, moreover, both restricted their study to $C^{\text{VN}}$. Thus, the natural question arises whether the inference strategies can be transferred to the other MCV variants from \eqref{eqn:MCV}, especially to be able to study also settings with non-regular covariance matrices $\boldsymbol{\Sigma}$. For this purpose, we will adopt the results of \cite{ditzhaus_smaga_2022} and, in particular, derive permutation versions with a better performance under small sample sizes, see Section~\ref{sec:simulation}. In this way, we add a further chapter to the success story of studentized permutation tests in complex factorial designs \citep{paulyETAL2015,friedrich2017permuting,harrar2019comparison, ditzhaus2021casanova}. While classical permutation tests for exchangeable data settings are well-known, it is less known that the studentized permutation versions are also valid beyond exchangeability.

Beside global null hypothesis testing in $k$-sample settings or inferring main/interaction effects in factorial designs, often a more in-depth analysis is wanted, e.g. multiple pairwise comparisons, to get a better picture of the underlying effects. Since Bonferroni correction leads partially to a significant power loss, strategies incorporating the concrete dependence structure are preferred. Therefor, multiple contrast tests and corresponding simultaneous confidence intervals are well established for means \citep{mukerjee1987comparison,bretz2001numerical}, the relative treatment effect \citep{umlauft2019wild,gunawardana2019nonparametric}, and the area under the receiving operating curve \citep{konietschke2018simultaneous,wechsung2021simultaneous}. A respective multiple strategy for the MCV is missing.
\vspace{0.2cm}

The remainder of this paper is organized as follows. In Section \ref{sec:setup}, we derive a central limit theorem for all MCV variants and their reciprocals in the one-sample setting. Moreover, we discuss assumptions such that the limit is not degenerated and how the limiting variances can be estimated consistently. All these results lie the foundation for the Wald-type statistics to infer main and interaction effects in terms of MCVs and standardized means in general factorial designs, see Section~\ref{sec:global}. Moreover, we develop respective permutation and bootstrap counterparts of these Wald-type statistics and prove their asymptotic validity. In Section~\ref{sec:multiple}, we discuss the issue of simultaneous inference and present max-type multiple contrast tests. Again, we complement this by an asymptotically valid resampling procedure. An exhaustive simulation study is presented in Section \ref{sec:simulation}.  The tests’ applicability are illustrated by analyzing data of external quality assessment in Section~\ref{sec:real_data_app}. Finally, Section \ref{sec:conclusion} summarizes the major results of the paper and discusses further research possibilities. All proofs and additional simulation results are in the supplementary materials.

\section{The nonparametric framework}\label{sec:setup}
Consider $n_1$ independent, identically distributed $d$-dimensional random variables
\begin{align*}
	\mbf{X}_{j} = (X_{j1}, \ldots, X_{jd})\trans \quad (j=1,\ldots,n_1).
\end{align*}
Hereby, we suppose no specific conditions on the distributions of $\mbf{X}_{j}$ except the following assumptions on the moments to ensure the well-definedness of $C^j$ and $B^j$:
\begin{assump}\label{ass:well_defined}
	Let $\mbf{\mu}\neq \mbf{0}$ and $E(X_{j\ell}^4)<\infty$ for all $j$ and $\ell$. Moreover, we suppose:
	\begin{enumerate}[(a)]
		\item\label{ass:well_defined_a} For $C^{\text{RR}}$, $C^{\text{VN}}$, $B^{\text{RR}}$, and $B^{\text{VN}}$, we consider only regular matrices $\mbf\Sigma$.
		\item\label{ass:well_defined_b} For $C^{\text{VV}}$ and $B^{\text{VV}}$, we assume $\mbf\Sigma \neq \mbf 0_{d\times d}$.
		\item\label{ass:well_defined_c} For $C^{\text{AZ}}$ and $B^{\text{AZ}}$, we suppose $\mbf\mu^{\top}\mbf\Sigma\mbf\mu>0$.
	\end{enumerate}
	Here and below, we denote by $\mbf 0_{d\times d}$ a $d\times d$-dimensional matrix consisting of zeros only.
\end{assump}
Assumption~\ref{ass:well_defined} ensures that $C^v$, $B^v$ are well-defined. Clearly, \ref{ass:well_defined_a} $\Rightarrow$ \ref{ass:well_defined_c} $\Rightarrow$ \ref{ass:well_defined_b}. Thus, $C^{\text{VV}}$ seems to be the most general variant. But, as mentioned in the introduction, $C^{\text{VV}}$ does not take the covariance structure into account and only combines the marginal variability into one standardized effect size. 
For statistical inference, we estimate the MCVs or their reciprocals, respectively, by plugging-in the sample mean $\widehat{\mbf\mu}$ and covariance matrix $\widehat{\mbf\Sigma}$:
\begin{equation}
	\label{mcv_est}
	{\widehat C}^{\text{RR}}=\sqrt{\frac{(\det\mbf{\widehat\Sigma})^{1/d}}{\mbf{\widehat\mu}^{\top}\mbf{\widehat\mu}}},\ 
	{\widehat C}^{\text{VV}}=\sqrt{\frac{\mathrm{tr}\mbf{\widehat\Sigma}}{\mbf{\widehat\mu}^{\top}\mbf{\widehat\mu}}},\ 
	{\widehat C}^{\text{VN}}=\sqrt{\frac{1}{\mbf{\widehat\mu}^{\top}\mbf{\widehat\Sigma}^{-1}\mbf{\widehat\mu}}},\ 
	{\widehat C}^{\text{AZ}}=\sqrt{\frac{\mbf{\widehat\mu}^{\top}\mbf{\widehat\Sigma}\mbf{\widehat\mu}}{(\mbf{\widehat\mu}^{\top}\mbf{\widehat\mu})^2}},
\end{equation}
and $\widehat B^v = 1/\widehat C^v$. By extending the results of \cite{ditzhaus_smaga_2022} for $C^{\text{VN}}$ and $B^{\text{VN}}$, we are able to derive central limit theorems for all these estimators. The respective asymptotic variances have a rather complex structure and depend on several quantities:
\begin{align*}
	&\mbf{A}_{\text{RR}}(\mbf{ \mu},\mbf{\Sigma})=\left(-2d\:\det(\mbf{\Sigma})\frac{\mbf{\mu}\trans}{(\mbf{\mu}\trans \mbf{\mu})^{d+1}}+\frac{\det(\mbf{\Sigma})\left(\text{vec}(\mbf{\Sigma}^{-1})\right)\trans}{(\mbf{\mu}\trans \mbf{\mu})^d}\mbf{\widetilde D}(\mbf{\mu}),\frac{\det(\mbf{\Sigma})\left(\text{vec}(\mbf{\Sigma}^{-1})\right)\trans}{(\mbf{\mu}\trans \mbf{\mu})^d}\right),\\
	&\mbf{A}_{\text{VV}}(\mbf{ \mu},\mbf{\Sigma})=\left(-2\:\textrm{tr}(\mbf{\Sigma})\frac{\mbf{\mu}\trans}{(\mbf{\mu}\trans \mbf{\mu})^2}+\frac{1}{\mbf{\mu}\trans \mbf{\mu}}\left(\text{vec}(\mbf{I}_d)\right)\trans \mbf{\widetilde D}(\mbf{\mu}),\frac{1}{\mbf{\mu}\trans \mbf{\mu}}\left(\text{vec}(\mbf{I}_d)\right)\trans\right),\\
	&\mbf{A}_{\text{VN}}(\mbf{ \mu},\mbf{\Sigma}) = \begin{pmatrix} 2\: \mbf{\mu}\trans \mbf{\Sigma}^{-1} - [(\mbf{\mu}\trans \mbf{\Sigma}^{-1}) \otimes (\mbf{\mu}\trans \mbf{\Sigma}^{-1})]\mbf{\widetilde D}(\mbf{\mu}),-(\mbf{\mu}\trans \mbf{\Sigma}^{-1}) \otimes (\mbf{\mu}\trans \mbf{\Sigma}^{-1})  \end{pmatrix}, \\
	&\mbf{A}_{\text{AZ}}(\mbf{ \mu},\mbf{\Sigma})=\left(-4\:\mbf{\mu}\trans\mbf{\Sigma}\mbf{\mu}\frac{\mbf{\mu}\trans}{(\mbf{\mu}\trans \mbf{\mu})^3}+2\frac{\mbf{\mu}\trans\mbf{\Sigma}}{(\mbf{\mu}\trans \mbf{\mu})^2} + \frac{\mbf{\mu}\trans\otimes\mbf{\mu}\trans}{(\mbf{\mu}\trans \mbf{\mu})^2}\mbf{\widetilde D}(\mbf{\mu}),\frac{\mbf{\mu}\trans\otimes\mbf{\mu}\trans}{(\mbf{\mu}\trans \mbf{\mu})^2}\right)
\end{align*}
where $\otimes$ is the Kronecker product, and the matrices $\mbf{\widetilde D}(\mbf{x})\in \R^{d^2\times d}$ for $\mbf{x}=(x_1,\ldots,x_d)\trans \in \R^d$, $\mbf{\Psi}_{i3}\in \R^{d^2\times d}$ as well as $\mbf{\Psi}_{i4}\in \R^{d^2\times d^2}$ are given by their entries
\begin{align}\label{eqn:def_tilde_D}
	&[\mbf{\widetilde D}(\mbf{x})]_{ad-d+r,s} = - x_r\I\{s=a\neq r\} - 2 x_s \I\{s=r=a\} - x_a \I\{r=s\neq a\}\nonumber\\
	&[\mbf{\Psi}_{3}]_{ad-d+r,s} = \E(X_{1a}X_{1r}X_{1s}) - \E(X_{1a}X_{1r})\E(X_{1s})  \\
	&[\mbf{\Psi}_{4}]_{ad-d+r,bd-d+s}  = \E(X_{1a}X_{1r}X_{1b}X_{1s}) - \E(X_{1a}X_{1r})\E(X_{1b}X_{1s}) \nonumber
\end{align}
for $a,b,r,s\in\{1,\ldots,d\}$. Now, we are able to formulate the central limit theorems. Here and subsequent, all limits are meant as $n_1\to\infty$ unless stated explicitly otherwise.
\begin{theorem}[Central limit theorem]\label{theo:conv_C+beta}
	Let $v\in\{\text{RR},\text{VV},\text{VN},\text{AZ}\}$ and Assumption~\ref{ass:well_defined} be fulfilled. The estimators $\widehat C^v$ and ${\widehat{B}}_i^v$ are asymptotically normal:
	\begin{align*}
		&n_1^{1/2} \Big( {\widehat C}^v - {C}^v \Big) \overset{ d}{\longrightarrow} Z_{C^v}\sim N(0,\sigma^2_{C^v}),\;\,\text{and}\,\,n_1^{1/2} \Big( {\widehat {B}}^v - {{B}}^v \Big) \overset{ d}{\longrightarrow} Z_{B^v}\sim N(0,\sigma^2_{B^v})
	\end{align*}
	with asymptotic variances $\sigma^2_{B^v}= (C^v)^{-4}\sigma^2_{C^v}$ and
	\begin{align*}
		&\sigma^2_{C^v}=\frac{S_v}{4}  \mbf{A}_v(\mbf{ \mu},\mbf{\Sigma})
		\begin{pmatrix}
			\mbf{ \Sigma} & \mbf{\Psi}_{3}\trans  \\
			\mbf{\Psi}_{3} & \mbf{\Psi}_{4}
		\end{pmatrix}
		\mbf{A}_v(\mbf{ \mu},\mbf{\Sigma})\trans,
	\end{align*}
	where $S_{\text{RR}} = d^{-2} (C^{\text{RR}})^{2-4d}, \,\, S_{\text{VV}} = (C^{\text{VV}})^{-2},\,\, S_{\text{VN}} = (C^{\text{VN}})^{6}, \,\, S_{\text{AZ}} = (C^{\text{AZ}})^{-2}$.
	\end{theorem}
	The typically unknown variances, $\sigma^2_{B^v}$ and $\sigma^2_{C^v}$, can be naturally estimated by replacing the expectations and covariances by their empirical counterparts, for instance:
	\begin{align*}
		[\widehat{\mbf{\Psi}}_{3}]_{ad-d+r,s} = \Big(n_1^{-1}\sum_{j=1}^{n_1}X_{ja}X_{jr}X_{js}\Big)-\Big(n_1^{-1}\sum_{j=1}^{n_1}X_{ja}X_{jr}\Big)\Big(n_1^{-1}\sum_{j=1}^{n_1}X_{js} \Big).
	\end{align*}
	In this way, we obtain
	\begin{align}\label{eqn:sigma_hat_def}
		&\widehat{\sigma}^2_{C^v}=\frac{\widehat S_v}{4}  \mbf{A}_v(\widehat{\mbf{ \mu}},\widehat{\mbf{\Sigma}})
		\begin{pmatrix}
			\widehat{\mbf{ \Sigma}} & \widehat{\mbf{\Psi}}_{3}\trans  \\
			\widehat{\mbf{\Psi}}_{3} & \widehat{\mbf{\Psi}}_{4}
		\end{pmatrix}
		\mbf{A}_v(\widehat{\mbf{ \mu}},\widehat{\mbf{\Sigma}})\trans,\quad \widehat \sigma^2_{B^v}= (\widehat C^v)^{-4}\widehat\sigma^2_{C^v},
	\end{align}
	where $\widehat S_{\text{RR}} = d^{-2} (\widehat C^{\text{RR}})^{2-4d},$ $\widehat S_{\text{VV}} = (\widehat C^{\text{VV}})^{-2},$ $\widehat S_{\text{VN}} = (\widehat C^{\text{VN}})^{6},$ $\widehat S_{\text{AZ}} = (\widehat C^{\text{AZ}})^{-2}$. A direct consequence of the continuous mapping theorem and the strong law of large numbers is
	\vspace{0.3cm}
	
	\begin{lemma}\label{lem:const_var_est}
		Under Assumption~\ref{ass:well_defined}, $\widehat{\sigma}^2_{C^v} \overset{p}{\rightarrow} \sigma^2_{C^v}$ and $\widehat{\sigma}^2_{B^v}\overset{p}{\rightarrow}\sigma^2_{B^v}$.
	\end{lemma}

	In general, there is no guarantee that the limits from Theorem~\ref{theo:conv_C+beta} are not degenerated, i.e. $\sigma_{C^v}^2=0$ and, equivalently, $\sigma_{B^v}^2=0$ might be possible. As in the univariate case \citep{PaulySmaga2020} and for $v=\text{VN}$ in the multivariate setting \citep{ditzhaus_smaga_2022}, degeneracy can just appear in rather unusual scenarios of the following kind:
	\begin{definition}[Conditional two-point distribution]
		Let $\mbf{Y}=(Y_1,\ldots,Y_d)\trans \in \R^d$ be a multivariate random variable. We call the $r$th coordinate $Y_{r}$ \textit{conditionally two-point distributed} if it is (conditionally) degenerated or it just takes (conditionally) two different values with positive probability, both given the remaining components $(Y_{s})_{s=1,\ldots,d;s\neq r}$.
	\end{definition}
	For example, the coordinates of $(Y_1,Y_2,Y_1+Y_2^2)$ for arbitrarily distributed $Y_1,Y_2$ are conditionally two-point distributed. The same is true for the coordinates of $(Y_1,\ldots,Y_d)$ with binomial distributed $Y_j$. These extreme cases need to be excluded:
	\begin{assump}\label{ass:two_point}
		No coordinate of $\mbf{X}_{1}$ is conditionally two-point distributed. 
	\end{assump}
	In fact, a weaker assumption is also enough as illustrated in the proofs. In detail, it is sufficient to suppose that the $\ell$th coordinate of $\mbf{X}_{1}$ is not conditionally two-point distributed for some $\ell=1,\ldots,d$. However, we then additionally require $[\mbf{\mu}]_{\ell}\neq 0$ in case of $v=\text{AZ}$ and $[\mbf{\mu}\trans\mbf{\Sigma}^{-1}]_{\ell}\neq 0$ for $v=\text{VN}$. From our point of view, Assumption~\ref{ass:two_point} is easier to check, in particular later for the resampling procedures, and we do not loose much of generality. 
	\begin{lemma}\label{lem:var_equal_zero}
		Under Assumptions~\ref{ass:well_defined} and \ref{ass:two_point} we have $\sigma^2_{C^v} > 0$ and, thus, $\sigma^2_{B^v} > 0$.
	\end{lemma}
	By Theorem~\ref{theo:conv_C+beta} and Lemma~\ref{lem:const_var_est}  $[\widehat C^v \pm n_1^{-1/2}\widehat{\sigma}_{C^v}z_{1-\alpha/2}]$ and $[\widehat B^v \pm n_1^{-1/2}\widehat{\sigma}_{B^v}z_{1-\alpha/2}]$ are asymptotically valid confidence intervals for $C^v$ and $B^v$, respectively, where $z_{1-\alpha/2}$ is the $(1-\alpha/2)$-quantile of $N(0,1)$. Moreover, these results serve as the foundation for inference methods in more complex models as discussed below.
	
	\section{Global testing for factorial designs}\label{sec:global}
	
	\subsection{Factorial designs with respective null hypotheses}
	
	From the easy one-sample scenario from the previous section, we immediately turn to general factorial designs covering two- and $k$-sample settings as special cases. Notationally, factorial designs can be incorporated in a $k$-sample framework by interpreting the groups as subgroups for different factor combinations. We explain this concept below in more detail but first start with introducing the concrete model. For this purpose, we add another index $i=1,\ldots,k$, $k\in\N$, to all quantities from Section~\ref{sec:setup}. In particular, let
	\begin{align*}
		\mbf{X}_{ij} = (X_{ij1}, \ldots, X_{ijd})\trans \quad (i=1,\ldots,k;j=1,\ldots,n_i),
	\end{align*}
	where $\mbf{X}_{i1}, \ldots,\mbf{X}_{in_i}$ are identically distributed for each $i=1,\ldots,k$ and all observations $\mbf{X}_{11}, \ldots,\mbf{X}_{kn_k}$ are mutually independent.
	Depending on the research question, we choose a contrast matrix $\mbf H\in\R^{r\times k}$, i.e. $\mbf H \mbf 1_{k\times 1} = \mbf 0_{r\times 1}$. We now like to infer the following null hypothesis in terms of the chosen $\mbf{H}$ and by this cover a huge variety of testing problems:
	\begin{align}\label{eqn:null}
		\mathcal H_{0,C^v}: \mbf{H}\mbf{C}^{v} = \mbf{0}_{r\times 1}, \quad \mathcal H_{0,B^v}: \mbf{H}\mbf{B}^v = \mbf{0}_{r\times 1}.
	\end{align}
	Here, $\mbf{C}^v=(C_1^v,\ldots,C_k^v)\trans $ and $\mbf{{B}}^v=(B_1^v,\ldots,B_k^v)\trans$. In the same way, we denote by $\mbf{\widehat C}^v$ and $\mbf{\widehat {B}}^v$ the respective estimators.  
	
	\textbf{Different choices for $\mbf{H}$:} To see the high flexibility of \eqref{eqn:null}, we like to discuss some specific cases. The most prominent one is the $k$-sample scenario $\mathcal H_{0,C^v}: \{ \mbf{P}_k \mbf{C}^v = \mbf{0}_{k\times 1} \} = \{C_1^v= \ldots = C_k^v\}$, where $\mbf{P}_k= \mbf{I}_k  - \mbf{1}_{k\times k}/k$ and $\mbf{I}_k$ is the $k\times k$-dimensional unity matrix. Turning to a more complex scenario, we next consider a two-way layout with two factors $A$ and $E$ possessing $a$ and $e$ levels, respectively. By splitting up the group index $i=(i_A, i_E)$ we incorporate this scenario in the aforementioned $k$-sample framework. In particular, we obtain $k=a\cdot e$ subgroups. Now, we divide the subgroup-specific MCV 
	\begin{align*}
		C^v_{i_A,i_E} = C^{v0} + C^{v\alpha}_{i_A} + C^{v\epsilon}_{i_E} + C^{v\alpha\epsilon}_{i_Ai_E}
	\end{align*}
	into a general effect $C^{v0}$, the two main effects $C^{v\alpha}_{i_A},\, C^{v\epsilon}_{i_E}$, and an interaction effect $C^{v\alpha\epsilon}_{i_Ai_E}$. Here, the usual side conditions $\sum_{i_A}C^{v\alpha}_{i_A} = \sum_{i_B} C^{v\epsilon}_{i_E} =\sum_{i_A} C^{v\alpha\epsilon}_{i_Ai_E} =\sum_{i_E} C^{v\alpha\epsilon}_{i_Ai_E} = 0$ ensure the identifiability of the aforementioned effects. Related null hypotheses are:
	\begin{itemize}
		\item $\mathcal H_{0,C^v}^{A} : \{\mbf{H}_A\mbf{C}^v = \mbf{0}\}  = \{C^{v\alpha}_{i_A} = 0\,\, \forall i_A\}$, $\mbf{H}_A= \mbf{P}_a\otimes (\mbf{1}_{1\times e}/e)$ (no main effect $A$).
		
		\item $\mathcal H_{0,C^v}^{E} : \{\mbf{H}_E\mbf{C}^v = \mbf{0}\}  = \{C^{v\epsilon}_{i_E} = 0\,\, \forall i_E\}$, $\mbf{H}_E= (\mbf{1}_{1\times a}/a) \otimes \mbf{P}_e$ (no main effect $E$).
		
		\item $\mathcal H_{0,C^v}^{AE} : \{\mbf{H}_{AE}\mbf{C}^v = \mbf{0}\}  = \{C^{v\alpha\epsilon}_{i_Ai_E} = 0\,\, \forall i_A,i_E\}$, $\mbf{H}_{AE}= \mbf{P}_a \otimes \mbf{P}_e$ (no interaction). 
	\end{itemize}
	Hereby, $\otimes$ is the Kronecker product. Clearly, we can replace $C^v$ by $B^v$ to get respective null hypotheses for the latter. The described strategy can be extended, in a straightforward manner, to higher-way layouts and hierarchical designs with nested factors, see e.g. Sec. S1.3 of the supplement from \cite{ditzhaus2021casanova} or Sec. 4 of \cite{paulyETAL2015}.
	
	\subsection{Wald-type tests}\label{sec:Waldtest}

	How to test global null hypothesis of the form \eqref{eqn:null} is discussed in several papers. In general, quadratic forms are built on the estimating vectors, here $\mbf{\widehat C}^v$ or $\mbf{\widehat B}^v$. Popular examples are (modified) ANOVA-type statistics \citep[e.g.][]{brunner:dette:munk:1997,friedrich2018mats,sattler2022testing} and Wald-type statistics \citep[e.g.][]{paulyETAL2015,smaga2015,smaga2017,ditzhaus2021casanova}. In particular, the latter is usually an asymptotically pivotal statistic which is beneficial for the resampling procedures proposed later. For the results, we require the following classical assumption of non-vanishing groups
	\begin{align}\label{eqn:non_vanish_groups}
		\frac{n_i}{n} \to \kappa_i \in (0,1) \text{ for all }i =1,\ldots,k.
	\end{align} 
	This limit and all following ones are meant as the total sample size $n=\sum_{i=1}^kn_i$ tends to $\infty$. This is in line with the prior convention from Section~\ref{sec:setup}, where we considered only one group, i.e. $n=n_1$. Now, we can formulate and motivate the Wald-type statistics for \eqref{eqn:null}:
	\begin{align*}
		&S_{n,C^v}(\mbf{H}) = n (\mbf{H}\mbf{\widehat C}^v)^{\top} (\mbf{H}\mbf{\widehat \Sigma}_{C^v} \mbf{H}^{\top} )^+\mbf{H}\mbf{\widehat C}^v,\ S_{n,{B}^v}(\mbf{H}) = n (\mbf{H}\mbf{\widehat {B}}^v)^{\top} (\mbf{H}\mbf{\widehat \Sigma}_{B^v} \mbf{H}^{\top} )^+\mbf{H}\mbf{\widehat {B}}^v,
	\end{align*}
	where $\mbf{\widehat\Sigma}_{C^v} = \text{diag}((n/n_1)\widehat\sigma^2_{1,C^v},\ldots,(n/n_k)\widehat\sigma^2_{k,C^v})$ and $\mbf{\widehat\Sigma}_{B^v} = \text{diag}((n/n_1)\widehat\sigma^2_{1,{B^v}},\ldots)$, and $\mbf{A}^+$ denotes the Moore–Penrose inverse. By Lemma~\ref{lem:const_var_est}, these estimators are consistent for $\mbf{\Sigma}_{C^v} = \text{diag}(\kappa_1^{-1}\sigma^2_{1,C^v},\ldots,\kappa_k^{-1}\sigma^2_{k,C^v})\ \ $ and $\mbf{\Sigma}_{B^v} = \text{diag}(\kappa_1^{-1}\sigma^2_{1,{B^v}},\ldots)$. Whenever Assumption~\ref{ass:two_point} holds for all $i$, the limiting covariance matrices $\mbf{\Sigma}_{C^v}$ and $\mbf{\Sigma}_{B^v}$ are regular. Finally, we can deduce from Theorem~\ref{theo:conv_C+beta}, Lemma~\ref{lem:const_var_est} and Theorem 9.2.2 of \cite{raoMitra1971} that the limits of $S_{n,C^v}(\mbf{H})$ and $S_{n,B^v}(\mbf{H})$ under $\mathcal H_{0,C^v}$ and $\mathcal H_{0,B^v}$, respectively, are chi-squared distributed with rank$(\mbf{H})$ degrees of freedom. Moreover, the same arguments yield that $n^{-1}S_{n,C^v}(\mbf{H})$ and $n^{-1}S_{n,B^v}(\mbf{H})$ always converge in probability to $(\mbf{H}\mbf{ C}^v)^{\top} (\mbf{H}\mbf{\Sigma}_{C^v} \mbf{H}^{\top} )^+\mbf{H}\mbf{ C}^v$ and $(\mbf{H}\mbf{{B}}^v)^{\top} (\mbf{H}\mbf{\Sigma}_{B^v} \mbf{H}^{\top} )^+\mbf{H}\mbf{ {B}}^v$. In the proofs, we show that these limits are positive under alternatives $\mathcal H_{1,C^v}: \mbf{H}\mbf{C}^v \neq \mbf{0}$ or $\mathcal H_{1,B^v}: \mbf{H}\mbf{B}^v \neq \mbf{0}$.
	
	\begin{theorem}\label{theo:wts_convergence}
		Let \eqref{eqn:non_vanish_groups} as well as Assumptions~\ref{ass:well_defined} and \ref{ass:two_point} be fulfilled for all subgroups $i$.
		\begin{enumerate}
			\item[(i)] Under $\mathcal H_{0,C^v}: \mbf{H}\mbf{C}^v = \mbf{0}$, $S_{n,C^v}(\mbf{H})$ tends in distribution to $Z\sim \chi^2_{\text{rank}(\mbf{H})}$.
			
			\item[(ii)] Under $\mathcal H_{1,C^v}: \mbf{H}\mbf{C}^v \neq \mbf{0}$, $S_{n,C^v}(\mbf{H})$ diverges, i.e. $S_{n,C^v}(\mbf{H})\overset{p}{\rightarrow}\infty$.
			
			\item[(iii)] Under $\mathcal H_{0,{B}^v}: \mbf{H}\mbf{B}^v = \mbf{0}$, $S_{n,B^v}(\mbf{H})$ tends in distribution to $Z\sim \chi^2_{\text{rank}(\mbf{H})}$.
			
			\item[(iv)] Under $\mathcal H_{1,B^v}: \mbf{H}\mbf{B}^v \neq \mbf{0}$, $S_{n,B^v}(\mbf{H})$ diverges, i.e. $S_{n,B^v}(\mbf{H})\overset{p}{\rightarrow}\infty$.
		\end{enumerate}
	\end{theorem}
	
	As a result of Theorem~\ref{theo:wts_convergence}, we obtain asymptotically valid tests $\varphi_{n,C^v}=\mathbf{1}\{S_{n,C^v}(\mbf{H}) > \chi^2_{\text{rank}(\mbf{H}),1-\alpha}\}$  for the testing problems $\mathcal H_{0,C^v}$ vs. $\mathcal H_{1,C^v}$, i.e. they have an asymptotic level $\alpha$ and an asymptotic power of $1$, similarly for $B^v$. Here, $\chi^2_{\text{rank}(\mbf{H}),1-\alpha}$ denotes the $(1-\alpha)$-quantile of a chi-square distribution with rank($\mbf{H})$ degrees of freedom. 
	
	The convergences rate of Wald-type statistics is known to be rather slow. The simulations of \cite{ditzhaus_smaga_2022} confirmed this general impression for the specific variant $v=\text{VN}$. The new simulation study in Section~\ref{sec:simulation} underpins that for the other variants. To tackle this problem, we follow a studentized permutation strategy. Moreover, we also consider a (pooled) bootstrap test, which is of particular interest for local null hypotheses testing, see Section~\ref{sec:multiple}.

	\subsection{Permutation and bootstrapping}\label{sec:perm+boot}
	
	Resampling procedures are popular and well-accepted tools to improve the tests' performance and, in particular, their control of the type-1 error. Due to our good experience with permutation procedures for Wald-type statistics \citep{ditzhausETAL2019,ditzhaus_smaga_2022,smaga2015,smaga2017}, we propose to follow this successful and powerful strategy also for the underlying problem. A remarkable advantage of permuting over other resampling strategies is its finite exactness under exchangeability, i.e. under the more restrictive null hypothesis $\widetilde {\mathcal H}_0: \mathbf{X}_{11} \overset{d}{=} \ldots \overset{d}{=} \mathbf{X}_{k1}$. For more general (potentially nonexchangeable) null hypotheses, it is not clear whether the permutation strategy leads indeed  to a valid testing procedure. But in case of studentized statistics, as Wald-type statistics, this desirable validity was proven in various other settings and we provide a proof in the underlying set-up. The (pooled) bootstrap, i.e. drawing from the pooled data with replacement, is closely related to the permutation approach. We consider it here as well because the permutation strategy is not appropriate for simultaneous testing as discussed in Section~\ref{sec:multiple}. 
	
	Let us become more specific. We first group all data together and denote the resulting pooled data by $\mbf{X}=(\mbf{X}_{ij})_{i=1,\ldots,k;j=1,\ldots,n_i}$. Now, we draw with or without replacement from $\mbf{X}$ to obtain a permutation ($\mbf{X}^\pi=(\mbf{X}_{ij}^\pi)_{i=1,\ldots,k;j=1,\ldots,n_i}$) or a bootstrap ($\mbf{X}^b = (\mbf{X}_{ij}^b)_{i=1,\ldots,k;j=1,\ldots,n_i}$) sample, respectively. Since no observation can be drawn twice for $\mbf{X}^\pi$, the permuted observations mutually depend from each. In contrast to that, we draw for the bootstrap sample with replacement and, hence, some individuals appear multiple times and some do not appear at all in $\mbf{X}^b$. In particular, the bootstrap observations are independent from each other as the original observations. In Section~\ref{sec:multiple}, we explain why this property is beneficial for the simultaneous testing and why the permutation fails there.
	
	To differentiate between the quantities from the previous Sections for the different samples, we add the superscript $^\pi$ or $^b$ to them when they rely on the permutation or bootstrap sample. For example, $S_{n,C^v}^\pi(\mbf{H})$ denotes the permuted test statistic. Since we draw from the pooled data, the assumptions need to be translated from the specific groups to the pooled distribution. Therefore, we introduce the expectation ${\mbf{\mu}_0} = \sum_{i=1}^k \kappa_i \mbf{\mu}_i$ and the covariance matrix ${\mbf{\Sigma}}_0$ for the (asymptotic) pooled distribution $P_0 = \sum_{i=1}^k\kappa_iP^{\mbf{X}_{i1}}$, where the matrix is given by its entries $[\mbf{\Sigma}_0]_{\ell m} = (\sum_{i=1}^k \kappa_i E(X_{\ell 1}X_{m 1})) - [\mbf{\mu}_0]_\ell [\mbf{\mu}_0]_m$. It is easy to check, e.g. via projections, that Assumption~\ref{ass:two_point} is true for the pooled distribution when this is the case for all (sub-)groups $i$. Consequently, we just need, in addition to the conditions from Theorem~\ref{theo:conv_C+beta}, that Assumption~\ref{ass:well_defined} is fulfilled for the pooled quantities:
	
	\begin{theorem}\label{theo:perm+bootstrap}
		In addition to the assumptions of Theorem~\ref{theo:conv_C+beta}, we suppose that  Assumption~\ref{ass:well_defined} is fulfilled for the pooled quantities $\mbf{\mu}_0$ and $\mbf{\Sigma}_0$. Then the following statements are valid under the null hypotheses $\mathcal H_{0,C^v}$, $\mathcal H_{0,B^v}$ and their respective alternatives:
		\begin{enumerate}[(a)]
			\item\label{enu:theo:perm} The permutation statistics $S_{n,C^v}^\pi(\mbf{H})$ and $S_{n,B^v}^\pi(\mbf{H})$ always mimic the null distribution limit of $S_{n,C^v(\mbf{H})}$ and $S_{n,B^v}(\mbf{H})$ asymptotically. In formulas (exemplarily for $C^v$): 
			\begin{align*}
				\sup_{x\in\mathbb{R}} \Big| \Pr\Big(S_{n,C^v}^\pi(\mbf{H}) \leq x \mid \mbf{X} \Big) - \chi^2_{\text{rank}(\mbf{H})}(x) \Big| \overset{p}{\rightarrow} 0.
			\end{align*}
			
			\item\label{enu:theo:boot} The statement in \eqref{enu:theo:perm} is also true for the bootstrap statistics $S_{n,C^v}^b(\mbf{H})$ and $S_{n,B^v}^b(\mbf{H})$.
		\end{enumerate}
	\end{theorem}
	
	Theorem~\ref{theo:perm+bootstrap} justifies the validity of the permutation and bootstrap method. To accept this, let $q_{n,1-\alpha,C^v}^\pi(\mbf{X})$ be the $(1-\alpha)$-quantile of the permutation distribution $\R\ni x \mapsto \Pr(S_{n,C^v}^\pi(\mbf{H}) \leq x \mid \mbf{X})$. Then Theorem~\ref{theo:perm+bootstrap} ensures that $q_{n,1-\alpha,C^v}^\pi(\mbf{X})$ approximates always (!) the quantile $\chi^2_{\text{rank}(\mbf{H}),1-\alpha}$ of the test $\varphi_{n,C^v}$ from Section~\ref{sec:Waldtest}. Consequently, the asymptotic properties of $\varphi_{n,C^v}$, namely asymptotic exactness under the null and consistency for all alternatives, can be transferred to its permutation counterpart $\varphi_{n,C^v}^\pi=\mathbf{1}\{S_{n,C^v}(\mbf{H}) > q_{n,1-\alpha,C^v}^\pi(\mbf{X})\}$ \citep[cf.][Lem. 1 and Theo. 7]{janssenPauls2003}. Clearly, the same is true when we consider ${B}$ instead of ${C}$ and/or the bootstrap instead of the permutation method. 
	
	\section{Multiple testing}\label{sec:multiple}
	\subsection{Local null hypotheses and the multiple contrast tests}
	While testing for main and interaction effects in factorial designs is of high interest, often a more in-depth analysis is wanted to check which part of the equation systems $\mbf{H}\mbf{C}^v = \mbf{0}$ or $\mbf{H}\mbf{B}^v = \mbf{0}$, respectively, is not true. This leads to the multiple testing problem
	\begin{align}\label{eqn:local_nulls}
		\mathcal H_{0,\ell, C^v}: \mbf{h}_\ell^\top \mbf{C}^v = 0 \qquad (\ell=1, \ldots, r)
	\end{align}
	for contrast vectors $\mbf{h}_\ell\in\R^k$, i.e. $\mbf{h}_\ell^\top \mbf{1}_{k\times 1} = 0$. The intersection $\bigcap_{\ell=1}^r\mathcal H_{0,\ell, C^v}$ of the local null hypotheses coincides with the global null hypothesis $\mathcal H_{0,C^v}$ from \eqref{eqn:null} with $\mbf{H} = (\mbf{h}_1,\ldots,\mbf{h}_r)^\top$. In the easiest case, we are interested in group differences, i.e. the global null hypotheses is $\mathcal H_{0,C^v}: C_1^v=\ldots=C_k^v$. Prominent examples to split the latter into local null hypotheses are Tukey's all-pairs comparison \citep{tukey1953problem}
	\begin{align*}
		\mathcal H_{0,C^v}: \begin{cases} C_1^v = C_2^v \\
			C_1^v = C_3^v \\
			\vdots \\
			C_1^v=C_k^v\\
			C_2^v = C_3^v \\
			\vdots\\
			C_{k-1}^v = C_k^v
		\end{cases}
		\Leftrightarrow \,\,
		\mathcal H_{0,C^v}: \begin{pmatrix} -1 & 1 & 0 & \ldots & \ldots & 0 & 0\\			
			-1 & 0 & 1 & 0 & \ldots & \ldots & 0 \\
			\vdots & \vdots & \vdots & \vdots & \vdots & \vdots & \vdots \\
			-1 & 0 & 0 & 0 & \ldots  & \ldots & 1 \\
			0 & -1 & 1 & 0 &\ldots & \ldots & 0 \\
			\vdots & \vdots & \vdots & \vdots & \vdots & \vdots & \vdots \\
			0 & \ldots & \ldots & \ldots & 0 & -1 &  1\\
		\end{pmatrix} \mbf{C}^v = \mathbf{0}		 
	\end{align*}
	or Dunnet's multiple-to-one comparison \citep{dunnett1955multiple}
	\begin{align*}
		\mathcal H_{0,C^v}: \begin{cases} C_1^v = C_2^v \\
			C_1^v = C_3^v \\
			\vdots \\
			C_1^v=C_k^v
		\end{cases}
		\Leftrightarrow \,
		\mathcal H_{0,C^v}: \begin{pmatrix} -1 & 1 & 0 & \ldots & \ldots  & 0\\			
			-1 & 0 & 1 & 0 & \ldots  & 0 \\
			\vdots & \vdots & \vdots & \vdots & \vdots & \vdots  \\
			-1 & 0 & \ldots & \ldots & 0  &  1\\
		\end{pmatrix} \mbf{C}^v = \mathbf{0}.	 
	\end{align*}
	Further proposals can be found in \cite{bretz2001numerical}. In principle, we can consider different strategies from multiple testing, e.g. the Bonferroni or Holm correction, to adjust the type-1 errors. However, this leads typically to a significant power loss. A more promising approach are multiple contrast tests \citep{bretz2001numerical,hothorn2008simultaneous,gunawardana2019nonparametric}. For them, we  test the single null hypotheses $\mathcal H_{0,\ell,C^v}$ by $S_{n,C^v}(\mbf{h}_\ell^\top)$ from Section~\ref{sec:Waldtest} and then incorporate the explicit dependence structure of them to obtain a valid testing procedure. In detail, we consider the following max-type statistic
	\begin{align*}
		S_{n,\text{max},C^v}(\mbf{H}) = \max_{\ell=1,\ldots,r} (S_{n,C^v}(\mbf{h}_\ell^\top))^{1/2} = \max_{\ell=1,\ldots,r} |T_{\ell,n}^v|,\qquad T_{\ell,n}^v = \sqrt{n}\frac{ \mbf{h}_\ell^\top \widehat{\mbf{C}}^v}{ \sqrt{\mbf{h}_\ell^\top \widehat{\mbf{\Sigma}}_{C^v} \mbf{h}_\ell}}.
	\end{align*}
	By Theorem~\ref{theo:conv_C+beta}, $(T_{1,n}^v,\ldots,T_{r,n}^v)$ converges in distribution to a multivariate normal distribution with standard normal distributed marginals and correlation matrix $\mbf{R}_{C^v}$ given by 
	\begin{align}\label{eqn:R}
		[\mbf{R}_{C^v}]_{\ell m} = \frac{ \mbf{h}_\ell^\top \mbf{\Sigma}_{C^v} \mbf{h}_m}{\sqrt{\mbf{h}_\ell^\top {\mbf{\Sigma}}_{C^v} \mbf{h}_\ell}\sqrt{\mbf{h}_m^\top {\mbf{\Sigma}}_{C^v} \mbf{h}_m}}.
	\end{align}
	The studentization of the $T_{\ell,n}^v$'s, i.e. dividision by an estimator for the asymptotic variance, ensures that each null hypothesis is (asymptotically) treated in the same way. Thus, the equicoordinate $(1-\alpha)$-quantile $q_{1-\alpha,\text{max},C^v}(\mbf{R})$ of a $N(\mbf{0},\mbf{R}_{C^v})$-distribution serves as a ``fair'' critical value. Such quantiles can be determined numerically by computer software, e.g. the function \textit{qmvnorm()} from the R-package \textit{mvtnorm} \citep{Rcore, Genzetal2021, GenzBretz2009}. In practice, $\mbf{R}_{C^v}$ needs to be estimated by $\widehat{\mbf{R}}_{C^v}$, where we replace $\mbf{\Sigma}_{C^v}$ in \eqref{eqn:R} by its estimator $\widehat{\mbf{\Sigma}}_{C^v}$. In summary, we obtain an asymptotically exact test $\varphi_{n,\text{max},C^v} = \mathbf{1}\{S_{n,\text{max},C^v}(\mbf{H}) > q_{1-\alpha,\text{max},C^v}(\widehat{\mbf{R}}_{C^v})\}$ for the global null hypothesis $\mathcal H_{0,C^v}$. In contrast to the test from Section~\ref{sec:Waldtest}, this test also provides additional information in case of a rejection, namely which local null hypotheses \eqref{eqn:local_nulls} caused this rejection. In detail, we reject the local null hypothesis $\mathcal H_{0,\ell,C^v}$ when $T_{\ell,n}^n>q_{1-\alpha,\text{max},C^v}(\widehat{\mbf{R}}_{C^v})$. Moreover, multiple max-type contrast tests can be inverted to obtain simultaneous confidence intervals for all contrasts $\mbf{h}_\ell^\top \mbf{C}^v$. All these theoretical properties are summarized in the following theorem:

	\begin{theorem}\label{theo:multi}
		Let \eqref{eqn:non_vanish_groups} as well as Assumptions~\ref{ass:well_defined} and \ref{ass:two_point} be fulfilled for all (sub-)groups $i$.
		\begin{enumerate}[(a)]
			\item\label{enu:theo:multi_exact} The test $\varphi_{n,\text{max},C^v}$ is asymptotically exact for $\mathcal H_{0,C^v}$, i.e. $E_{\mathcal H_{0,C^v}}(\varphi_{n,\text{max},C^v})\to\alpha$. 
			
			\item\label{enu:theo:multi_r'} Suppose that the first $r'\leq r$ null hypotheses and the remaining $r-r'$ alternatives, i.e. $\mathcal H_{1,\ell,C^v}:  \mbf{h}_\ell^\top \mbf{C}^v \neq 0$ for $\ell=r'+1,\ldots,r$, are true. Then
			\begin{align*}
				&\limsup_{n\to\infty} \Pr \Bigl( \bigcup_{\ell=1}^{\:r'}\{|T_{\ell,n}^v| > q_{1-\alpha,\text{max},C^v}(\widehat{\mbf{R}}_{C^v}) \} \Bigr) \leq \alpha \quad\\ \text{and}\quad &
				\lim_{n\to\infty} \Pr \Bigl( \bigcap_{\ell=r'+1}^{r}\{|T_{\ell,n}^v| > q_{1-\alpha,\text{max},C^v}(\widehat{\mbf{R}}_{C^v}) \} \Bigr) = 1.
			\end{align*}
			
			\item\label{enu:theo:multi_SCI} An asymptotically valid simultaneous confidence interval for $\mbf{h}_\ell^\top \mbf{C}^v$ is given by 
			\begin{align*}
				\Pr\Bigl( \bigcap_{\ell=1}^r\Bigl\{ \mbf{h}_\ell^\top \widehat{\mbf{C}}^v \in \Big[ \mbf{h}_\ell^\top \mbf{C}^v \pm n^{-1/2}\sqrt{\mbf{h}_\ell^\top \widehat{\mbf{\Sigma}}_{C^v}\mbf{h}_\ell}\, q_{1-\alpha,\text{max},C^v}(\widehat{\mbf{R}}_{C^v}) \Big] \Bigr\} \Bigr) \to 1-\alpha.
			\end{align*}
			
			\item\label{enu:theo:multi_B} All statements are true for $\mbf{B}^v$ instead of $\mbf{C}^v$ when adjusting the estimators properly.
		\end{enumerate}
	\end{theorem}
	
	For a performance improvement for small $n$, we modify the bootstrap from Section~\ref{sec:perm+boot}. 
	
	\subsection{Bootstrapping}\label{sec:boot_multple}
	
	It is not straightforward how or even whether the resampling strategies from Section~\ref{sec:perm+boot} can also be used for multiple contrast tests. Let us first have a closer look on the bootstrap strategy.  In the proofs, we show that given the data $\mbf{X}$ almost surely
	\begin{align*}
		&n^{1/2} \Bigl( \widehat C_1^{vb} - \widehat C_0^v,\ldots, \widehat C_k^{vb} - \widehat C_0^v \Bigr)^{\top} 
		\overset{ d}{\rightarrow} \mbf{G}_{C^v}^b \sim N(\mbf{0}_{k\times 1}, \mbf{\widetilde \Sigma}_{C^v}),
	\end{align*}
	where $\widehat C_0^v$ is the MCV estimator based on the pooled data and $\mbf{\widetilde \Sigma}_{C^v} =\text{diag}(\widetilde\sigma_{1,C^v}^{2},\ldots,\widetilde\sigma_{k,C^v}^{2})$. In general, the covariance matrices $\mbf{\widetilde \Sigma}_{C^v}$ and $\mbf{\Sigma}_{C^v}$ differ. That is why we cannot approximate the limiting null distribution of $(T_{1,n}^{v},\ldots,T_{r,n}^{v})$ by $(T_{1,n}^{v,b},\ldots,T_{r,n}^{v,b})$ directly. For the Wald-type statistic, we faced a similar problem and solved it by studentization, i.e. by eliminating
	the dependence of the limit distribution on the covariance structure $\mbf{\Sigma}_{C^v}$ or $\mbf{\Sigma}^b_{C^v}$, respectively. Translated to the present setting, we would   studentize $(T_{1,n}^v,\ldots,T_{r,n}^{v})$ first, and then taking the maximum of its entries. In formulas, we would end up with $\max_{\ell=1}^r [\widehat{\mbf{\Sigma}}_{C^v}^{-1/2}(T_{1,n}^v,\ldots,T_{r,n}^v)^\top]_\ell$. This is indeed a valid testing procedure for the global $\mathcal H_{0,C^v}$. But we cannot match the entries $[\widehat{\mbf{\Sigma}}_{C^v}^{-1/2}(T_{1,n}^v,\ldots,T_{r,n}^v)^\top]_\ell$ with the respective local null hypothesis $\mathcal H_{0,\ell,C^v}$ anymore. That is why we need to approximate $(T_{1,n}^v,\ldots,T_{r,n}^v)$ directly and not just a transformation of it. For this purpose, we can find different strategies for other testing problems in the literature, e.g. wild bootstrapping \citep[c.f.][]{umlauft2019wild,konietschke2021small} or group-wise bootstrapping \citep[c.f.][]{wechsung2021simultaneous}. However, we like to exemplify that the pooled bootstrap with a certain modified studentization can also be applied. In detail, we approximate $n^{1/2}(\widehat {\mbf{C}}^v - \mbf{C}^v)$ by
	\begin{align*}
		n^{1/2}\widehat{\boldsymbol{\Sigma}}_{C^v}^{1/2} (\widehat{\mbf{\Sigma}}_{C^v}^b)^{-1/2} ( \widehat{\mbf{C}}^{vb} - \widehat{\mbf{C}}^v_{0}),\qquad \widehat{\mbf{C}}^v_{0} = \widehat{C}_0^v \cdot\mbf{1}_{k\times 1}.
	\end{align*}
	In words, we first studentize $(\widehat {\mbf{C}}^{vb} - \widehat{\mbf{C}}_0^v)$ and then multiple the result by $\widehat{\boldsymbol{\Sigma}}_{C^v}^{1/2}$  leading  to the correct asymptotic covariance structure. The respective multiple contrast statistic becomes 
	\begin{align*}
		S^b_{n,\text{max},C^v}(\mbf{H}) = n^{1/2}\max_{\ell=1,\ldots, r} | T_{\ell,n}^{v,b} |,\qquad T_{\ell,n}^{v,b} = \frac{\mbf{h}_\ell^\top \widehat{\mbf{\Sigma}}_{C^v}^{1/2} (\widehat{ \mbf \Sigma}^b_{C^v})^{-1/2} ( \widehat{\mbf{C}}^{vb} - \widehat{\mbf{C}}^v_{0})}{ \sqrt{\mbf{h}_\ell^\top \widehat{\mbf \Sigma}_{C^v}\mbf{h}_\ell}}.	
	\end{align*}
	Now, let $q_{1-\alpha,\max,C^v}^b(\mbf{X})$ be the conditional, equicoordinate $(1-\alpha)$-quantile of $n^{1/2}(T_{1,n}^{v,b},\ldots,T_{r,n}^{v,b})^\top$ given the data $\mbf{X}$ and $\varphi_{n,\text{max},C^v}^b = \mathbf{1}\{S_{n,\text{max},C^v}(\mbf{H}) > q_{1-\alpha,\text{max},C^v}^b(\mbf{X})\}$ be the bootstrap multiple contrast test. Then we can transfer indeed all asymptotic properties from $\varphi_{n,\text{max},C^v}$ to $\varphi_{n,\text{max},C^v}^b$, and, moreover, obtain bootstrap-based simultaneous confidence intervals:
	
	\begin{theorem}\label{theo:multi_bootstrap}
		In addition to the assumptions of Theorem~\ref{theo:conv_C+beta}, we suppose that  Assumption~\ref{ass:well_defined} is fulfilled for the pooled quantities $\mbf{\mu}_0$ and $\mbf{\Sigma}_0$.
		Then the statements of Theorem~\ref{theo:multi}\ref{enu:theo:multi_exact}--\ref{enu:theo:multi_SCI} remain true when we replace $\varphi_{n,\text{max},C^v}$ and $q_{1-\alpha,\text{max},C^v}(\widehat{\mbf{R}}_{C^v})$ by their bootstrap counterparts $\varphi_{n,\text{max},C^v}^b$ and $q_{1-\alpha,\text{max},C^v}^b(\mbf{X})$, respectively. The analogue for $B$ is true.
	\end{theorem}
	
	At a first glance, a similar result might also be reachable for the permutation procedure but it is not. The asymptotic covariance structure of $n^{1/2}(\widehat{\mbf{C}}^{v\pi}-\widehat{\mbf{C}}_0^v)$ is more complicated than of the bootstrap one due to the strong dependence within the permutation sample. In particular, the permutation covariance matrix is neither diagonal nor regular.

	\section{Simulation study}
	\label{sec:simulation}
	
	Complementing the theoretical findings, we conducted an extensive simulation study to investigate the type-1 error level and power of the 40 tests proposed in the previous sections:
	
	
	\begin{itemize}
		\item the sixteen asymptotic tests: $\varphi_{C^{\text{RR}}}$, $\varphi_{B^{\text{RR}}}$, $\varphi_{C^{\text{VV}}}$, $\varphi_{B^{\text{VV}}}$, $\varphi_{C^{\text{VN}}}$, $\varphi_{B^{\text{VN}}}$, $\varphi_{C^{\text{\text{AZ}}}}$, $\varphi_{B^{\text{\text{AZ}}}}$, $\varphi_{\max,C^{\text{RR}}}$, $\varphi_{\max,B^{\text{RR}}}$, $\varphi_{\max,C^{\text{VV}}}$, $\varphi_{\max,B^{\text{VV}}}$, $\varphi_{\max,C^{\text{VN}}}$, $\varphi_{\max,B^{\text{VN}}}$, $\varphi_{\max,C^{\text{\text{AZ}}}}$, $\varphi_{\max,B^{\text{\text{AZ}}}}$,
		
		\item the eight permutation tests: $\varphi_{C^{\text{RR}}}^\pi$, $\varphi_{B^{\text{RR}}}^\pi$, $\varphi_{C^{\text{VV}}}^\pi$, $\varphi_{B^{\text{VV}}}^\pi$, $\varphi_{C^{\text{VN}}}^\pi$, $\varphi_{B^{\text{VN}}}^\pi$, $\varphi_{C^{\text{\text{AZ}}}}^\pi$, $\varphi_{B^{\text{AZ}}}^\pi$, 
		\item the sixteen bootstrap tests: $\varphi_{C^{\text{RR}}}^b$, $\varphi_{B^{\text{RR}}}^b$, $\varphi_{C^{\text{VV}}}^b$, $\varphi_{B^{\text{VV}}}^b$, $\varphi_{C^{\text{VN}}}^b$, $\varphi_{B^{\text{VN}}}^b$, $\varphi_{C^{\text{AZ}}}^b$, $\varphi_{B^{\text{AZ}}}^b$, $\varphi_{\max,C^{\text{RR}}}^b$, $\varphi_{\max,B^{\text{RR}}}^b$, $\varphi_{\max,C^{\text{VV}}}^b$, $\varphi_{\max,B^{\text{VV}}}^b$, $\varphi_{\max,C^{\text{VN}}}^b$, $\varphi_{\max,B^{\text{VN}}}^b$, $\varphi_{\max,C^{\text{AZ}}}^b$, $\varphi_{\max,B^{\text{AZ}}}^b$.
	\end{itemize}
	We omit the subscript $n$ here for the sake of clarity. For the multiple contrast tests, we use the Tukey's all-pairs comparison (see Section~\ref{sec:multiple}). The final simulation results shall serve as a first guideline for practical use of the R-package GFDmcv consisting of all these methods.  
	
	\subsection{Simulation setup}
	
	We considered inferring the null hypotheses in \eqref{eqn:null} in a multivariate one-way layout. The 5-dimensional data ($d=5$) were generated for $k=4$ groups. The mean vector was generated once from the normal distribution $\mathcal{N}(0,1)$, and $\mbf{\mu}_1,\dots,\mbf{\mu}_k$ were set to this vector. The covariance matrices $\mbf{\Sigma}_i$ were based on the completely symmetric matrix $(1-\rho)\mbf{I}_d+\rho \mbf{1}_d\mbf{1}_d^{\top}$. We considered $\rho=0.1,0.4,0.7$ for small, moderate, and large correlation, respectively. Of course, the MCVs usually measure the variability differently, see Section~\ref{sec:intro} or \cite{AlbertZhang2010} for a more detailed discussion on the differences. Thus, to compare the tests' properties in a unified way, we set the same values of all $C_i^v$, $v\in\{\text{RR},\text{VV},\text{VN},\text{AZ}\}$, i.e. for given $i=1,\dots,k$, $C_i^{\text{RR}}=C_i^{\text{VV}}=C_i^{\text{VN}}=C_i^{\text{AZ}}$. To obtain this, the matrices $\mbf{\Sigma}_i$ were multiplied by appropriate constants $a_i^v$, dependent on $v$. The observations were generated from three distributions: the normal ($\mathcal{N}$), the Student ($t_5$), and the chi-square ($\chi^2_{10}$), i.e. the symmetric, heavily tailed, and skewed distributions, respectively. For simplicity, we consider equal sample sizes in all groups. Namely, we set $n_1=\dots=n_k=30,50,70,100,150,200$ for the type-1 error control investigation, and $n_1=\dots=n_k=30,50$ for power comparison.
	
	The significance level was set to $\alpha=5\%$. For generating the data under $\mathcal{H}_{0,C^v}$, i.e. to investigate the type-1 error control, we set $C_i^v=0.1,0.5,1,1.5$ for $i$ and all variants $v$. For power comparison, we consider the following three alternative hypotheses:
	\begin{itemize}
		\item $\mathcal{H}_{1,C^v}^{(1)}:C_1^v=C_2^v=C_3^v=0.1\neq 0.15=C_4^v$;
		\item $\mathcal{H}_{1,C^v}^{(2)}:C_1^v=C_2^v=C_3^v=0.5\neq 0.7=C_4^v$; 
		\item $\mathcal{H}_{1,C^v}^{(3)}:C_1^v=C_2^v=C_3^v=1\neq 1.5=C_4^v$.
	\end{itemize}
	Empirical sizes and powers of the tests were computed as the proportion of rejections of the null hypothesis based on 1000 simulation replications. The $p$-values of the permutation and bootstrap tests were estimated by 1000 resampling samples. The simulation experiments and real data example of Section~\ref{sec:real_data_app} were performed in the R programming language \citep{Rcore}. Due to the twenty-four resampling tests and 270 simulation scenarios, a part of the calculations was made at the Poznań Supercomputing and Networking Center.
	
	\subsection{Simulation results}
	The simulation results are summarized by the box-and-whisker plots in Figures~\ref{fig_1}-\ref{fig_3} and Figures~S1-S13 in the supplementary materials. The complete list of empirical sizes and powers is presented in Tables~S3-S8 in the supplementary materials. First, we focus on the type-1 error level of the tests, and then we consider their power.
	
	\begin{figure}[!t]
		\centering
		\includegraphics[width=
		0.95\textwidth,height=0.3\textheight]{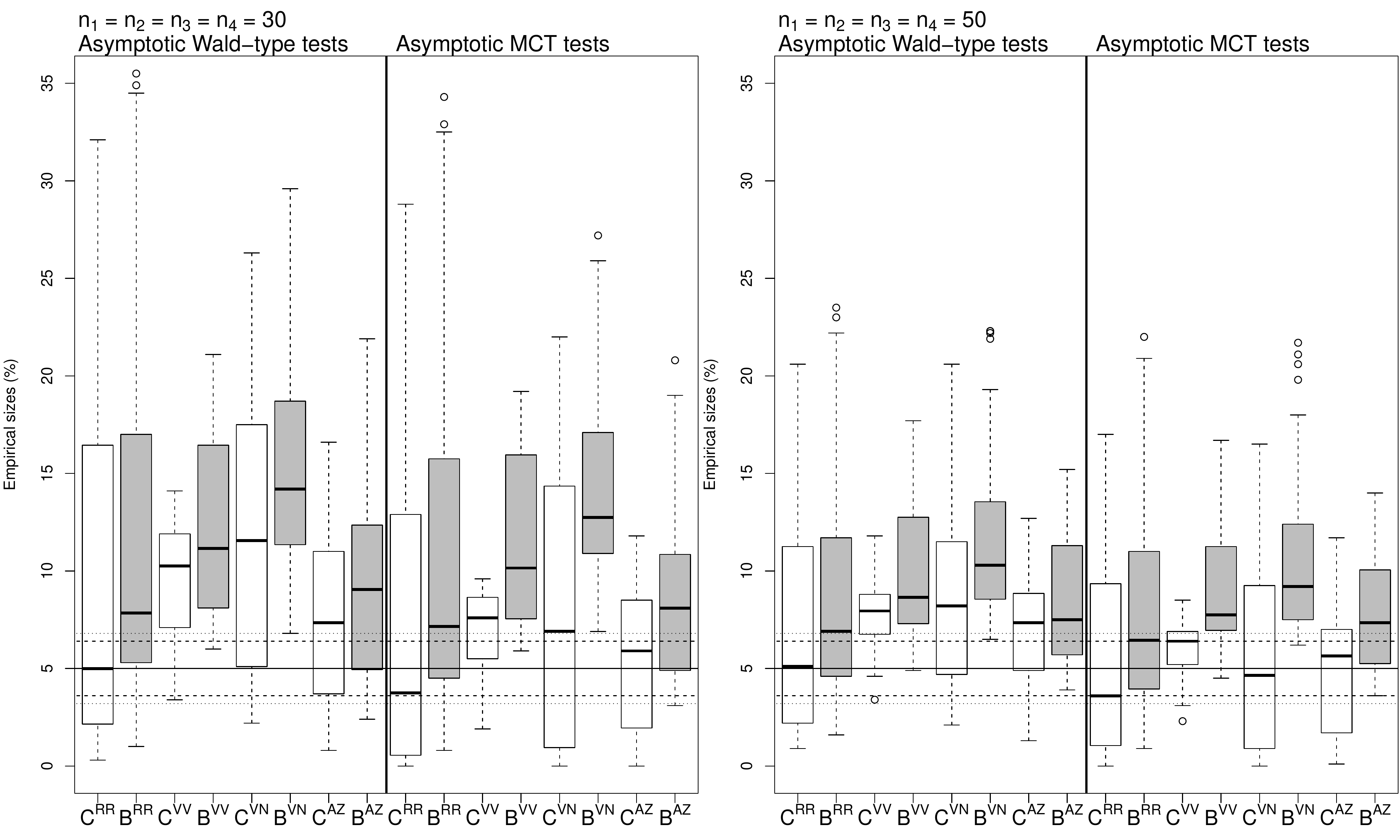}
		\caption{Box-and-whisker plots for the empirical sizes (in $\%$) of the asymptotic tests. The solid, dashed, and dotted lines represent the significance level $\alpha=5\%$ and the $95\%$ and $99\%$ binomial proportion confidence intervals $[3.6\%,6.4\%]$ and $[3.2\%,6.8\%]$, respectively.}
		\label{fig_1}
	\end{figure}
	
	\textit{Type-1 error level} For proper maintaining the type-1 error, the empirical sizes should belong to the binomial proportion 95\% and 99\% confidence intervals $[3.6\%,6.4\%]$ and $[3.2\%,6.8\%]$, respectively \citep{DuchesneFrancq2015}. In the figures, these limits are presented by the horizontal lines. It is apparent that the type-1 error control of the asymptotic tests is unstable and primarily lead to liberal decisions, i.e. their empirical sizes are (much) greater than $6.8\%$ in most cases (see Figure~\ref{fig_1} and Figure~S1 in the supplement). The record-breaking empirical sizes about $35\%$ are obtained for the $\varphi_{B^{\text{RR}}}$. For larger values of MCVs, the asymptotic tests tend to be quite conservative. Overall, the performance of the asymptotic tests improves with increasing sample sizes, but the converge speed is rather slow (see Figure~S1 in the supplement). The fastest convergence is observed for $\varphi_{\max,C^{\text{VV}}}$. 
	
	Contrary, the permutation and bootstrap Wald-type procedures from Section~\ref{sec:perm+boot} for the global null hypotheses perform very well (Figure~\ref{fig_2}) and control the type-1 error level accurately in the majority of settings. In particular, their empirical sizes are always smaller than the upper limits of the binomial proportion confidence intervals, even for small sample sizes. Only the bootstrap strategy based on the MCV $C^v$ lead partly to slight conservative decisions for $n_i=50,70$. Moreover, the boxs' for the permutation tests are slightly narrower than ones for bootstrapping in case of smaller sizes. This slight visible advantage of permuting can be explained by its finite exactness under exchangeability.
	
	Turning now to the bootstrap multiple tests, we can observe that the type-1 error control is still satisfactory for the standardized mean vectors $B$. However, the decisions based on the MCVs become very conservative reaching down to values below $1\%$. The latter type-1 error rates improve very slowly and for the largest sample size of $n_i=200$ the conservativness is still clearly present. In the supplement, all results are shown in tables and, in particular, a detailed comparison between the bootstrap and asymptotic multiple contrast test can be made. In summary, the bootstrap procedures converge much slower to the desired $5\%$-benchmark than the asymptotic test. But, under the small sample sizes, the bootstrap procedure is conservative while the asymptotic strategy is rather liberal, and, thus, only the first controls the type-1 error rate. Nevertheless, the overall performance of the bootstrap is not satisfactory, which opens the door for future research. Here, we like to point out that we already studied a group-wise bootstrap procedure, i.e. drawing just from the respective groups, a wild bootstrap procedure, a Bayesian bootstrap procedure, and a parametric bootstrap procedure without much more success. Despite the unsatisfactory results for the MCV, we like to highlight the good performance of this (pooled) bootstrap strategy for the standardized means $B^v$. To the best of our knowledge, this was the first time that such a pooled bootstrap procedure was applied for multiple contrast tests.

	
	\textit{Power} Let us now discuss the results of the power comparison. The resulting empirical powers are presented in Figure~\ref{fig_3} and Figures~S8-S13 in the supplement. Since the too liberal character of the asymptotic tests is unacceptable, we do not consider them in power investigation to avoid unfair comparisons and the potential of results' misinterpretation.
	
	First, let us consider the comparison for Wald-type and multiple contrast testing procedures for a given definition of MCV, since the conclusions presented below are the same for each $C^v$ and $B^v$, $v\in\{\text{RR},\text{VV},\text{VN},\text{AZ}\}$. From Figure~\ref{fig_3}, we can observe that the Wald-type permutation and bootstrap tests have very similar power, but the bootstrap ones seem to be slightly less powerful. The $\varphi_{\max,B^v}^b$ tests characterize similar power to their Wald-type counterparts. Unfortunately, the $\varphi_{\max,C^v}^b$ tests show their conservative character and have considerably smaller power than the other tests. In general, each test based on a given MCV is at least slightly less powerful than its analog based on the reciprocal of MCV. For each test, the power is similar under the normal and $\chi_{10}^2$ distributions, while for the $t_5$-distribution, it is much smaller (Figures~S8-S10). Thus, for the heavy-tailed distribution, the convergence may be slower. This is especially evident for the $\varphi_{\max,C^v}^b$ tests due to their extremely conservative character in such a case. Let us also notice that the power of all tests decreases with the increase of the value of the MCV, which was also observed for earlier methods for variability comparison proposed in \cite{AertsHaesbroeck2017}, \cite{ditzhaus_smaga_2022}, and \cite{PaulySmaga2020}.
	
	Now, we want to consider how the behavior of the tests depends on the MCV definition. We can observe that the power of all tests for different coefficients is not the same, which was expected due to various conceptions of measuring multivariate variability. The general observation is as follows: The tests based on the Van Valen coefficient are usually the most powerful and outperform the procedures for Rayment's MCV, which are followed by the tests for MCVs proposed by Albert and Zhang, and Voinov and Nikulin. We can also observe that, for different amounts of correlation, the power of most tests is stable, but sometimes greater differences appear for the Van Valen coefficient. A possible reason for the latter is that the VV variant does not take the dependence of variables into account.
	
	
	\begin{figure}[!h]
		\centering
		\includegraphics[width=
		0.95\textwidth,height=0.85\textheight]{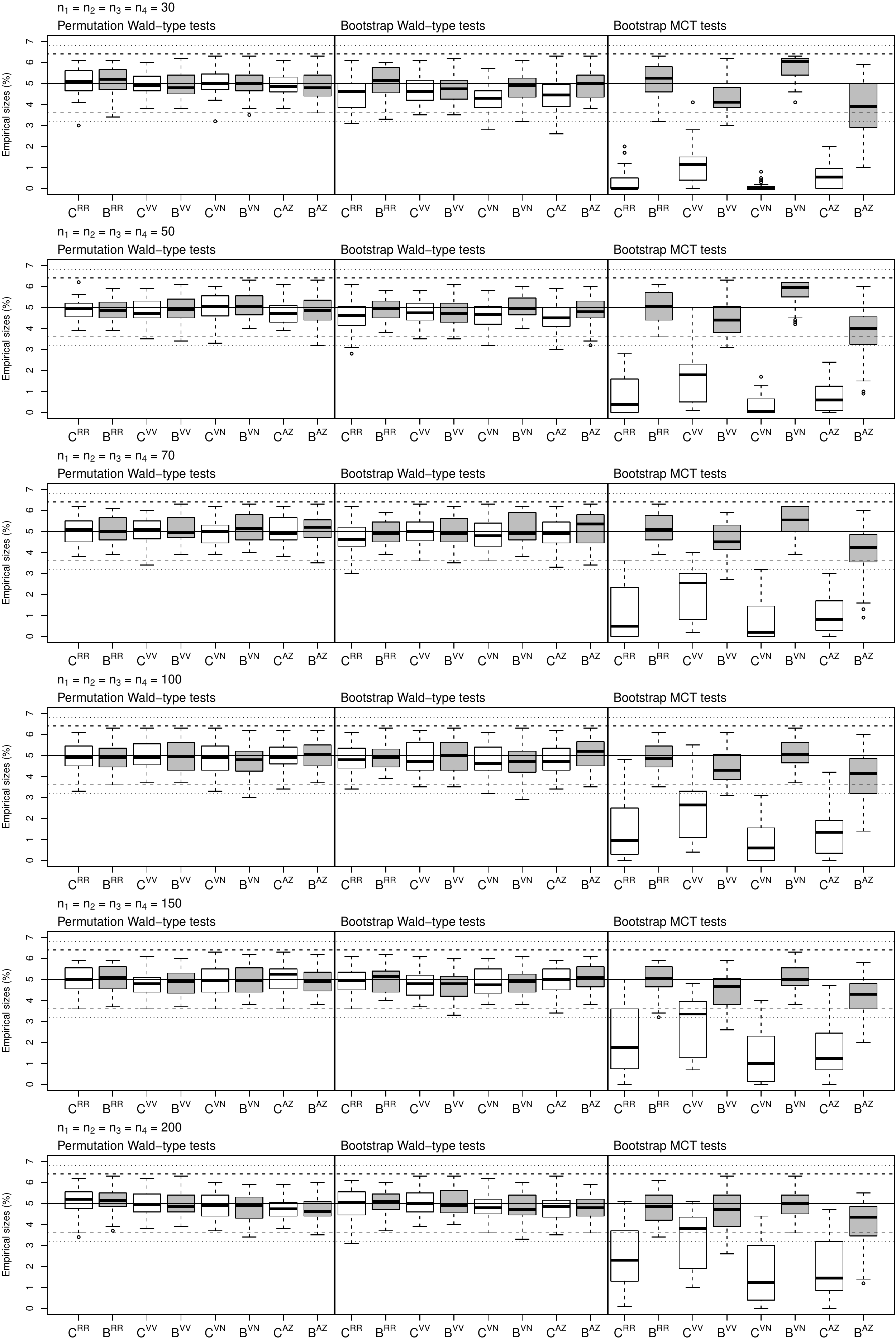}
		\caption{Box-and-whisker plots for the empirical sizes (as percentages) of the permutation and bootstrap tests obtained from all cases considered. The solid, dashed, and dotted lines represent the significance level $\alpha=5\%$ and the $95\%$ and $99\%$ binomial proportion confidence intervals $[3.6\%,6.4\%]$ and $[3.2\%,6.8\%]$, respectively.}
		\label{fig_2}
	\end{figure}

	\textit{Recommendation} To sum up, the Wald-type permutation and bootstrap tests as well as the multiple bootstrap tests based on reciprocals of MCVs control the type-1 error level even for small sample sizes. Moreover, these testing procedures have sensible power, which however depends on the coefficient used. For these reasons, these tests can be recommended for practical use, and thus, it is now available to infer about each of the four multivariate coefficients of variation. Note however that the advantage of the $\varphi_{\max,B^v}^b$ tests over the Wald-type tests is that they simultaneously verify the contrasts' significance, in particular, can perform post hoc testing. Unfortunately, the $\varphi_{\max,C^v}^b$ tests are very conservative and hence less powerful, which gives us a direction for future research.

	\begin{figure}[!t]
		\centering
		\includegraphics[width=
		0.95\textwidth,height=0.33\textheight]{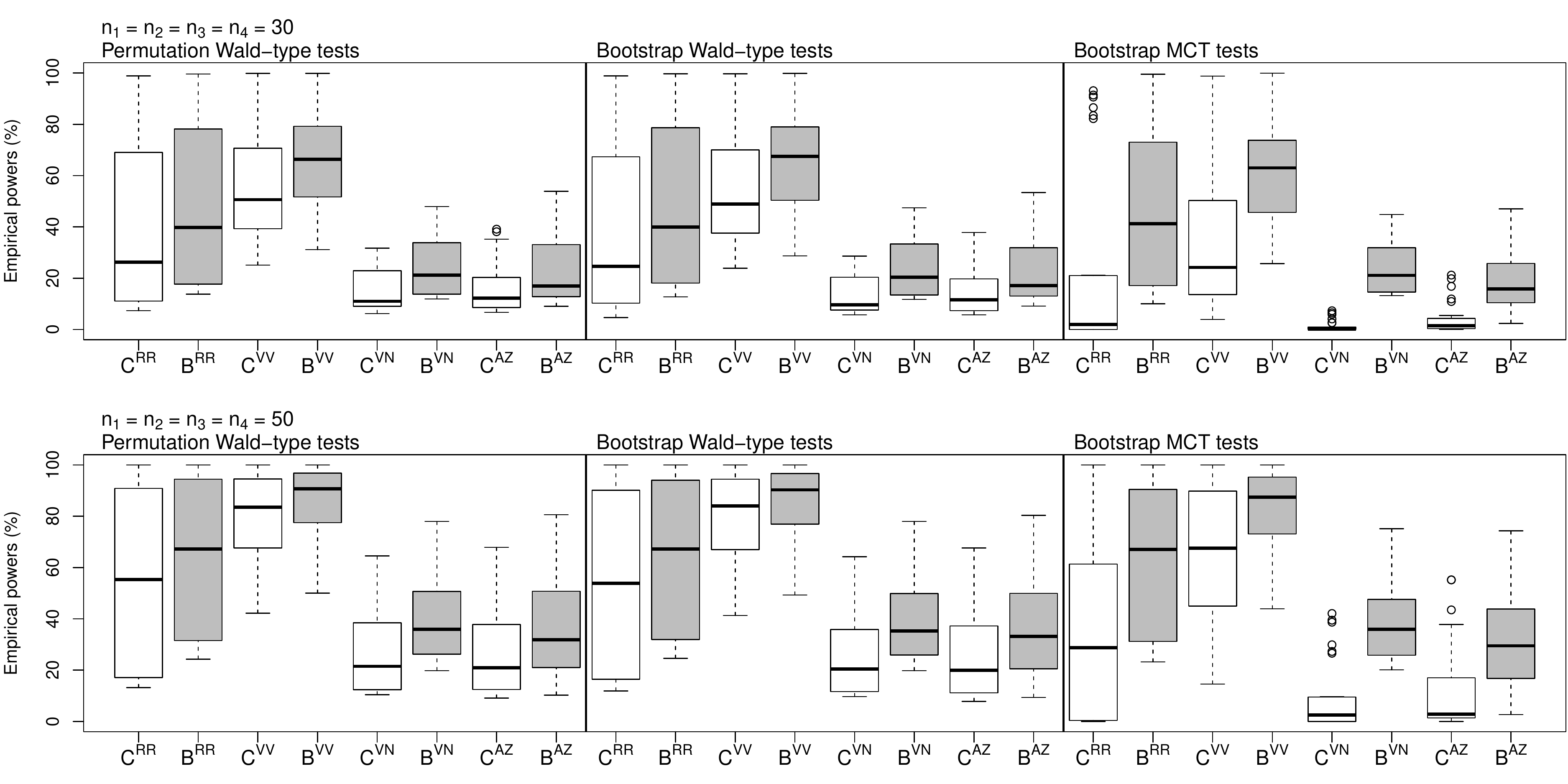}
		\caption{Box-and-whisker plots for the empirical powers (in $\%$) of the permutation and bootstrap tests obtained from all cases considered.}
		\label{fig_3}
	\end{figure}
	
	\section{Real data application}
	\label{sec:real_data_app}
	In this section, we consider, as an illustrative data example, the external quality assessment for clinical laboratories \citep{Libeer1993, Sciacovellietal2018, WHO2022}. Moreover, we present further simulation results, which mimic the set-up from this data example.
	
	\subsection{Analysis of EQA data set}
	In clinical laboratories, controlling analytical performance and maintaining inter-laboratory variability within acceptable limits are important issues that are even a concern in External Quality Assessment (EQA) schemes. They are organized nationally or internationally by government health agencies or private companies. In the case when the available data are multivariate, \cite{ZhangEtAl2010} proposed to use the multivariate coefficient of variation for comparing the inter-laboratory reproducibility of assay techniques used by clinical laboratories. Naturally, the lower the MCV, the better the analytical performance. However, the simple values of particular MCV do not mean significant differences in the techniques considered. That is why we illustrate the use of tests proposed for comparing four techniques based on electrophoretic data sets from the French and Belgian national EQA programs, which was also considered by \cite{ZhangEtAl2010} and \cite{AertsHaesbroeck2017}.
	
	The serum protein electrophoresis (SPE) is a laboratory test profile consisting of five fractions summing up to 100\% of total proteins. The fractions are albumin, $\alpha_1$, $\alpha_2$, $\beta$, and $\gamma$ globulins. The SPE can be assayed in different ways depending on the media or the analytical principle. In the experiment, the following four techniques were compared: HT cellulose acetate (CH), HT agarose gel (acid blue) (EH), HT agarose gel (amido black) (JH), and BCP capillary zone (GB). These four SPE techniques use distinct support mediums, staining colors, or analytical principles. Thus, we want to compare the techniques by testing  for equal MCVs. This was first done by \cite{AertsHaesbroeck2017}, but they considered the MCV of Voinov and Nikulin only. We extend this study using the new Wald-type and multiple contrast tests for all four variants. But first, the data need to be transformed and cleaned. Due to the compositional nature of the electrophoretic data, a one-to-one transformation from the 5-dimensional to the 4-dimensional space is required. For this purpose, the well-known isometric log-ratio transformation is used. Second, similarly to \cite{AertsHaesbroeck2017}, we remove the outliers from the data set, which we detected by computing the robust Mahalanobis distances of the observations. After that, we have four samples of 4-dimensional observations with sizes $n_1=133$, $n_2=112$, $n_3=74$, $n_4=62$.
	
	First of all, we calculate the estimators~\eqref{mcv_est} of MCVs for each technique and each MCV variant, see Table~\ref{tab_1}. It is apparent that the techniques do not perform equally well. In particular, the GB technique leads to the smallest MCV value for each variant and, thus, seems to be the most stable technique in terms of measurement variability. To check whether this first impression can be underpinned with statistical certainty, we first performed the different tests for the global hypotheses~\eqref{eqn:null} of technique-wise equality of MCVs. Hereby, we considered Tukey's contrast matrix for the different multiple contrast tests. For each MCV, all tests clearly rejected the global null hypotheses with $p$-values very close to zero (results are not shown). However, these rejections are just partially informative and we are rather interested in a more in-depth analysis with pairwise comparisons of the different techniques. Therefore, we can invert, as explained in Section~\ref{sec:multiple}, the multiple contrast tests into simultaneous confidence intervals, see Table~\ref{tab:sim_int}. These intervals indeed confirm our first impression and the GB technique leads to a significant lower MCV and grater value of its reciprocal compared to all three other techniques. Furthermore, it can be seen that the confidence intervals for the bootstrap approach are slightly wider than for the asymptotic one. This can be explained by the liberality of the asymptotic strategy which we observed in the simulation study. Another interesting observation is that a significant difference between the CH and EH techniques can only be detected by the AZ variant. Here, we like to point out again that the four variants are, in fact, different measures and, thus, opposite inference results may appear. \cite{AertsHaesbroeck2017} also considered pairwise tests for the underlying data set (see Table~6 in their paper) but for the specific variant $v=$VN only. In their analysis, they could detect a significant difference of CH and EH as well. However, they did not adjusted for multiplicity. Thus, their result is not contradicting ours for $v=$VN and need to be taken with a pinch of salt.

	\begin{table}[!t]
		\caption{Values of MCVs' estimators for four techniques.}
		\centering
		\begin{tabular}{rrrrr}
			\hline
			MCV&CH&EH&JH&GB\\
			\hline
			$\widehat C_i^{\text{RR}}$&0.0559 & 0.0525 & 0.0489 & 0.0249\\
			$\widehat C_i^{\text{VV}}$&0.1421 & 0.1235 & 0.1244 & 0.0627\\
			$\widehat C_i^{\text{VN}}$&0.0688 & 0.0558 & 0.0619 & 0.0238\\
			$\widehat C_i^{\text{AZ}}$&0.0985 & 0.0719 & 0.0811 & 0.0343\\
			\hline
		\end{tabular}
		\label{tab_1}
	\end{table}

	\begin{table}
		\caption{Estimates of the contrasts $\mbf{h}_{\ell}^\top\widehat{\mbf{C}}^v$ and $\mbf{h}_{\ell}^\top\widehat{\mbf{B}}^v$ with respective simultaneous $95\%$-confidence intervals [95\%-L,95\%-U] based on the asymptotic and bootstrap procedures of Section~\ref{sec:multiple}. The bold values represent the cases, where significant differences are obtained.
		}\label{tab:sim_int}
		\centering
		\small{
			\begin{tabular}{lllrrrrrr}
				\hline
				Comparison&Variant $v$&Method&$\mbf{h}_{\ell}^\top\widehat{\mbf{C}}^v$&95\%-L&95\%-U&$\mbf{h}_{\ell}^\top\widehat{\mbf{B}}^v$&95\%-L&95\%-U\\\hline
				CH-EH&RR&asym&-0.003&-0.010&0.003&1.181&-1.102&3.464\\
				&&boot&&-0.011&0.004&&-1.290&3.652\\
				&VV&asym&-0.019&-0.040&0.002&1.059&-0.114&2.231\\
				&&boot&&-0.041&0.004&&-0.195&2.312\\
				&VN&asym&-0.013&-0.028&0.002&3.392&-0.418&7.202\\
				&&boot&&-0.031&0.005&&-0.797&7.581\\
				&AZ&asym&\textbf{-0.027}&\textbf{-0.046}&\textbf{-0.007}&\textbf{3.751}&\textbf{0.969}&\textbf{6.534}\\
				&&boot&&\textbf{-0.047}&\textbf{-0.006}&&\textbf{0.862}&\textbf{6.640}\\\hline
				CH-JH&RR&asym&-0.007&-0.016&0.002&2.587&-0.729&5.903\\
				&&boot&&-0.017&0.003&&-1.002&6.176\\
				&VV&asym&-0.018&-0.042&0.007&0.999&-0.392&2.390\\
				&&boot&&-0.043&0.008&&-0.488&2.486\\
				&VN&asym&-0.007&-0.027&0.013&1.625&-3.248&6.499\\
				&&boot&&-0.031&0.017&&-3.733&6.984\\
				&AZ&asym&-0.017&-0.041&0.006&2.173&-0.920&5.266\\
				&&boot&&-0.042&0.007&&-1.039&5.385\\\hline
				CH-GB&RR&asym&\textbf{-0.031}&\textbf{-0.037}&\textbf{-0.025}&\textbf{22.217}&\textbf{17.301}&\textbf{27.132}\\
				&&boot&&\textbf{-0.038}&\textbf{-0.024}&&\textbf{16.896}&\textbf{27.537}\\
				&VV&asym&\textbf{-0.079}&\textbf{-0.098}&\textbf{-0.061}&\textbf{8.922}&\textbf{6.549}&\textbf{11.294}\\
				&&boot&&\textbf{-0.100}&\textbf{-0.059}&&\textbf{6.386}&\textbf{11.458}\\
				&VN&asym&\textbf{-0.045}&\textbf{-0.057}&\textbf{-0.033}&\textbf{27.480}&\textbf{17.938}&\textbf{37.021}\\
				&&boot&&\textbf{-0.060}&\textbf{-0.030}&&\textbf{16.989}&\textbf{37.971}\\
				&AZ&asym&\textbf{-0.064}&\textbf{-0.081}&\textbf{-0.047}&\textbf{19.034}&\textbf{12.492}&\textbf{25.577}\\
				&&boot&&\textbf{-0.082}&\textbf{-0.046}&&\textbf{12.241}&\textbf{25.827}\\\hline
				EH-JH&RR&asym&-0.004&-0.012&0.005&1.406&-1.913&4.724\\
				&&boot&&-0.013&0.006&&-2.186&4.997\\
				&VV&asym&0.001&-0.021&0.023&-0.059&-1.463&1.345\\
				&&boot&&-0.022&0.024&&-1.560&1.442\\
				&VN&asym&0.006&-0.013&0.025&-1.767&-7.032&3.499\\
				&&boot&&-0.017&0.029&&-7.556&4.023\\
				&AZ&asym&0.009&-0.012&0.031&-1.578&-5.104&1.947\\
				&&boot&&-0.013&0.032&&-5.239&2.082\\\hline
				EH-GB&RR&asym&\textbf{-0.028}&\textbf{-0.033}&\textbf{-0.022}&\textbf{21.035}&\textbf{16.118}&\textbf{25.953}\\
				&&boot&&\textbf{-0.033}&\textbf{-0.022}&&\textbf{15.713}&\textbf{26.358}\\
				&VV&asym&\textbf{-0.061}&\textbf{-0.076}&\textbf{-0.045}&\textbf{7.863}&\textbf{5.483}&\textbf{10.243}\\
				&&boot&&\textbf{-0.078}&\textbf{-0.044}&&\textbf{5.319}&\textbf{10.408}\\
				&VN&asym&\textbf{-0.032}&\textbf{-0.043}&\textbf{-0.021}&\textbf{24.088}&\textbf{14.340}&\textbf{33.835}\\
				&&boot&&\textbf{-0.045}&\textbf{-0.019}&&\textbf{13.370}&\textbf{34.805}\\
				&AZ&asym&\textbf{-0.038}&\textbf{-0.052}&\textbf{-0.023}&\textbf{15.283}&\textbf{8.525}&\textbf{22.041}\\
				&&boot&&\textbf{-0.053}&\textbf{-0.023}&&\textbf{8.266}&\textbf{22.300}\\\hline
				JH-GB&RR&asym&\textbf{-0.024}&\textbf{-0.031}&\textbf{-0.016}&\textbf{19.630}&\textbf{14.156}&\textbf{25.104}\\
				&&boot&&\textbf{-0.032}&\textbf{-0.015}&&\textbf{13.705}&\textbf{25.554}\\
				&VV&asym&\textbf{-0.062}&\textbf{-0.081}&\textbf{-0.042}&\textbf{7.922}&\textbf{5.427}&\textbf{10.417}\\
				&&boot&&\textbf{-0.083}&\textbf{-0.041}&&\textbf{5.255}&\textbf{10.589}\\
				&VN&asym&\textbf{-0.038}&\textbf{-0.056}&\textbf{-0.021}&\textbf{25.854}&\textbf{15.644}&\textbf{36.065}\\
				&&boot&&\textbf{-0.059}&\textbf{-0.017}&&\textbf{14.628}&\textbf{37.081}\\
				&AZ&asym&\textbf{-0.047}&\textbf{-0.066}&\textbf{-0.028}&\textbf{16.861}&\textbf{9.970}&\textbf{23.753}\\
				&&boot&&\textbf{-0.067}&\textbf{-0.027}&&\textbf{9.706}&\textbf{24.017}\\\hline
		\end{tabular}}
	\end{table}

	\subsection{Simulation study based on EQA data set}
	To check the appropriateness of the results for the EQA data set, we conduct an additional simulation study based on this set. To mimic the observation given in the data example, we generated the simulation data with four samples using:
	\begin{itemize}
		\item the samples sizes ($n_1=133$, $n_2=112$, $n_3=74$, $n_4=62$) from the data example,
		\item for checking the type-1 control, in each group, the mean and covariance matrix were set to the sample mean and sample covariance matrix of the pooled data,
		\item for power investigation, the mean and covariance matrix in the $i$-th group was equal to the quantities of the $i$-th sample from the data set.
	\end{itemize}
	The 4-dimensional data were generated from the same three distributions as in Section~\ref{sec:simulation}, i.e. the normal, $t_5$, and $\chi_{10}^2$ distributions. We have calculated the empirical sizes and powers for the global hypotheses~\eqref{eqn:null} and the same for multiple contrast tests.
	
\begin{figure}[!t]
		\centering
		\includegraphics[width=
		0.95\textwidth,height=0.42\textheight]{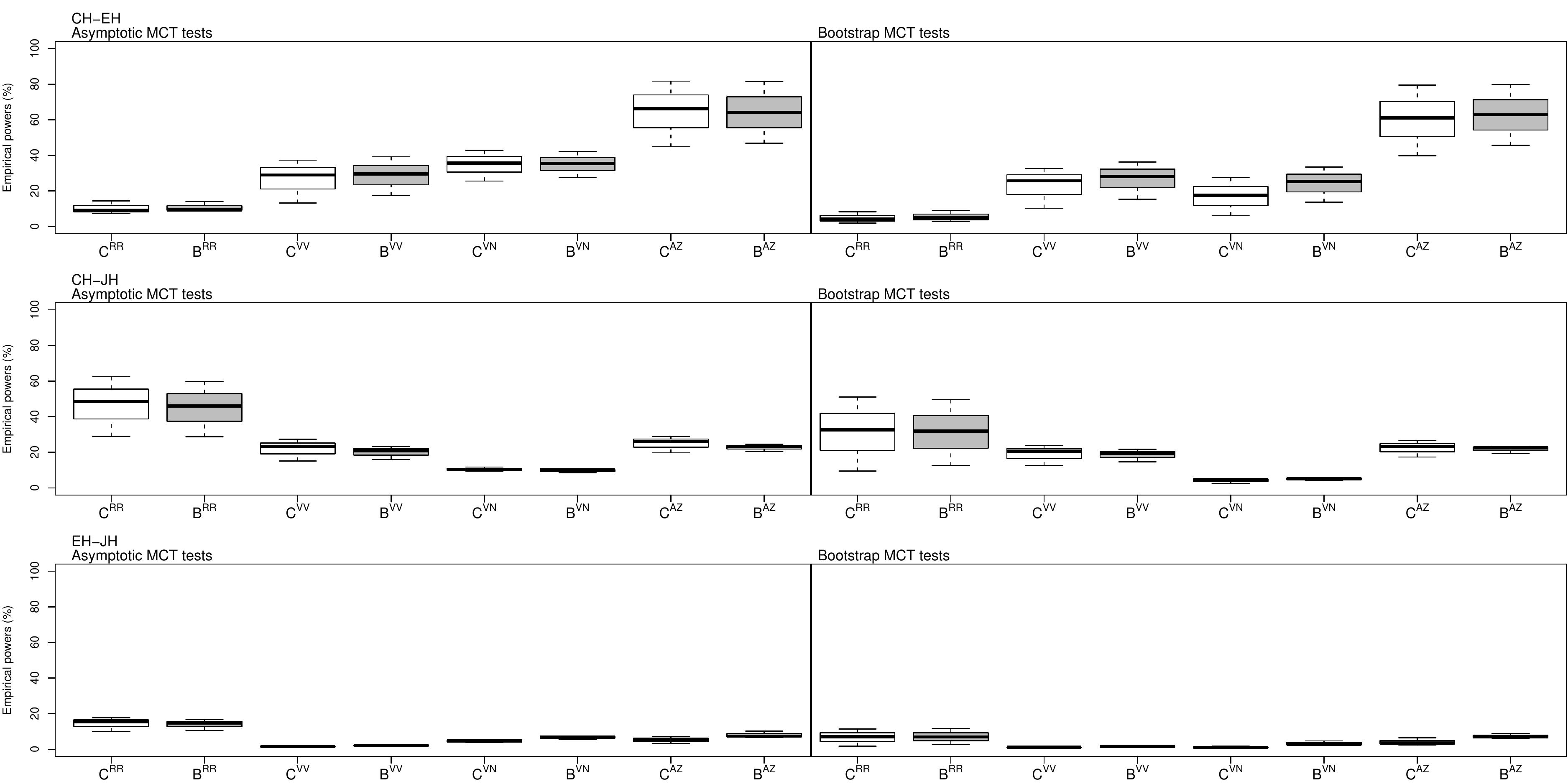}
		\caption{Box-and-whisker plots for the empirical powers (in $\%$) of the multiple contrast tests obtained for three ``nontrivial'' contrasts for simulation based on the EQA data.}
		\label{fig_5}
	\end{figure}

	The results regarding the type-1 error control of the (global) tests from Sections~\ref{sec:global} and~\ref{sec:multiple} (see Table~S1 in the supplement) are comparable to the ones from the prior simulation study in Section~\ref{sec:simulation} and, thus, not discussed in detail again. Moreover, all these tests have a power close to 100\% (results not shown). This confirms the appropriately of rejection of the global null hypotheses for the EQA data set. Let us turn to the post hoc testing problem and have a closer look on the asymptotic and bootstrap tests from Section~\ref{sec:multiple}. The power results are summarized in Figure~\ref{fig_5} and are shown more detailed in Table~S2 from the supplement. Although the bootstrap $\varphi_{\max,C^v}^b$ tests may still be conservative (Table~S1 in the supplement), the empirical powers of $\varphi_{\max,C^v}^b$ and $\varphi_{\max,B^v}^b$ tests are very similar for given $v\in\{\text{RR},\text{VV},\text{VN},\text{AZ}\}$. For CH-EH, CH-JH, and EH-JH comparisons, they are summarised in Figure~\ref{fig_5}. In the first case, the power of the tests is varied. The tests based on the MCV of Albert and Zhang are the most powerful, while the other tests have much smaller power. This clearly explains the rejections and nonrejections for the CH-EH comparison. On the other hand, for EH vs. JH, all tests have almost trivial power, which follows from the smallest differences in estimators of MCV's among all comparisons (see Table~\ref{tab_1}). In the case of CH-JH, the empirical power of all testing procedures increases (even up to about 60\% for the $\varphi_{\max,C^{\text{RR}}}^b$ and $\varphi_{\max,B^{\text{RR}}}^b$ tests). This increase of power in comparison to EH-JH can be explained by the increase in the differences in estimators (Table~\ref{tab_1}). For CH-GB, EH-GB, and JH-GB, the empirical powers of all tests were very close to 100\%, so we do not draw them. However, of course, they justify the recognition of significant differences for these contrasts using all multiple contrast tests proposed.

	\section{Summary and discussion}
	\label{sec:conclusion}
	
	In this paper, we discussed testing procedures for the multivariate coefficients of variation. To the best of our knowledge, such testing procedures were just developed for the special MCV version of \cite{voinovNikulin1996}. To be more concrete, \cite{AertsHaesbroeck2017} proposed (semi-)parametric procedures for the $k$-sample setting and, recently, \cite{ditzhaus_smaga_2022} suggested a nonparametric approach for general factorial designs. However, also the other MCV variants  \citep{Reyment1960,VanValen1974,AlbertZhang2010} are relevant for applications. That is why we followed the successful path of \cite{ditzhaus_smaga_2022} and adopted their results to the other MCV variants as well as their reciprocals, the so-called standardized means. In particular, we developed respective permutation procedure with a more satisfactory type-1 error control in small sample sizes. 
	
	Furthermore, we provided a post hoc strategy for a more in-depth analysis. Therefore, we combined the well-known tool of multiple contrast tests \citep{mukerjee1987comparison,bretz2001numerical,konietschke2018simultaneous,umlauft2019wild,gunawardana2019nonparametric,wechsung2021simultaneous} with our theoretical results. This resulted into asymptotically valid tests, which are rather liberal for small sample sizes, and a novel pooled bootstrap strategy. To the best of our knowledge, this pooled bootstrap idea was not used in the context of multiple testing procedures before. The simulation results confirm its usage for the reciprocals of the MCVs, by a more convincing type-1 error control under small sample sizes. However, for the MCVs itself the bootstrap procedures become quite conservative and we cannot completely recommend its usage. Here, further research is required to obtain a better resampling strategy for MCV post hoc testing. 
	
	All procedures are shown to be asymptotically valid and consistent by empirical process theory \citep{vaartWellner1996} and are extensively studied in simulations. To motivate their use, we provide the R package GFDmcv. Currently, the package can be requested via email and will be soon publicly available on CRAN. 
	
	A further interesting aspect is  inferring paired data, as in the recent investigation regarding the repeatability and reliability of GABA (Gamma-aminobutyric acid) measurements taken from the same patient group \citep{duda2021repeatability}. Under the assumption of normality, \cite{shoukri2008comparison} already discussed this issue for comparing two univariate CV. Conceptionally, this situation is not different to our set-up and, in more detail, corresponds to the one-sample scenario with $k=1$ and $d=2$. Consequently, the nonparametric methodology of comparing two (or more) CVs or even MCVs in paired data settings can be directly deduced from the theory presented in Section~\ref{sec:setup}. However, a detailed investigation would overload this paper and is, thus, postponed to future research.  Moreover, as already raised by \cite{ditzhaus_smaga_2022}, the problem of outliers, as in the EQA data set, shall be tackled nonparametrically by considering robust estimators for $\mbf{\mu}$ and $\mbf{\Sigma}$. For first (semi-)parametric solutions in this context, we refer to \cite{AertsHaesbroeck2017}.
	
	\section*{Acknowledgement}
	The authors would like to thank Professor Adelin Albert (Facult\'{e} de M\'{e}decine, University of Li\`{e}ge) for sharing the electrophoretic data sets from the French and Belgian national EQA programs used in \cite{ZhangEtAl2010}. A part of calculations for the simulation study was made at the Pozna\'n Supercomputing and Networking Center (grant no. 382).

\section*{Supplementary materials}
In the supplementary material, all results of simulation studies of Sections \ref{sec:simulation} and \ref{sec:real_data_app} are presented.

\appendix

\section{Proofs}

\subsection{Proof of Theorem \ref{theo:conv_C+beta}}
From the proof of Theorem~1 in \cite{ditzhaus_smaga_2022}, we obtain 
\begin{align}\label{eqn:conv_mu+sigma}
	n_1^{1/2} \begin{pmatrix} \mbf{\widehat \mu} - \mbf{\mu} \\ \text{vec}( \mbf{\widehat \Sigma}) - \text{vec}( \mbf{\Sigma}) \end{pmatrix} 
	\overset{\mathrm d}{\longrightarrow} \mbf{D}_\psi(\mbf{\mu}) \mbf{G},
\end{align}
where
\begin{align*}
	\mbf{G} \sim & N\Bigl( \mbf{0}, \begin{pmatrix}
	\mbf{ \Sigma} & \mbf{\Psi}_{3}\trans  \\
	\mbf{\Psi}_{3} & \mbf{\Psi}_{4}
\end{pmatrix} \Bigr),\ \mbf{D}_\psi(\mbf{x}) =  \begin{pmatrix} \mbf{I}_d & \mbf{0}_{d\times d^2} \\ \mbf{\widetilde D}(\mbf{x})  & \mbf{I}_{d^2} \end{pmatrix},
\end{align*}
$\mbf{\Psi}_{3}$, $\mbf{\Psi}_{4}$, $\mbf{\widetilde D}$ are defined in \eqref{eqn:def_tilde_D}, $\mbf{I}_d$ is the $d\times d$-dimensional unity matrix and $\mbf{0}_{d\times d^2}$ is the $d\times d^2$-dimensional zero matrix. Here and subsequently, we consider a matrix as an element of $\R^\ell$, $\ell\in\N$, by vectorization $\text{vec}(\mbf{A})=(A_{11},\ldots,A_{d1},A_{21},\ldots,A_{dd})\trans$ for $\mbf{A}=(A_{ij})_{i,j=1,\ldots,d}$. In this way, we can equip  $\text{GL}_{\text{sym}}(\R^d)$ with the Euclidean norm.  

To verify the first statement for the different choices of $\mbf{C}$, we apply the $\delta$-method and, later, its permutation/bootstrap analogue. For the latter, we require differentiability in a stronger sense \citep[see e.g.][Theorem 3.9.5]{vaartWellner1996}, so-called uniform differentiability. In our specific multivariate set-up, we call a map $\Phi:(\R^d\setminus\{0\})\times \text{GL}_{\text{sym}}(\R^d) \to\R $ uniformly differentiable when
\begin{align}\label{eqn:uniform_diff}
	t_n^{-1} \Bigl( \Phi[\mbf{a}_n + t_n \mbf{b}_n, \mbf{A}_n + t_n \mbf{B}_n] - \Phi[\mbf{a}_n , \mbf{A}_n ] \Bigr) \to \mbf{D}_\Phi(\mbf{a},\mbf{A}) \begin{pmatrix} \mbf{b} \\ \text{vec}(\mbf{B}) \end{pmatrix} \textrm{ as }n=n_1\to\infty
\end{align}
for a (Jacobi) matrix $\mbf{D}_\Phi(\mbf{a},\mbf{A})\in \R^{d+d^2}$ as well as for all $\mbf{A}_n \to \mbf{A}\in \text{GL}_{\text{sym}}(\R^d)$, $\mbf{B}_n\to \mbf{B}\in\R^{d\times d}$, $\mbf{a}_n\to \mbf{a}\in\R^d\setminus\{0\}$, $\mbf{b}_n\to \mbf{b}\in\R^d$ and $t_n\to 0$. Setting $\mbf{a}_n=\mbf{a}$ and $\mbf{A}_n=\mbf{A}$, we get (standard) differentiability.

\vspace{0.3cm}
\begin{lemma}\label{lem:differentiable_maps}
	The maps $\Phi_{\text{RR}}$, $\Phi_{\text{VV}}$, $\Phi_{\text{AZ}}:(\R^d\setminus\{0\})\times \text{GL}_{\text{sym}}(\R^d)\to\R$ defined as
	\begin{align*}
		\Phi_{\text{RR}}(\mbf{a}, \mbf{A}) = \sqrt{\frac{\det(\mbf{A})^{1/d}}{(\mbf{a}\trans \mbf{a})}},\qquad
		\Phi_{\text{VV}}(\mbf{a}, \mbf{A}) = \sqrt{\frac{\mathrm{tr}(\mbf{A})}{\mbf{a}\trans \mbf{a}}}\quad\text{ and }\quad
		\Phi_{\text{AZ}}(\mbf{a}, \mbf{A}) = \sqrt{\frac{\mbf{a}\trans\mbf{A}\mbf{a}}{(\mbf{a}\trans \mbf{a})^2}}
	\end{align*}
	are uniformly differentiable in the sense of \eqref{eqn:uniform_diff} with (Jacobi) matrices
	\begin{align*}
		&\mbf{D}_{\Phi_{\text{RR}}}(\mbf{a},\mbf{A}) = \frac{1}{2d} \left(\frac{\det(\mbf{A})}{(\mbf{A}\trans \mbf{a})^{d}}\right)^{1/(2d)-1}\Big(-2d\det(\mbf{A})\frac{\mbf{a}\trans}{(\mbf{a}\trans \mbf{a})^{d+1}},\frac{\det(\mbf{A})\left(\mathrm{vec}(\mbf{A}^{-1})\right)\trans}{(\mbf{a}\trans \mbf{a})^d}\Big),\\
		&\mbf{D}_{\Phi_{\text{VV}}}(\mbf{a}, \mbf{A})=\frac{1}{2}\left(\frac{\textrm{tr}(\mbf{A})}{\mbf{a}\trans \mbf{a}}\right)^{-1/2}\left(-2\,\mathrm{tr}(\mbf{A})\frac{\mbf{a}\trans}{(\mbf{a}\trans \mbf{a})^2},\frac{1}{\mbf{a}\trans \mbf{a}}\left(\mathrm{vec}(\mbf{I}_d)\right)\trans\right),\\
		&\mbf{D}_{\Phi_{\text{AZ}}}(\mbf{a}, \mbf{A})=\frac{1}{2}\left(\frac{\mbf{a}\trans\mbf{A}\mbf{a}}{(\mbf{a}\trans \mbf{a})^2}\right)^{-1/2}\left(-4\mbf{a}\trans\mbf{A}\mbf{a}\frac{\mbf{a}\trans}{(\mbf{a}\trans \mbf{a})^3}+2\frac{\mbf{a}\trans\mbf{A}}{(\mbf{a}\trans \mbf{a})^2},\frac{\mbf{a}\trans\otimes\mbf{a}\trans}{(\mbf{a}\trans \mbf{a})^2}\right).
	\end{align*}
\end{lemma}

\textbf{Proof of Lemma~\ref{lem:differentiable_maps}}: We first consider the maps 
\begin{align*}
	\widetilde\Phi_{\text{RR}}(\mbf{a}, \mbf{A}) = \frac{\det(\mbf{A})}{(\mbf{a}\trans \mbf{a})^d},\qquad
	\widetilde\Phi_{\text{VV}}(\mbf{a}, \mbf{A}) = \frac{\mathrm{tr}(\mbf{A})}{\mbf{a}\trans \mbf{a}}\quad\text{ and }\quad
	\widetilde\Phi_{\text{AZ}}(\mbf{a}, \mbf{A}) = \frac{\mbf{a}\trans\mbf{A}\mbf{a}}{(\mbf{a}\trans \mbf{a})^2}.
\end{align*}
 Let $\mbf{A}_n \to \mbf{A}\in \text{GL}_{\text{sym}}(\R^d)$, $\mbf{B}_n\to \mbf{B}\in\R^{d\times d}$, $\mbf{a}_n\to \mbf{a}\in\R^d\setminus\{\mbf{0}\}$, $\mbf{b}_n\to \mbf{b}\in\R^d$ and $t_n\to 0$. Then we first observe that
\begin{align*}
	t_n^{-1}\Bigl[ \det(\mbf{A}_n+t_n\mbf{B}_n) - \det(\mbf{A}_n) \Bigr] \to \det(\mbf{A})\textrm{tr}(\mbf{A}^{-1}\mbf{B}).
\end{align*}
For the fixed choice $\mbf{A}_n=\mbf{A}$, this follows from the differentiability of the determinant, e.g. shown in Theorem 8.1 from \cite{magnusNeudeckerBOOK2019}. For arbitrary $\mbf{A}_n$, we recall that the determinant can be expressed as a polynomial of the matrix's entries and polynomials are uniformly differentiable in the sense of \eqref{eqn:uniform_diff}. 

Secondly, we obtain from the linearity of the trace
\begin{align*}
		t_n^{-1}\Bigl[ \textrm{tr}(\mbf{A}_n+t_n\mbf{B}_n) - \textrm{tr}(\mbf{A}_n) \Bigr] = \textrm{tr}(\mbf{B}_n) \to \textrm{tr}(\mbf{\mbf{B}}).
\end{align*}
Thirdly, we recall that $\mbf{A}$ is symmetric and, thus,
\begin{align*}
	&t_n^{-1}\Bigl[ (\mbf{a}_n+t_n\mbf{b}_n)\trans(\mbf{A}_n+t_n\mbf{B}_n)(\mbf{a}_n+t_n\mbf{b}_n) - \mbf{a}_n\trans\mbf{A}_n\mbf{a}_n \Bigr] \\
	&= 2 \mbf{b}_n\trans\mbf{A}_n\mbf{a}_n + \mbf{a}_n\mbf{B}_n\mbf{a}_n  + O(t_n) \\
	&\to 2 \mbf{b}\trans\mbf{A}\mbf{a} + \mbf{a}\mbf{B}\mbf{a}.
\end{align*}
In particular, setting $\mbf{A}_n = \mbf{I}_d$ and $\mbf{B}_n=\mbf{0}_{d\times d}$ we obtain
\begin{align*}
	&t_n^{-1}\Bigl[ (\mbf{a}_n+t_n\mbf{b}_n)\trans(\mbf{a}_n+t_n\mbf{b}_n) - \mbf{a}_n\trans\mbf{a}_n \Bigr] \to 2 \mbf{b}\trans\mbf{a}.
\end{align*}
Consequently, applying the chain rule yields the results. In detail, we get	
\begin{align*}
	& t_n^{-1} \Bigl( \widetilde\Phi_{\text{RR}}[\mbf{a}_n + t_n \mbf{b_n}, \mbf{A}_n + t_n \mbf{B_n} ] - \widetilde\Phi_{\text{RR}}[\mbf{a}_n , \mbf{A}_n ] \Bigr)\\
	&\rightarrow \frac{\det(\mbf{A})\textrm{tr}(\mbf{A}^{-1}\mbf{B}) (\mbf{a}\trans\mbf{a})^d - d (\mbf{a}\trans\mbf{a})^{d-1} ( 2 \mbf{b}\trans\mbf{a}) \det(\mbf{A})}{ (\mbf{a}\trans\mbf{a})^{2d}}\\
	&= \frac{\det(\mbf{A})\textrm{tr}(\mbf{A}^{-1}\mbf{B}) (\mbf{a}\trans\mbf{a})^d - d (\mbf{a}\trans\mbf{a})^{d-1} ( 2 \mbf{b}\trans\mbf{a}) \det(\mbf{A})}{ (\mbf{a}\trans\mbf{a})^{2d}}\\
	&=-2d\det(\mbf{A})\frac{\mbf{a}\trans\mbf{b}}{(\mbf{a}\trans \mbf{a})^{d+1}}+\frac{\det(\mbf{A})\textrm{tr}(\mbf{A}^{-1}\mbf{B})}{(\mbf{a}\trans \mbf{a})^d}\\
	&=\left(-2d\det(\mbf{A})\frac{\mbf{a}\trans}{(\mbf{a}\trans \mbf{a})^{d+1}},\frac{\det(\mbf{A})\left(\text{vec}(\mbf{A}^{-1})\right)\trans}{(\mbf{a}\trans \mbf{a})^d}\right)\left(\begin{array}{c}
		\mbf{b}\\
		\text{vec}(\mbf{B})\\
	\end{array}\right),
\end{align*}
where we used the (easy-to-see) equation $\textrm{tr}(\mbf{E}\trans\mbf{F})=\left(\text{vec}(\mbf{E})\right)\trans\text{vec}(\mbf{F})$ for respective matrices $\mbf{E},\mbf{F}$ in the last equation. Moreover,
\begin{align*}
	 &t_n^{-1} \Bigl( \widetilde\Phi_{\text{VV}}[\mbf{a}_n + t_n \mbf{b_n}, \mbf{A}_n + t_n \mbf{B_n} ] - \widetilde\Phi_{\text{VV}}[\mbf{a}_n , \mbf{A}_n ] \Bigr)\\
	&\rightarrow 
	\frac{\textrm{tr}(\mbf{B}) (\mbf{a}\trans\mbf{a}) - 2 (\mbf{b}\trans\mbf{a})\textrm{tr}(\mbf{A})}{(\mbf{a}\trans\mbf{a})^2}\\
	&=\left(-2\textrm{tr}(\mbf{A})\frac{\mbf{a}\trans}{(\mbf{a}\trans \mbf{a})^2},\frac{1}{\mbf{a}\trans \mbf{a}}\left(\text{vec}(\mbf{I}_d)\right)\trans\right)\left(\begin{array}{c}
	\mbf{b}\\
			\text{vec}(\mbf{B})\\
		\end{array}\right).
\end{align*}
Finally,
\begin{align*}
	&t_n^{-1} \Bigl( \widetilde\Phi_{\text{AZ}}[\mbf{a}_n + t_n \mbf{b_n}, \mbf{A}_n + t_n \mbf{B_n} ] - \widetilde\Phi_{\text{AZ}}[\mbf{a}_n , \mbf{A}_n ] \Bigr)\\
	&\rightarrow \frac{(2 \mbf{b}\trans\mbf{A}\mbf{a} + \mbf{a}\mbf{B}\mbf{a})(\mbf{a}\trans\mbf{a})^2 - 2(\mbf{a}\trans\mbf{a}) (2\mbf{b}\trans\mbf{a})\mbf{a}\trans \mbf{A} \mbf{a} }{(\mbf{a}\trans\mbf{a})^4 } \\
	&= \left(-4\mbf{a}\trans\mbf{A}\mbf{a}\frac{\mbf{a}\trans}{(\mbf{a}\trans \mbf{a})^3}+2\frac{\mbf{a}\trans\mbf{A}}{(\mbf{a}\trans \mbf{a})^2},\frac{\mbf{a}\trans\otimes\mbf{a}\trans}{(\mbf{a}\trans \mbf{a})^2}\right)\left(\begin{array}{c}
		\mbf{b}\\
		\text{vec}(\mbf{B})\\
	\end{array}\right),
\end{align*}
where we used the following equality for general matrices $\mbf{E}$, $\mbf{F}$ and $\mbf{G}$ with appropriate dimensions such that the respective multiplications are well-defined:
\begin{align*}
	\text{vec}(\mbf{E}\mbf{F} \mbf{G})=( \mbf{G}\trans \otimes  \mbf{E}) \text{vec}( \mbf{F}).
\end{align*}
The statements regarding the maps $\Phi_{\text{RR}}$, $\Phi_{\text{VV}}$ and $\Phi_{\text{AZ}}$ follow easily from the chain rule, when the (clearly uniformly differentiable) maps $\varphi_1,\varphi_2:(0,\infty) \to \R$ defined by $\varphi_1(x)=x^{1/(2d)}$ and $\varphi_2(x)=x^{1/2}$ are applied to them.
\hfill $\square$

Combining Lemma~\ref{lem:differentiable_maps}, \eqref{eqn:conv_mu+sigma}, the $\delta$-method and the definition of $\mbf{A}_v(\mbf{\mu},\mbf{\Sigma})$ yields
\begin{align*}
	&n_1^{1/2} \Big( \widehat C^v - C^v \Big) \overset{\mathrm d}{\longrightarrow}  \mbf{D}_{\Phi_v}(\mbf{\mu}, \mbf{\Sigma}) \mbf{D}_\psi(\mbf{\mu}) \mbf{G} \sim N(0,\sigma_{C^v}^2), 
\end{align*}
The results for $B^v=1/C^v$ follows by the chain rule with the map $\varphi_3(x)=x^{-1}$, $x\neq 0$.

\subsection{Proof of Lemma \ref{lem:var_equal_zero}}
We first like to note that the proof for the case $v=\text{VN}$ can be found in the appendix of \cite{ditzhaus_smaga_2022}. In fact, we follow their arguments and adopt them to the different variants $v=\text{RR},\text{VV},\text{AZ}$.

First, we define, for abbreviation,
\begin{align}\label{eqn:def_tildeXi}
\mbf{\widetilde X} = ( X_{11}, \ldots, X_{1d}, X_{11}X_{11}, \ldots,  X_{11}X_{1d}, X_{12}X_{11}, \ldots, X_{1d}X_{1d} )\trans .
\end{align}
It is easy to see that the covariance matrix of $\mbf{\widetilde X}_i$ equals
\begin{align*}
\mbf{\Sigma}_{\mbf{\widetilde X}} = \begin{pmatrix}
\mbf{ \Sigma} & \mbf{\Psi}_{3}\trans  \\
\mbf{\Psi}_{3} & \mbf{\Psi}_{4}
\end{pmatrix}.
\end{align*}
We observe that $\sigma^2_{C^v} = 0$ implies degeneracy of $\mbf{A}_v(\mbf{\mu},\mbf{\Sigma})\mbf{\widetilde X}$. For $v=\text{RR}$, the latter translates to: There exists a constant $\widetilde c\in\R$ such that
\begin{align}\label{eqn:var=0_mat_r}
\begin{pmatrix} -2d\det(\mbf{\Sigma})\frac{\mbf{\mu}_i\trans}{(\mbf{\mu}\trans \mbf{\mu})^{d+1}}+\frac{\det(\mbf{\Sigma})\left(\text{vec}(\mbf{\Sigma}^{-1})\right)\trans}{(\mbf{\mu}\trans \mbf{\mu})^d}\mbf{\widetilde D}(\mbf{\mu}),&\frac{\det(\mbf{\Sigma})\left(\text{vec}(\mbf{\Sigma}^{-1})\right)\trans}{(\mbf{\mu}\trans \mbf{\mu})^d}  \end{pmatrix}  \mbf{\widetilde X}
= \widetilde c
\end{align}
with probability one. Define for $m,s\in\{1,\ldots,d\}$
\begin{align}
	&a_s = \Big[-2d\det(\mbf{\Sigma})\frac{\mbf{\mu}\trans}{(\mbf{\mu}\trans \mbf{\mu})^{d+1}}+\frac{\det(\mbf{\Sigma})\left(\text{vec}(\mbf{\Sigma}^{-1})\right)\trans}{(\mbf{\mu}\trans \mbf{\mu})^d}\mbf{\widetilde D}(\mbf{\mu})\Big]_s \in \R, \nonumber \\
	& b_{s m} = \Big[\frac{\det(\mbf{\Sigma})\left(\text{vec}(\mbf{\Sigma}^{-1})\right)\trans}{(\mbf{\mu}\trans \mbf{\mu})^d}\Big]_{sd-d+m} = \frac{\det(\mbf{\Sigma})[\mbf{\Sigma}^{-1}]_{m,s}}{(\mbf{\mu}\trans \mbf{\mu})^d}\in\R.\label{eqn:def_b}
\end{align}
Now, we can simplify \eqref{eqn:var=0_mat_r} to
\begin{align}\label{eqn:var=0_poly}
\sum_{m=1}^d( a_m X_{1m} + b_{mm}X_{1m}^2) + \sum_{s,m=1;s\neq m}^d b_{s m}X_{1s}X_{1m} = \widetilde c 
\end{align}
with probability one. Since $\mbf{\Sigma}^{-1}$ is nonsingular covariance matrix itself, the diagonal entries need to be positive and, thus, $b_{mm}\neq 0$ holds for all $m\in \{1,\ldots,d\}$. Now, fix some $m$. Given the other components $(X_{1s})_{s\neq m}$, the left hand side of \eqref{eqn:var=0_poly} is a polynomial in $X_{1m}$ of degree two and, thus, $X_{1m}$ can take at most two different values to solve \eqref{eqn:var=0_poly}. This contradicts Assumption~\ref{ass:two_point}.

This proof strategy can be used, in the same way, for the other two variants $v\in\{\text{VV},\text{AZ}\}$. For this purpose, we just need to consider the respective counterparts of $b_{s m}$ from \eqref{eqn:def_b} and show $b_{mm}\neq 0$ for some $m$. Let us start with $v=\text{VV}$:
\begin{align*}
	b^{\text{VV}}_{mm} = \Big[\frac{1}{\mbf{\mu}\trans \mbf{\mu}}\left(\text{vec}(\mbf{I}_d)\right)\trans \Big]_{m d-d+m} = \frac{1}{\mbf{\mu}\trans \mbf{\mu}} \neq 0\,\, \text{ for all }m=1,\ldots,d.
\end{align*}
Finally, we consider $v=\text{AZ}$:
\begin{align*}
		b^{\text{AZ}}_{mm} = \Big[\frac{\mbf{\mu}\trans\otimes\mbf{\mu}\trans}{(\mbf{\mu}\trans \mbf{\mu})^2} \Big]_{m d-d+m} = \frac{[\mbf{\mu}]_{m}^2}{(\mbf{\mu}\trans \mbf{\mu})^2} \neq 0
\end{align*}
for all $m=1,\ldots,d$ with $[\mbf{\mu}]_{m}\neq 0$. By Assumption~\ref{ass:well_defined}, we have $\mbf{\mu}\neq \mbf{0}$ and, thus, $b^{\text{AZ}}_{mm}\neq 0$ for some $m$.

\subsection{Proof of Theorem~\ref{theo:wts_convergence}} As already mentioned in the paper, it remains to show that $(\mbf{H}\mbf{ C}^v)^{\top} (\mbf{H}\mbf{\Sigma}_{C^v} \mbf{H}^{\top} )^+\mbf{H}\mbf{ C}^v$ as well as $(\mbf{H}\mbf{{B}}^v)^{\top} (\mbf{H}\mbf{\Sigma}_{B^v} \mbf{H}^{\top} )^+\mbf{H}\mbf{ {B}}^v$ are positive under any alternative $\mathcal H_{1,C^v}: \mbf{H}\mbf{C}^v \neq \mbf{0}$ or $\mathcal H_{1,B^v}: \mbf{H}\mbf{B}^v \neq \mbf{0}$, respectively. We just give the proof for $C$ and note that the statement for $B$ follows by simply interchanging the letters $C$ and $B$. 

For the proof, we need some (well-known) properties of the Moore--Penrose inverse for a matrix $\mbf{A}$, which can be found e.g. in \cite{raoMitra1971}: (I) $(\mbf{A}^{\top} )^+ = (\mbf{A}^+)^{\top} $, (II) $(\mbf{A}^{\top} \mbf{A})^+=\mbf{A}^+(\mbf{A}^{\top} )^+$, and (III) $\mbf{A}\mbf{A}^+\mbf{A} = \mbf{A}$. Now, suppose that the alternative $\mathcal H_{1,C^v}: \mbf{H}\mbf{C}^v \neq \mbf{0}$ is true. By Lemma~\ref{lem:var_equal_zero}, the root $\mbf{\Sigma}_{C^v}^{1/2}=\text{diag}(\kappa_1^{-1/2}\sigma_{1,C^v},\ldots,\kappa_k^{-1/2}\sigma_{k,C^v})$ of the covariance matrix is regular. Thus, there is some $\mbf{v}\in\mathbb{R}^{k}\setminus\{\mbf{0}_{k\times 1}\}$ such that $\mbf{C}^v=\mbf{\Sigma}_{C^v}^{1/2}\mbf{v}$. From this and (I)--(III) we obtain
\begin{align*}
	\mbf{0}_{k\times 1} \neq \mbf{H}\mbf{C}^v = \mbf{H}\mbf{\Sigma}_{C^v}^{1/2}\mbf{v} = \mbf{H}\mbf{\Sigma}_{C^v}^{1/2}  (\mbf{H}\mbf{ \Sigma}_{C^v}^{1/2})^+\mbf{H} \mbf{\Sigma}_C^{1/2}\mbf{v} 
	= \mbf{H}\mbf{ \Sigma}_{C^v}^{1/2} \Bigl[ (\mbf{H}\mbf{ \Sigma}_{C^v}^{1/2})^+\mbf{H}\mbf{C}^v \Bigr].
\end{align*}
This implies $ (\mbf{H}\mbf{ \Sigma}_{C^v}^{1/2})^+\mbf{H}\mbf{C}^v\neq \mbf{0}$. Consequently, 
\begin{align*}
	( \mbf{H} \mbf{C}^v)^{\top}  ( \mbf{H} \mbf{\Sigma}_{C^v} \mbf{H}^{\top}  )^+ \mbf{H} \mbf{C}^v &= ( \mbf{H} \mbf{C}^v)^{\top}  ( \mbf{\Sigma}_{C^v}^{1/2} \mbf{H}^{\top}  )^+ ( \mbf{H} \mbf{  \Sigma}_{C^v}^{1/2} )^+ \mbf{H} \mbf{C}^v \\
	&= \Bigl[ (\mbf{H}\mbf{\Sigma}_{C^v}^{1/2})^+\mbf{H}\mbf{C}^v \Bigr]^{\top} \Bigl[ (\mbf{H}\mbf{\Sigma}_{C^v}^{1/2})^+\mbf{H}\mbf{C}^v \Bigr] >0.
\end{align*}

\subsection{Proof of Theorem~\ref{theo:perm+bootstrap}\eqref{enu:theo:perm}}
One difficulty and significant difference to the asymptotic approach is that the permutation sample is dependent and, thus, the groups need to be considered simultaneously. The reason for the latter is that we pool the data and then draw without (!) replacement. The action of pooling the data is also important for deriving the theory. Thus, we introduce, on the one hand, the pooled estimators $\mbf{\widehat \mu}_0$, $\mbf{\widehat \Sigma}_0$, $\widehat C_0^v$ and $\widehat B_0^v$ depending on all observations and not just observations from a specific group. On the other hand, let $\mbf{Y}\sim P_0$ be a random, $d$-dimensional vector following the asymptotic pooled distribution $P_0=\sum_{i=1}^k\kappa_i P^{\mbf{X}_{i1}}$. Moreover, we denote by $\mbf{\mu}_Y,\,\mbf{ \Sigma}_Y,\, \mbf{\Psi}_{Y3},\, \mbf{\Psi}_{Y4}$ the respective theoretical quantities from the main paper but for $\mbf{Y}$ instead of $\mbf{X}_{i1}$. In particular, $\mbf{ \Sigma}_Y$ is the covariance matrix of $\mbf{Y}$.

 Again, we benefit from the preparatory work of \cite{ditzhaus_smaga_2022}, who already verified (see their (12)) that given the data almost surely
\begin{align}\label{eqn:conv_perm_mu+Sigma}
	n^{1/2} \begin{pmatrix} \mbf{\widehat \mu}_1^\pi - \mbf{\widehat\mu}_0 \\ \text{vec}( \mbf{\widehat \Sigma}_1^\pi) - \text{vec}( \mbf{\widehat\Sigma}_0) \\
	\vdots \\
	\mbf{\widehat \mu}_k^\pi - \mbf{\widehat\mu}_0 \\ \text{vec}( \mbf{\widehat \Sigma}_k^\pi) - \text{vec}( \mbf{\widehat\Sigma}_0)
 \end{pmatrix} 
	\overset{\mathrm d}{\longrightarrow} \mbf{G}^\pi = \begin{pmatrix} \mbf{G}_1^\pi\\ \vdots \\ \mbf{G}_k^\pi \end{pmatrix}.
\end{align}
Furthermore, they showed that $\mbf{G}^\pi$ is centered, $kd(d+1)$-dimensional normal distributed with covariance structure
\begin{align}\label{eqn:perm_cov+matrix}
	\begin{pmatrix}
		\gamma(1,1)\mbf{\Sigma}^\pi & \ldots & \gamma(1,k)\mbf{\Sigma}^\pi \\
		\vdots & \ddots & \vdots \\
		\gamma(k,1)\mbf{\Sigma}^\pi & \ldots & \gamma(k,k)\mbf{\Sigma}^\pi
	\end{pmatrix} 
	=&\begin{pmatrix}
		\kappa_1^{-1}\mbf{\Sigma}^\pi & \mbf{0}_{d'\times(k-2)d'} & \mbf{0}_{d'\times d'}\\[0.3em]
		\mbf{0}_{(k-2)d'\times d'} & \ddots & \mbf{0}_{(k-2)d'\times d'} \\[0.3em]
		\mbf{0}_{d'\times d'} & \ldots & \kappa_k^{-1}\mbf{\Sigma}^\pi
	\end{pmatrix}- 
	\begin{pmatrix}
		\mbf{\Sigma}^\pi & \ldots & \mbf{\Sigma}^\pi\\
		\vdots & \ddots & \vdots \\
		\mbf{\Sigma}^\pi & \ldots & \mbf{\Sigma}^\pi
	\end{pmatrix},
\end{align} 
where $d'=d(d+1)$,
\begin{align*}
	\mbf{\Sigma}^\pi =  \begin{pmatrix}
		\mbf{ \Sigma}_Y & \mbf{\Psi}_{Y3}^{\top}  \\
		\mbf{\Psi}_{Y3} & \mbf{\Psi}_{Y4}
	\end{pmatrix}\quad \text{and}\quad \gamma(i,i') = \kappa_i^{-1}\mathbf{1}\{i=i'\} - 1 \quad(i,i'=1,\ldots,k).
\end{align*}
To obtain the asymptotic normality of the permuted MCVs, we apply the $\delta$-method, similar to the proof of Theorem~\ref{theo:conv_C+beta}. However, we need to respect that we center the permutation quantities $\widehat C_i^{v\pi}$ and $\widehat B_i^{v\pi}$ by $\widehat C_0^v$ and $\widehat B_0^v$, respectively, which both change with growing sample sizes. That is why we need a stronger form of the $\delta$-method \citep[see e.g.][Theorem 3.9.5]{vaartWellner1996} which requires uniform differentiability in the sense of \eqref{eqn:uniform_diff}. By Lemma~\ref{lem:differentiable_maps} the maps $\Phi_{\text{RR}}$, $\Phi_{\text{VV}}$, $\Phi_{\text{AZ}}$ fulfil this requirement and, hence, we get that given the observations almost surely
\begin{align*}
	n^{1/2} \Bigl( \widehat C_1^{v\pi} - \widehat C_0^v,\ldots, \widehat C_k^{v\pi} - \widehat C_0^v \Bigr)^{\top} 
	& \overset{\mathrm d}{\longrightarrow} \begin{pmatrix}
		\mbf{D}_{\Phi_v}(\mbf{\mu}_Y,\mbf{\Sigma}_Y) &
		\mbf{0}_{1\times (k-2)d'} &	\mbf{0}_{1\times d'} \\[0.2em]
		\mbf{0}_{(k-2)\times d'} & \ddots & 	\mbf{0}_{(k-2)\times d'}  \\[0.4em]
		\mbf{0}_{1\times d'} & \mbf{0}_{1\times (k-2)d'} & \mbf{D}_{\Phi_v}(\mbf{\mu}_Y,\mbf{\Sigma}_Y)
	\end{pmatrix}
	\mbf{G}^\pi = \mbf{G}_{C^v}^\pi.
\end{align*}
To get the analogue results for $\widehat{B}_i^{v\pi}$ ($v\in\{\text{RR},\text{VV},\text{AZ}\}$) we apply the (uniform) $\delta$-method to $\varphi_3\circ \Phi_v$ instead. We leave the additional writing effort to the interested readers and proceed just with the $C$'s. Clearly, $\mbf{G}_{C^v}^\pi$ follows a centered, multidimensional normal distribution and, in regard to \eqref{eqn:perm_cov+matrix}, we can simplify its covariance matrix to
$$\mbf{\widetilde \Sigma}_{C^v} - (\mbf{D}_{\Phi_v}(\mbf{\mu}_Y,\mbf{\Sigma}_Y)\mbf{\Sigma}^\pi\mbf{D}_{\Phi_v}(\mbf{\mu}_Y,\mbf{\Sigma}_Y)^{\top} ) \mbf{1}_{k\times k},$$
where 
\begin{align*}
	\mbf{\widetilde \Sigma}_{C^v} &= \text{diag}(\kappa_1^{-1}\mbf{D}_{\Phi_v}(\mbf{\mu}_Y,\mbf{\Sigma}_Y)\mbf{\Sigma}^\pi\mbf{D}_{\Phi_v}(\mbf{\mu}_Y,\mbf{\Sigma}_Y)^{\top} , \ldots,\kappa_k^{-1}\mbf{D}_{\Phi_v}(\mbf{\mu}_Y,\mbf{\Sigma}_Y)\mbf{\Sigma}^\pi\mbf{D}_{\Phi_v}(\mbf{\mu}_Y,\mbf{\Sigma}_Y)^{\top} )\\
	&=\text{diag}(\sigma_{1,C^v,Y}^{2},\ldots,\sigma_{k,C^v,Y}^{2})
\end{align*}
and $\sigma_{1,C^v,Y}^{2}$ is the pooled counterpart of $\sigma_{1,C^v}^{2}$. Now, it becomes clear why we require the matrix $\mbf{H}$ to be a contrast matrix. Namely, this implies $\mbf{H}(\widehat C_0^v\cdot\mbf{1}_{k\times 1}) = \mbf{0}$ as well as $\mbf{H}\mbf{1}_{k\times k} = \mbf{0}$, and, consequently, it follows
\begin{align*}
	n^{1/2} \mbf{H} \widehat{\mbf{C}}^{\pi} = n^{1/2} \mbf{H} \Bigl( \widehat C_1^{v\pi} - \widehat C_0^v,\ldots, \widehat C_k^{v\pi} - \widehat C_0^v \Bigr)^{\top} 
	& \overset{\mathrm d}{\longrightarrow} \mbf{H}\mbf{G}_{C^v}^\pi \sim N(\mbf{0},\mbf{H}\mbf{\widetilde \Sigma}_{C^v}\mbf{H}^{\top})
\end{align*}
given the observations almost surely. As in Section~\ref{sec:Waldtest}, this is the first important ingredient for the asymptotic convergence of the Wald-type statistic. The second is the convergence of  $(\mbf{H}\widehat{\mbf{\Sigma}}_{C^v}^\pi\mbf{H}^{\top})^+$ to $(\mbf{H}\widetilde{\mbf{\Sigma}}_{C^v}\mbf{H}^{\top})^+$. Therefore, it remains to argue (1) $\sigma_{i,C^v,Y}^{2}>0$ for all $i=1,\ldots,k$ which implies the regularity of $\widetilde{\mbf{\Sigma}}_{C^v}$, and (2) $\widehat{\mbf{\Sigma}}_{C^v}^\pi$ converges (conditionally) in probability to $\widetilde{\mbf{\Sigma}}_{C^v}$ given the observations almost surely. 

Hereby, (1) follows directly from Lemma~\ref{lem:var_equal_zero} and the simple observation that Assumption~\ref{ass:two_point} can directly be transferred to the pooled distribution. For (2), we recall from \cite{ditzhaus_smaga_2022} that
\begin{align*}
	\begin{pmatrix}
		\mbf{\widehat \mu}_i^\pi  \\ \text{vec}( \mbf{\widehat \Sigma}_i^\pi) 
	\end{pmatrix}
	\overset{p}{\longrightarrow}
	\begin{pmatrix}
		\mbf{\mu}_Y  \\ \text{vec}( \mbf{\Sigma}_Y)  \end{pmatrix} ,\ \  i\in\{1,\ldots,k\},
\end{align*}
given the observations almost surely. Then (conditional) convergence in probability of $\widehat{\mbf{\Sigma}}_{C^v}^\pi$ to $\widetilde{\mbf{\Sigma}}_{C^v}$ follows from the continuous mapping theorem.

\subsection{Proof of Theorem~\ref{theo:perm+bootstrap}\eqref{enu:theo:boot}}
In principle, the proof follows the same argumentation as for Theorem~\ref{theo:perm+bootstrap}\eqref{enu:theo:perm}. However, the bootstrap procedure is slightly easier to handle because we draw with (!) replacement and, thus, the groups are still independent. The proof of \cite{ditzhaus_smaga_2022} for our \eqref{eqn:conv_perm_mu+Sigma} can be adopted to get an analogous result for the bootstrap. Here, one just need to replace Theorems 3.7.1 and 3.7.2 of \cite{vaartWellner1996} for the permutation procedure by the bootstrap counterparts Theorem 3.7.6 and 3.7.7 in their argumentation via empirical processes. All these results originally cover only the two-sample case ($k=2$) but can be directly extended to $k\geq 3$, as argued e.g. by \cite{ditzhausETAL2019} in their Lemma 9 and Remark 1. Finally, we can obtain that given the observations almost surely
\begin{align}
		n^{1/2} \begin{pmatrix} \mbf{\widehat \mu}_1^b - \mbf{\widehat\mu}_0 \\ \text{vec}( \mbf{\widehat \Sigma}_1^b) - \text{vec}( \mbf{\widehat\Sigma}_0) \\
			\vdots \\
			\mbf{\widehat \mu}_k^b - \mbf{\widehat\mu}_0 \\ \text{vec}( \mbf{\widehat \Sigma}_k^b) - \text{vec}( \mbf{\widehat\Sigma}_0)
		\end{pmatrix} 
		\overset{\mathrm d}{\longrightarrow} \mbf{G}^b = \begin{pmatrix} \mbf{G}_1^b\\ \vdots \\ \mbf{G}_k^b \end{pmatrix},
\end{align}
where $\mbf{G}^b$ is centered, $kd(d+1)$-dimensional normal distributed with covariance structure
\begin{align}\label{eqn:boot_cov+matrix}
	\begin{pmatrix}
		\kappa_1^{-1}\mbf{\Sigma}^\pi & \mbf{0}_{d'\times(k-2)d'} & \mbf{0}_{d'\times d'}\\[0.3em]
		\mbf{0}_{(k-2)d'\times d'} & \ddots & \mbf{0}_{(k-2)d'\times d'} \\[0.3em]
		\mbf{0}_{d'\times d'} & \ldots & \kappa_k^{-1}\mbf{\Sigma}^\pi
	\end{pmatrix}.
\end{align} 
Applying the (uniform) $\delta$-method we get, given the observations almost surely,
\begin{align}\label{eqn:C_boot_conv}
	&n^{1/2} \Bigl( \widehat C_1^{vb} - \widehat C_0^v,\ldots, \widehat C_k^{vb} - \widehat C_0^v \Bigr)^{\top} \nonumber \\
		& \overset{\mathrm d}{\longrightarrow} \begin{pmatrix}
			\mbf{D}_{\Phi_v}(\mbf{\mu}_Y,\mbf{\Sigma}_Y) &
			\mbf{0}_{1\times (k-2)d'} &	\mbf{0}_{1\times d'} \\[0.2em]
			\mbf{0}_{(k-2)\times d'} & \ddots & 	\mbf{0}_{(k-2)\times d'}  \\[0.4em]
			\mbf{0}_{1\times d'} & \mbf{0}_{1\times (k-2)d'} & \mbf{D}_{\Phi_v}(\mbf{\mu}_Y,\mbf{\Sigma}_Y)
		\end{pmatrix}
		\mbf{G}^b = \mbf{G}_{C^v}^b \sim N(\mbf{0}_{k\times 1}, \mbf{\widetilde \Sigma}_{C^v}),
\end{align}
where $\mbf{\widetilde \Sigma}_{C^v} =\text{diag}(\sigma_{1,C^v,Y}^{2},\ldots,\sigma_{k,C^v,Y}^{2})$. The rest of the proof follows the arguments from Section~\ref{sec:Waldtest} and from the proof's end of Theorem~\ref{theo:perm+bootstrap}\eqref{enu:theo:perm}. To avoid unnecessary repetition, we leave the details to the interested reader.  As an intermediate result, we just like to mentioned that given the data almost surely
\begin{align}\label{eqn:sigma_boot_conv}
	\widehat{\mbf{\Sigma}}_{C^v}^b \overset{p}{\rightarrow} \widetilde{\mbf{\Sigma}}_{C^v}
\end{align}
while $\widetilde{\mbf{\Sigma}}_{C^v}$ is regular as already argued in the proof of Theorem~\ref{theo:perm+bootstrap}\eqref{enu:theo:perm}.

\subsection{Proof of Theorem~\ref{theo:multi}}
The statements in \eqref{enu:theo:multi_exact} and \eqref{enu:theo:multi_SCI} follow immediately from Theorem~\ref{theo:conv_C+beta} as discussed briefly before Theorem~\ref{theo:multi}. The key step is to follow from Theorem~\ref{theo:conv_C+beta} that
\begin{align}\label{eqn:joint_conv}
	&\Big(
		\sqrt{n}\frac{ \mbf{h}_1^\top \widehat{\mbf{C}}^v - \mbf{h}_1^\top {\mbf{C}}^v}{ \sqrt{\mbf{h}_1^\top \widehat{\mbf{\Sigma}}_{C^v} \mbf{h}_1}}, \ldots,
		\sqrt{n}\frac{ \mbf{h}_r^\top \widehat{\mbf{C}}^v - \mbf{h}_r^\top {\mbf{C}}^v}{ \sqrt{\mbf{h}_r^\top \widehat{\mbf{\Sigma}}_{C^v} \mbf{h}_r}}\Big)^\top
	 \nonumber \\
	 &= \sqrt{n}\text{diag}((\mbf{h}_1^\top \widehat{\mbf{\Sigma}}_{C^v} \mbf{h}_1)^{-1/2},\ldots,(\mbf{h}_r^\top \widehat{\mbf{\Sigma}}_{C^v} \mbf{h}_r)^{-1/2}) \mbf{H} ( \widehat{\mbf{C}}^v - \mbf{C}^v)
	\overset{ d}{\rightarrow} \mbf{Z}\sim N(\mbf{0}_{r\times 1}, \mbf{R}_{C^v}).
\end{align}
Then \eqref{enu:theo:multi_SCI} follows by a simple inversion of this convergence statement and for \eqref{enu:theo:multi_exact} we just need to remind ourselves that $\mbf{h}_\ell^\top {\mbf{C}}^v=0$ for all $\ell=1,\ldots,r$ under $\mathcal H_{0,C^v}$.

Now, suppose $\mathcal H_{1,\ell,C^v}: \mbf{h}_\ell^\top \mbf{C}^v \neq 0$ is true. Then we can deduce from Slutzky's Lemma and \eqref{eqn:joint_conv}
\begin{align*}
	|T_{\ell,n}^v| \geq \sqrt{n}\frac{ |\mbf{h}_\ell^\top {\mbf{C}}^v|}{ \sqrt{\mbf{h}_\ell^\top \widehat{\mbf{\Sigma}}_{C^v} \mbf{h}_\ell}} - \Bigl | \sqrt{n}\frac{ \mbf{h}_\ell^\top \widehat{\mbf{C}}^v - \mbf{h}_\ell^\top {\mbf{C}}^v}{ \sqrt{\mbf{h}_\ell^\top \widehat{\mbf{\Sigma}}_{C^v} \mbf{h}_\ell}} \Bigr| \overset{p}{\rightarrow} \infty.
\end{align*}
From this, we obtain the second statement of \eqref{enu:theo:multi_r'}.

Now, let $\mathcal H_{0,1,C^v},\ldots, \mathcal H_{0,r',C^v}$ be true for $r'\leq r$ and define $\widehat{\mbf{R}}'=([\widehat{\mbf{R}}]_{\ell m})_{\ell,m=1,\ldots, r'}$. By \eqref{enu:theo:multi_exact}
\begin{align*}
	\Pr\Big(\max_{\ell = 1,\ldots,r'} |T_{\ell,n}^v|> q_{1-\alpha}(\widehat{\mbf{R}}')\Big) \to \alpha.
\end{align*}
Moreover, it is easy to see that $q_{1-\alpha,\text{max},C^v}(\widehat{\mbf{R}})\geq q_{1-\alpha}(\widehat{\mbf{R}}')$, e.g. by noting  $S_{n,\max,C^v}(\mbf{H}) \geq \max_{\ell = 1,\ldots,r'} |T_{\ell,n}^v|$. In summary, we obtain
\begin{align*}
	\Pr \Bigl( \bigcup_{\ell=1}^{\:r'}\{|T_{\ell,n}^v| > q_{1-\alpha,\text{max},C^v}(\widehat{\mbf{R}})\} \Bigr) =& \Pr\Bigl( \max_{\ell = 1,\ldots,r'} |T_{\ell,n}^v|> q_{1-\alpha,\text{max},C^v}(\widehat{\mbf{R}}) \Bigr) \\
	\leq & \Pr\Bigl( \max_{\ell = 1,\ldots,r'} |T_{\ell,n}^v|> q_{1-\alpha}(\widehat{\mbf{R}}') \Bigr)\\
	 \to & \alpha
\end{align*}
proving the first statement of \eqref{enu:theo:multi_r'}.

Clearly, an analogue of \eqref{eqn:joint_conv} is true for $B$ instead of $C$, see Theorem~\ref{theo:conv_C+beta}. Consequently, \eqref{enu:theo:multi_B} follows from the same arguments as used for \eqref{enu:theo:multi_exact}--\eqref{enu:theo:multi_SCI}.

\subsection{Proof of Theorem~\ref{theo:multi_bootstrap}}
To prove the statement, we need a bootstrap analogue of \eqref{eqn:joint_conv}. Therefore, we repeat from Lemma~\ref{lem:const_var_est} that $\widehat{\mbf{\Sigma}}_{C^v}^{1/2}$ converges in probability to $\mbf{\Sigma}_{C^v}^{1/2}$. Moreover, the assumptions ensure that $\mbf{\Sigma}_{C^v}^{1/2}$ and $\widetilde{\mbf{\Sigma}}_{C^v}$ are regular, for the later we refer to the proof of Theorem~\ref{theo:perm+bootstrap}\eqref{enu:theo:perm}.  Having classical subsequence arguments in mind, we fix the data and assume without a loss of generality that the aforementioned convergence as well as \eqref{eqn:C_boot_conv} and \eqref{eqn:sigma_boot_conv} hold. Then we obtain from Slutzky's Lemma, \eqref{eqn:C_boot_conv} and \eqref{eqn:sigma_boot_conv} that
\begin{align*}
	n^{1/2} \widehat{\mbf{\Sigma}}_{C^v}^{1/2} (\widehat{\mbf{\Sigma}}_{C^v}^b)^{-1/2} \Bigl( \widehat{\mbf{C}}^{vb} - \widehat{\mbf{C}}_0^v \Bigr)^\top
	& \overset{\mathrm d}{\longrightarrow} \mbf{\Sigma}_{C^v}^{1/2} (\widetilde{\mbf{\Sigma}}_{C^v})^{-1/2} \mbf{G}_{C^v}^b \sim N(\mbf{0}_{k\times 1}, \mbf{\Sigma}_{C^v}).
\end{align*}
This is the desired bootstrap analogue of  \eqref{eqn:joint_conv} and the rest follows by the same arguments as in Theorem~\ref{theo:multi}.

\bibliographystyle{spbasic} 
\bibliography{other_mcv}

\begin{thebibliography}{51}
\providecommand{\natexlab}[1]{#1}
\providecommand{\url}[1]{{#1}}
\providecommand{\urlprefix}{URL }
\expandafter\ifx\csname urlstyle\endcsname\relax
  \providecommand{\doi}[1]{DOI~\discretionary{}{}{}#1}\else
  \providecommand{\doi}{DOI~\discretionary{}{}{}\begingroup
  \urlstyle{rm}\Url}\fi
\providecommand{\eprint}[2][]{\url{#2}}

\bibitem[{Aerts and Haesbroeck(2017)}]{AertsHaesbroeck2017}
Aerts S, Haesbroeck G (2017) Robust asymptotic tests for the equality of
  multivariate coefficients of variation. TEST 26:163--187

\bibitem[{Albert and Zhang(2010)}]{AlbertZhang2010}
Albert A, Zhang L (2010) A novel definition of the multivariate coefficient of
  variation. Biom J 52:667--675

\bibitem[{Baigent et~al.(1998)Baigent, Collins, Appleby, Parish, Sleight, and
  Peto}]{baigent:etal:1998}
Baigent C, Collins R, Appleby P, Parish S, Sleight P, Peto R (1998) {ISIS}-2:
  10 year survival among patients with suspected acute myocardial infarction in
  randomised comparison of intravenous streptokinase, oral aspirin, both, or
  neither. BMJ 316:1337

\bibitem[{Bretz et~al.(2001)Bretz, Genz, and A.~Hothorn}]{bretz2001numerical}
Bretz F, Genz A, A~Hothorn L (2001) On the numerical availability of multiple
  comparison procedures. Biometrical Journal: Journal of Mathematical Methods
  in Biosciences 43(5):645--656

\bibitem[{Brunner et~al.(1997)Brunner, Dette, and
  Munk}]{brunner:dette:munk:1997}
Brunner E, Dette H, Munk A (1997) Box-type approximations in nonparametric
  factorial designs. J Amer Statist Assoc 92:1494--1502

\bibitem[{Cassidy et~al.(2008)Cassidy, Clarke, D{\'\i}az-Rubio, Scheithauer,
  Figer, Wong, Koski, Lichinitser, Yang, and Rivera}]{cassidy:etal:2008}
Cassidy J, Clarke S, D{\'\i}az-Rubio E, Scheithauer W, Figer A, Wong R, Koski
  S, Lichinitser M, Yang TS, Rivera F (2008) Randomized phase {III} study of
  capecitabine plus oxaliplatin compared with fluorouracil/folinic acid plus
  oxaliplatin as first-line therapy for metastatic colorectal cancer. Journal
  of Clinical Oncology 26:2006--2012

\bibitem[{Ditzhaus and Smaga(2022)}]{ditzhaus_smaga_2022}
Ditzhaus M, Smaga {\L} (2022) Permutation test for the multivariate coefficient
  of variation in factorial designs. J Multivariate Anal 187:104848

\bibitem[{Ditzhaus et~al.(2021{\natexlab{a}})Ditzhaus, Fried, and
  Pauly}]{ditzhausETAL2019}
Ditzhaus M, Fried R, Pauly M (2021{\natexlab{a}}) {QANOVA}: {Q}uantile-based
  permutation methods for general factorial designs. TEST 30(4):960--979

\bibitem[{Ditzhaus et~al.(2021{\natexlab{b}})Ditzhaus, Genuneit, Janssen, and
  Pauly}]{ditzhaus2021casanova}
Ditzhaus M, Genuneit J, Janssen A, Pauly M (2021{\natexlab{b}}) {CASANOVA}:
  {P}ermutation inference in factorial survival designs. Biometrics

\bibitem[{Duchesne and Francq(2015)}]{DuchesneFrancq2015}
Duchesne P, Francq C (2015) Multivariate hypothesis testing using generalized
  and $\{2\}$-inverses - with applications. Statistics 49:475--496

\bibitem[{Duda et~al.(2021)Duda, Moser, Zuo, Du, Chen, Perlo, Richards,
  Nascimento, Ironside, Crowley et~al.}]{duda2021repeatability}
Duda JM, Moser AD, Zuo CS, Du F, Chen X, Perlo S, Richards CE, Nascimento N,
  Ironside M, Crowley DJ, et~al. (2021) Repeatability and reliability of {GABA}
  measurements with magnetic resonance spectroscopy in healthy young adults.
  Magnetic Resonance in Medicine 85(5):2359--2369

\bibitem[{Dunnett(1955)}]{dunnett1955multiple}
Dunnett C (1955) A multiple comparison procedure for comparing several
  treatments with a control. Journal of the American Statistical Association
  50(272):1096--1121

\bibitem[{Ferri and Jones(1979)}]{Ferri1979}
Ferri M, Jones W (1979) Determinants of financial structure: A new
  methodological approach. The Journal of Finance 34:631--644

\bibitem[{Friedrich and Pauly(2018)}]{friedrich2018mats}
Friedrich S, Pauly M (2018) {MATS}: Inference for potentially singular and
  heteroscedastic {MANOVA}. Journal of Multivariate Analysis 165:166--179

\bibitem[{Friedrich et~al.(2017)Friedrich, Brunner, and
  Pauly}]{friedrich2017permuting}
Friedrich S, Brunner E, Pauly M (2017) Permuting longitudinal data in spite of
  the dependencies. J Multivariate Anal 153:255--265

\bibitem[{Genz and Bretz(2009)}]{GenzBretz2009}
Genz A, Bretz F (2009) Computation of Multivariate Normal and t Probabilities.
  Lecture Notes in Statistics, Springer-Verlag, Heidelberg

\bibitem[{Genz et~al.(2021)Genz, Bretz, Miwa, Mi, Leisch, Scheipl, and
  Hothorn}]{Genzetal2021}
Genz A, Bretz F, Miwa T, Mi X, Leisch F, Scheipl F, Hothorn T (2021) {mvtnorm}:
  Multivariate Normal and t Distributions.
  \urlprefix\url{https://CRAN.R-project.org/package=mvtnorm}, r package version
  1.1-3

\bibitem[{GISSI-2(1990)}]{gissi:1990}
GISSI-2 TISG (1990) In-hospital mortality and clinical course of 20,891
  patients with suspected acute myocardial infarction randomized between
  alteplase and streptokinase with or without heparin. Lancet 336:71--75

\bibitem[{Gunawardana and Konietschke(2019)}]{gunawardana2019nonparametric}
Gunawardana A, Konietschke F (2019) Nonparametric multiple contrast tests for
  general multivariate factorial designs. Journal of Multivariate Analysis
  173:165--180

\bibitem[{Harrar et~al.(2019)Harrar, Ronchi, and
  Salmaso}]{harrar2019comparison}
Harrar S, Ronchi F, Salmaso L (2019) A comparison of recent nonparametric
  methods for testing effects in two-by-two factorial designs. J Appl Stat
  46:1649--1670

\bibitem[{Hothorn et~al.(2008)Hothorn, Bretz, and
  Westfall}]{hothorn2008simultaneous}
Hothorn T, Bretz F, Westfall P (2008) Simultaneous inference in general
  parametric models. Biometrical Journal: Journal of Mathematical Methods in
  Biosciences 50(3):346--363

\bibitem[{Jalilibal et~al.(2021)Jalilibal, Amiri, Castagliola, and
  Khoo}]{jalilibal2021monitoring}
Jalilibal Z, Amiri A, Castagliola P, Khoo MB (2021) Monitoring the coefficient
  of variation: {A} literature review. Computers \& Industrial Engineering
  161:107600

\bibitem[{Janssen and Pauls(2003)}]{janssenPauls2003}
Janssen A, Pauls T (2003) How do bootstrap and permutation tests work? Ann
  Statist 31(3):768--806

\bibitem[{Konietschke et~al.(2018)Konietschke, Aguayo, and
  Staab}]{konietschke2018simultaneous}
Konietschke F, Aguayo RR, Staab W (2018) Simultaneous inference for factorial
  multireader diagnostic trials. Statistics in Medicine 37(1):28--47

\bibitem[{Konietschke et~al.(2021)Konietschke, Schwab, and
  Pauly}]{konietschke2021small}
Konietschke F, Schwab K, Pauly M (2021) Small sample sizes: A big data problem
  in high-dimensional data analysis. Statistical Methods in Medical Research
  30(3):687--701

\bibitem[{Kurz et~al.(2015)Kurz, Fleischmann, Sessler, Buggy, Apfel,
  Ak{\c{c}}a, Investigators, Fleischmann, Erdik, and Eredics}]{kurz:etal:2015}
Kurz A, Fleischmann E, Sessler D, Buggy D, Apfel C, Ak{\c{c}}a O, Investigators
  FT, Fleischmann E, Erdik E, Eredics K (2015) Effects of supplemental oxygen
  and dexamethasone on surgical site infection: a factorial randomized trial.
  British Journal of Anaesthesia 115:434--443

\bibitem[{Libeer(1993)}]{Libeer1993}
Libeer J (1993) External quality assessment in clinical laboratories. european
  perspectives: today and tomorrow. PhD thesis, Higher Education Doctoral
  Thesis, Antwerpen

\bibitem[{Lubsen and Pocock(1994)}]{lubsen:pocock:1994}
Lubsen J, Pocock S (1994) Factorial trials in cardiology: pros and cons.
  European Heart Journal 15:585--588

\bibitem[{Magnus and Neudecker(2019)}]{magnusNeudeckerBOOK2019}
Magnus J, Neudecker H (2019) Matrix differential calculus with applications in
  statistics and econometrics. John Wiley \& Sons

\bibitem[{Mukerjee et~al.(1987)Mukerjee, Robertson, and
  Wright}]{mukerjee1987comparison}
Mukerjee H, Robertson T, Wright FT (1987) Comparison of several treatments with
  a control using multiple contrasts. Journal of the American Statistical
  Association 82(399):902--910

\bibitem[{Neumann et~al.(2021)Neumann, Schidlowski, G{\"u}nther, St{\"o}cker,
  and D{\"u}zel}]{neumann2021reliability}
Neumann K, Schidlowski M, G{\"u}nther M, St{\"o}cker T, D{\"u}zel E (2021)
  Reliability and reproducibility of {H}adamard encoded pseudo-continuous
  arterial spin labeling in healthy elderly. Frontiers in Neuroscience
  15:711898

\bibitem[{Pauly and Smaga(2020)}]{PaulySmaga2020}
Pauly M, Smaga {\L} (2020) Asymptotic permutation tests for coefficients of
  variation and standardized means in general one-way {ANOVA} models. Stat
  Methods Med Res \doi{10.1177/0962280220909959}

\bibitem[{Pauly et~al.(2015)Pauly, Brunner, and Konietschke}]{paulyETAL2015}
Pauly M, Brunner E, Konietschke F (2015) Asymptotic permutation tests in
  general factorial designs. J R Stat Soc Ser B Stat Methodol 77:461--473

\bibitem[{{R Core Team}(2022)}]{Rcore}
{R Core Team} (2022) R: A Language and Environment for Statistical Computing. R
  Foundation for Statistical Computing, Vienna, Austria,
  \urlprefix\url{https://www.R-project.org/}

\bibitem[{Rao and Mitra(1971)}]{raoMitra1971}
Rao C, Mitra S (1971) Generalized inverse of matrices and its applications.
  John Wiley \& Sons, Inc., New York-London-Sydney

\bibitem[{Reyment(1960)}]{Reyment1960}
Reyment RA (1960) Studies on Nigerian Upper Cretaceous and Lower Tertiary
  Ostracoda: part 1. Senonian and Maastrichtian Ostracoda, Stockholm
  Contributions in Geology, vol~7

\bibitem[{Sattler et~al.(2022)Sattler, Bathke, and Pauly}]{sattler2022testing}
Sattler P, Bathke AC, Pauly M (2022) Testing hypotheses about covariance
  matrices in general {MANOVA} designs. Journal of Statistical Planning and
  Inference 219:134--146

\bibitem[{Sciacovelli et~al.(2018)Sciacovelli, Secchiero, Padoan, and
  Plebani}]{Sciacovellietal2018}
Sciacovelli L, Secchiero S, Padoan A, Plebani M (2018) External quality
  assessment programs in the context of iso 15189 accreditation. Clinical
  Chemistry and Laboratory Medicine (CCLM) 56(10):1644--1654

\bibitem[{Shoukri et~al.(2008)Shoukri, Colak, Kaya, and
  Donner}]{shoukri2008comparison}
Shoukri MM, Colak D, Kaya N, Donner A (2008) Comparison of two dependent within
  subject coefficients of variation to evaluate the reproducibility of
  measurement devices. BMC Medical Research Methodology 8(1):1--11

\bibitem[{Smaga(2015)}]{smaga2015}
Smaga {\L} (2015) Wald-type statistics using \{2\}-inverses for hypothesis
  testing in general factorial designs. Statist Probab Lett 107:215--220

\bibitem[{Smaga(2017)}]{smaga2017}
Smaga {\L} (2017) Diagonal and unscaled {W}ald-type tests in general factorial
  designs. Electron J Stat 11:2613--2646

\bibitem[{Tukey(1953)}]{tukey1953problem}
Tukey JW (1953) The problem of multiple comparisons. Multiple Comparisons

\bibitem[{Umlauft et~al.(2019)Umlauft, Placzek, Konietschke, and
  Pauly}]{umlauft2019wild}
Umlauft M, Placzek M, Konietschke F, Pauly M (2019) Wild bootstrapping
  rank-based procedures: Multiple testing in nonparametric factorial repeated
  measures designs. Journal of Multivariate Analysis 171:176--192

\bibitem[{van~der Vaart and Wellner(1996)}]{vaartWellner1996}
van~der Vaart A, Wellner J (1996) Weak convergence and empirical processes.
  Springer Series in Statistics, Springer-Verlag, New York, with applications
  to statistics

\bibitem[{Van~Valen(1974)}]{VanValen1974}
Van~Valen L (1974) Multivariate structural statistics in natural history.
  Journal of Theoretical Biology 45:235--247

\bibitem[{Voinov and Nikulin(1996)}]{voinovNikulin1996}
Voinov V, Nikulin M (1996) Unbiased Estimators and Their Applications, Vol. 2,
  Multivariate Case. Kluwer, Dordrecht

\bibitem[{Weber et~al.(2004)Weber, Shafir, and Blais}]{Weber2004}
Weber E, Shafir S, Blais A (2004) Predicting risk sensitivity in humans and
  lower animals: risk as variance or coefficient of variation. Psychological
  Review 111:430--445

\bibitem[{Wechsung and Konietschke(2021)}]{wechsung2021simultaneous}
Wechsung M, Konietschke F (2021) Simultaneous inference for partial areas under
  receiver operating curves--with a view towards efficiency. arXiv preprint
  arXiv:210409401

\bibitem[{WHO(2022)}]{WHO2022}
WHO (2022) Overview of External Quality Assessment (EQA).
  \urlprefix\url{https://cdn.who.int/media/docs/default-source/
  essential-medicines/norms-and-standards/10-b-eqa-contents.pdf?sfvrsn=181d9a32\_4\&download=true},
  accessed on 14 December 2022

\bibitem[{Yeong et~al.(2016)Yeong, Khoo, Teoh, and
  Castagliola}]{yeong2016control}
Yeong WC, Khoo MBC, Teoh WL, Castagliola P (2016) A control chart for the
  multivariate coefficient of variation. Quality and Reliability Engineering
  International 32(3):1213--1225

\bibitem[{Zhang et~al.(2010)Zhang, Albar\`ede, Dumont, {Van Campenhout},
  Libeer, and Albert}]{ZhangEtAl2010}
Zhang L, Albar\`ede S, Dumont G, {Van Campenhout} C, Libeer J, Albert A (2010)
  The multivariate coefficient of variation for comparing serum protein
  electrophoresis techniques in {External} {Quality} {Assessment} schemes.
  Accreditation and Quality Assurance 15:351--357

\end{thebibliography}

\end{document}


\title{\Large \bf
Supplementary materials to\\
Inference for all variants of the multivariate coefficient of variation\\ in factorial designs}
\author[1]{Marc Ditzhaus}
\author[2,$^*$]{\L ukasz Smaga}

\affil[1]{Faculty of Mathematics, Otto-von-Guericke University Magdeburg, Germany}
\affil[2]{Faculty of Mathematics and Computer Science, Adam Mickiewicz University, Poland}

\maketitle

\begin{figure}[!t]
\centering
\includegraphics[width=
0.95\textwidth,height=0.85\textheight]{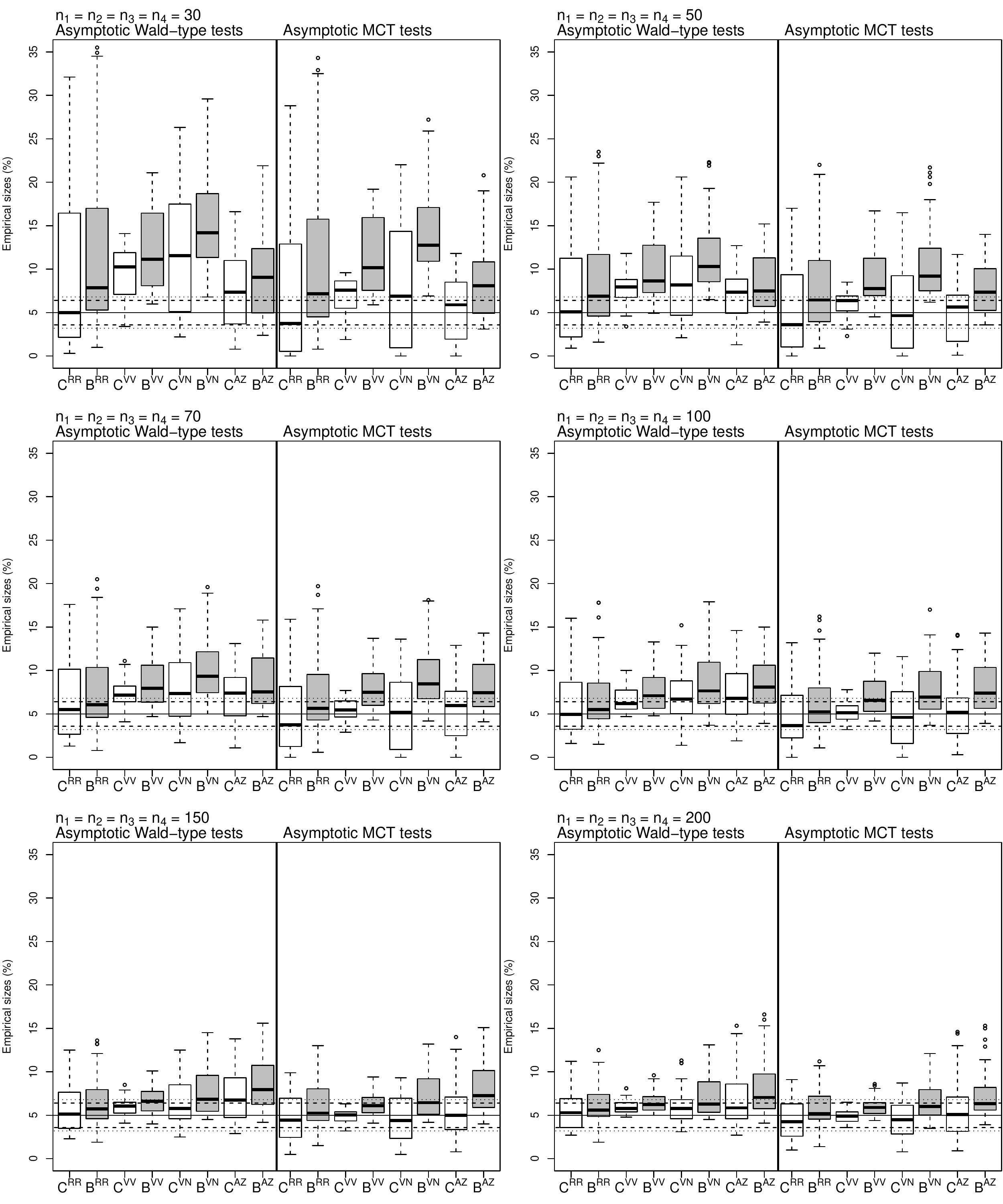}
\caption[Box-and-whisker plots for the empirical sizes of the asymptotic tests obtained from all cases considered]{Box-and-whisker plots for the empirical sizes (as percentages) of the asymptotic tests obtained from all cases considered. The solid, dashed, and dotted lines represent the significance level $\alpha=5\%$ and the $95\%$ and $99\%$ binomial proportion confidence intervals $[3.6\%,6.4\%]$ and $[3.2\%,6.8\%]$, respectively.}
\label{fig_sm_1}
\end{figure}
\begin{figure}[!t]
\centering
\includegraphics[width=
0.95\textwidth,height=0.85\textheight]{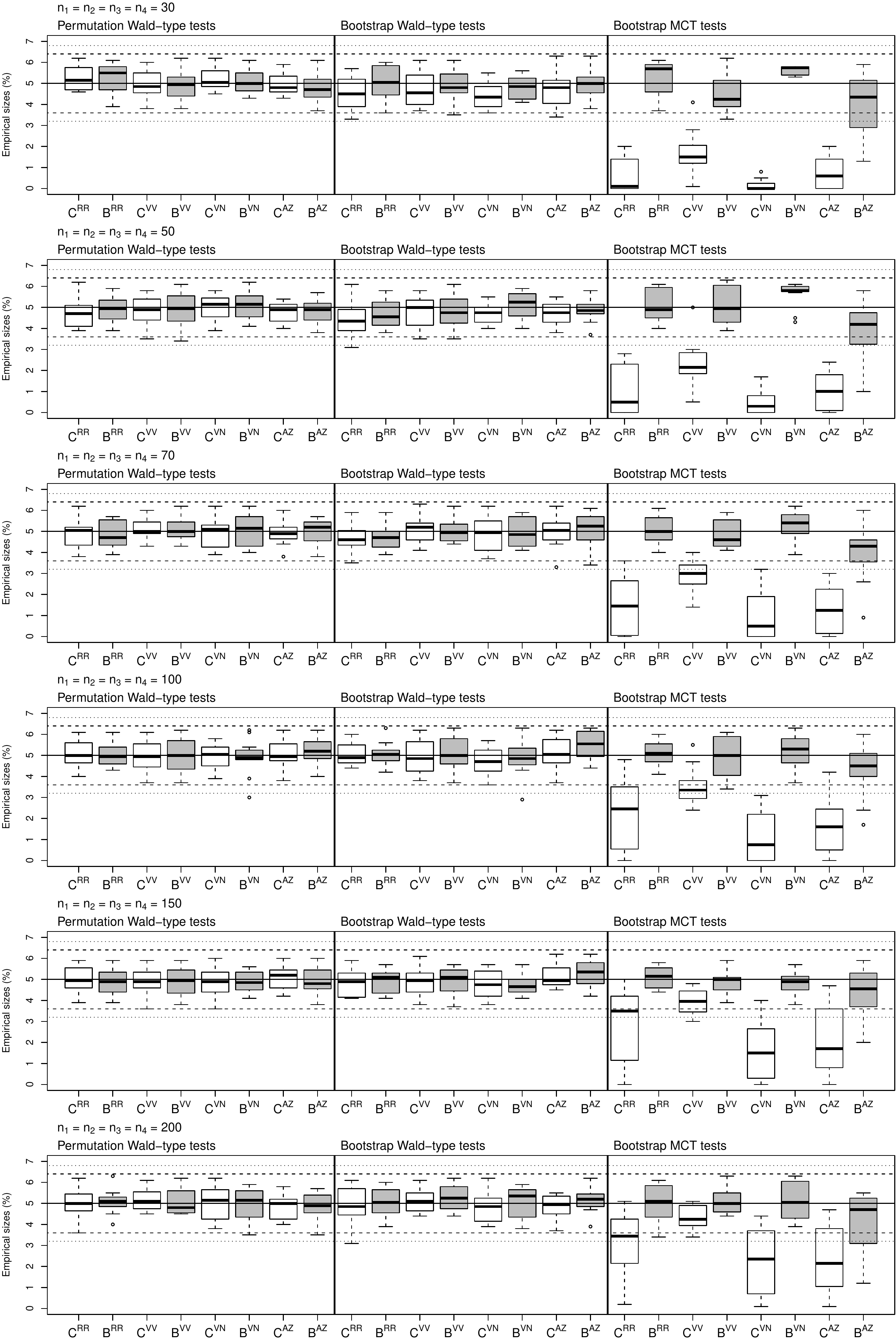}
\caption[Box-and-whisker plots for the empirical sizes of the permutation and bootstrap tests obtained under normal distribution]{Box-and-whisker plots for the empirical sizes (as percentages) of the permutation and bootstrap tests obtained under normal distribution. The solid, dashed, and dotted lines represent the significance level $\alpha=5\%$ and the $95\%$ and $99\%$ binomial proportion confidence intervals $[3.6\%,6.4\%]$ and $[3.2\%,6.8\%]$, respectively.}
\label{fig_sm_2}
\end{figure}
\begin{figure}[!t]
\centering
\includegraphics[width=
0.95\textwidth,height=0.85\textheight]{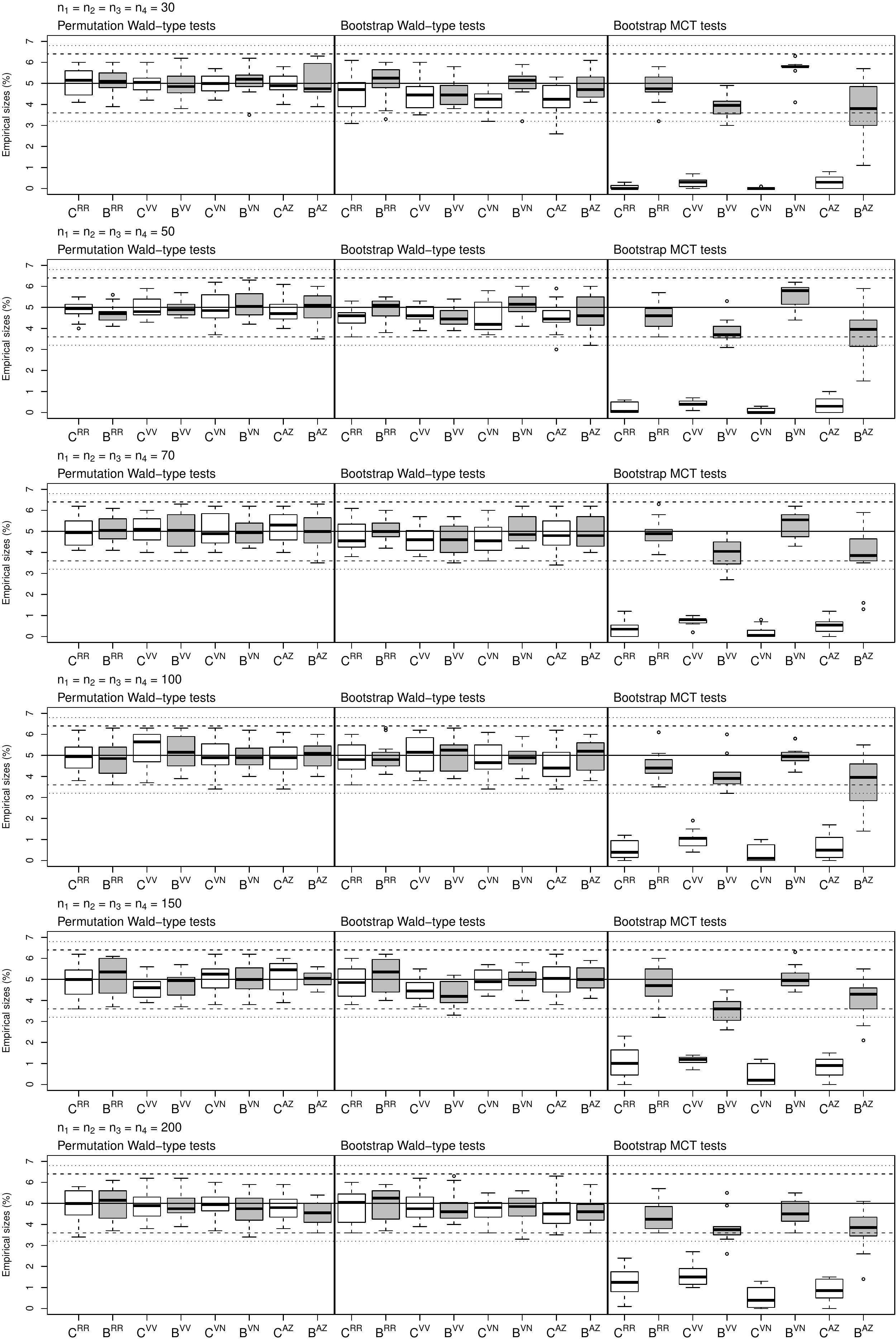}
\caption[Box-and-whisker plots for the empirical sizes of the permutation and bootstrap tests obtained under $t_5$-distribution]{Box-and-whisker plots for the empirical sizes (as percentages) of the permutation and bootstrap tests obtained under $t_5$-distribution. The solid, dashed, and dotted lines represent the significance level $\alpha=5\%$ and the $95\%$ and $99\%$ binomial proportion confidence intervals $[3.6\%,6.4\%]$ and $[3.2\%,6.8\%]$, respectively.}
\label{fig_sm_3}
\end{figure}
\begin{figure}[!t]
\centering
\includegraphics[width=
0.95\textwidth,height=0.85\textheight]{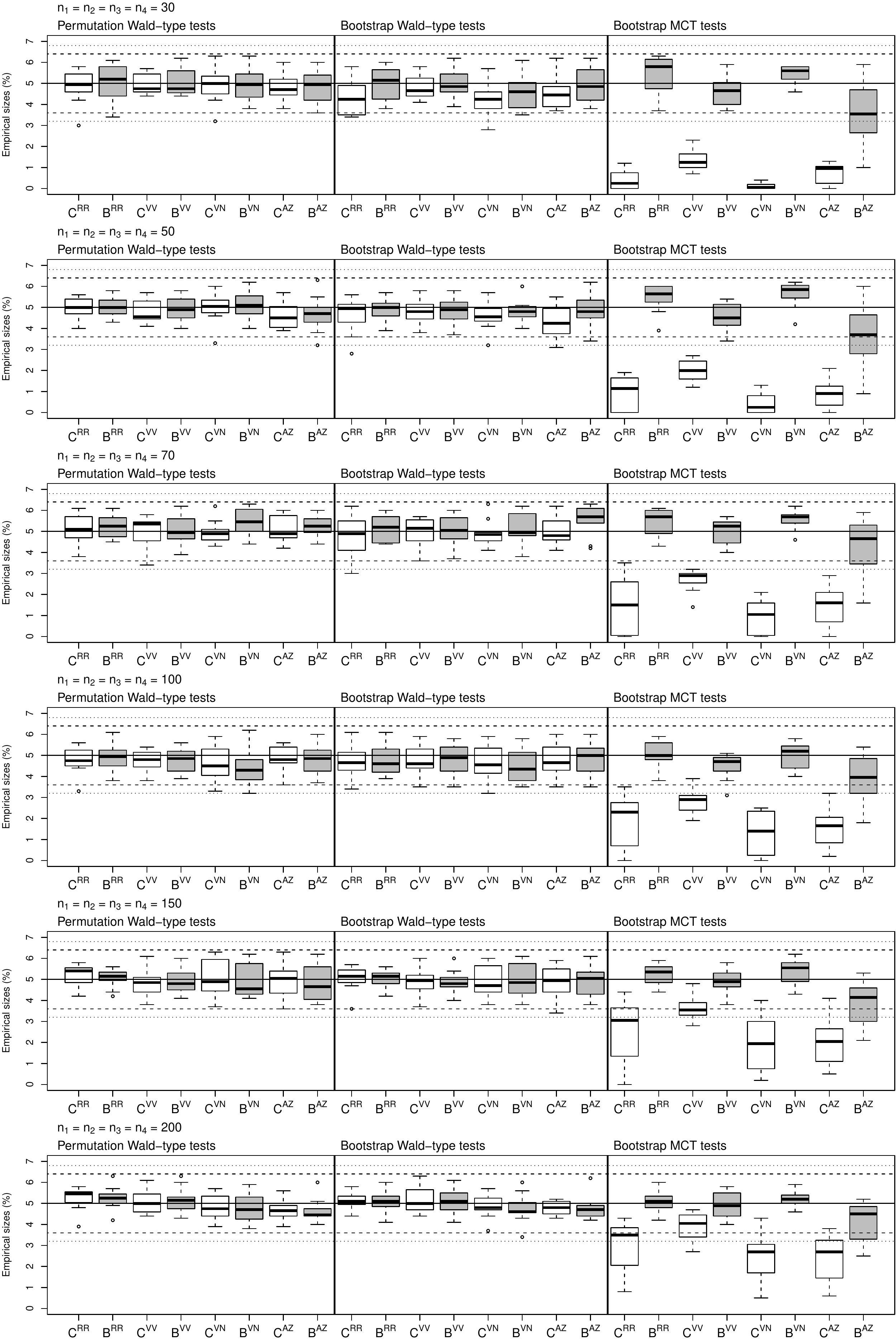}
\caption[Box-and-whisker plots for the empirical sizes of the permutation and bootstrap tests obtained under $\chi_{10}^2$-distribution]{Box-and-whisker plots for the empirical sizes (as percentages) of the permutation and bootstrap tests obtained under $\chi_{10}^2$-distribution. The solid, dashed, and dotted lines represent the significance level $\alpha=5\%$ and the $95\%$ and $99\%$ binomial proportion confidence intervals $[3.6\%,6.4\%]$ and $[3.2\%,6.8\%]$, respectively.}
\label{fig_sm_4}
\end{figure}
\begin{figure}[!t]
\centering
\includegraphics[width=
0.95\textwidth,height=0.85\textheight]{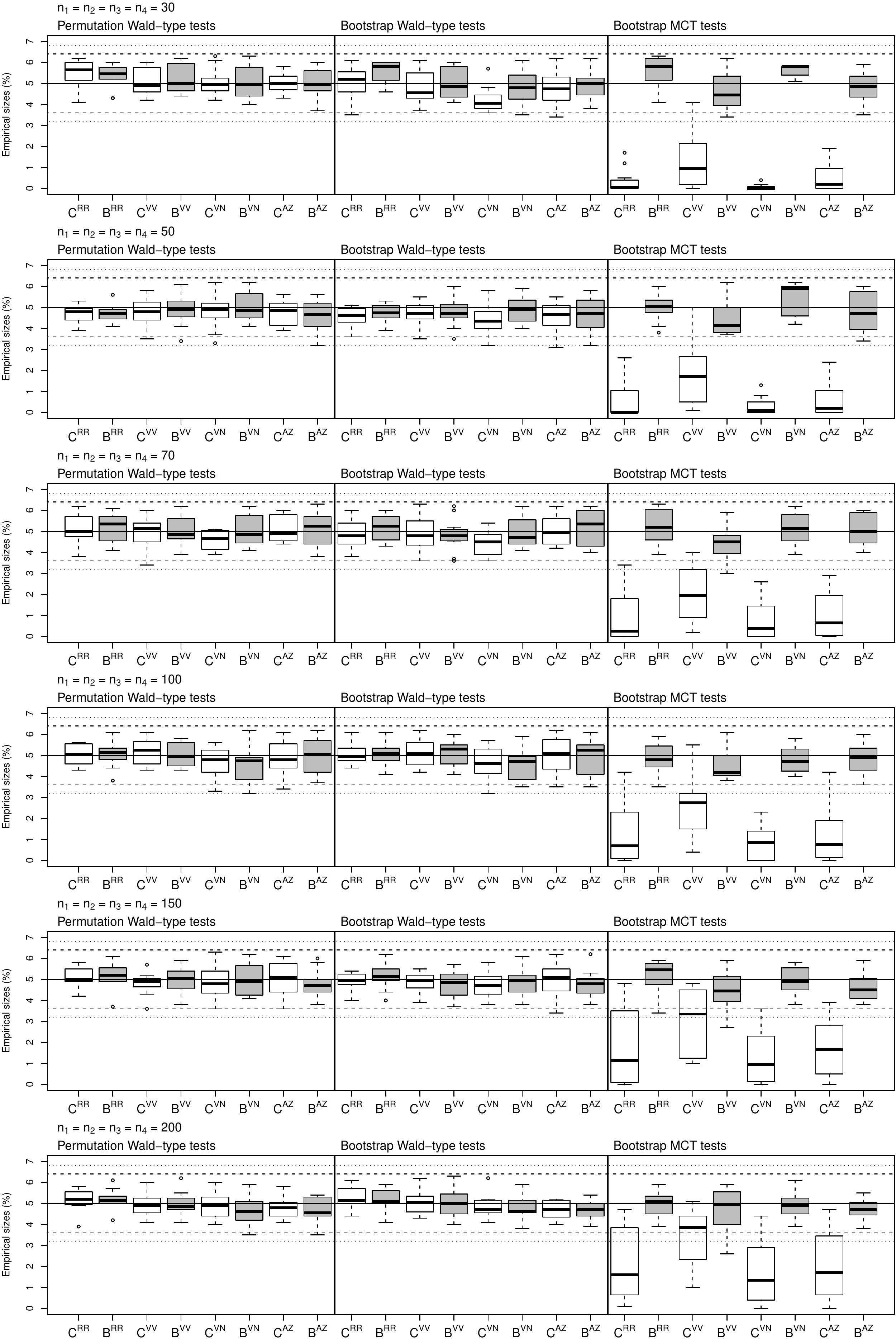}
\caption[Box-and-whisker plots for the empirical sizes of the permutation and bootstrap tests obtained for $\rho=0.1$]{Box-and-whisker plots for the empirical sizes (as percentages) of the permutation and bootstrap tests obtained for $\rho=0.1$. The solid, dashed, and dotted lines represent the significance level $\alpha=5\%$ and the $95\%$ and $99\%$ binomial proportion confidence intervals $[3.6\%,6.4\%]$ and $[3.2\%,6.8\%]$, respectively.}
\label{fig_sm_5}
\end{figure}
\begin{figure}[!t]
\centering
\includegraphics[width=
0.95\textwidth,height=0.85\textheight]{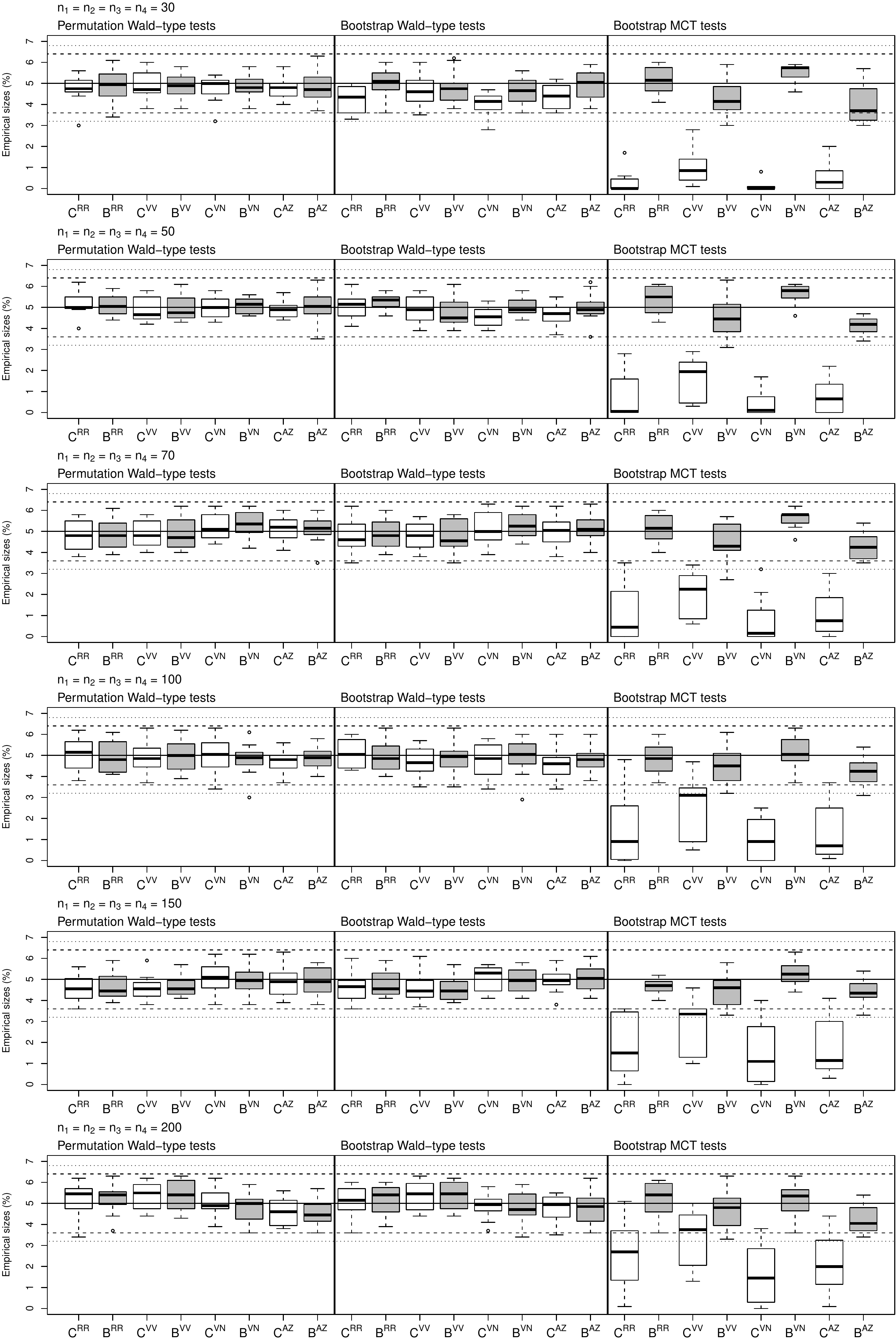}
\caption[Box-and-whisker plots for the empirical sizes of the permutation and bootstrap tests obtained for $\rho=0.4$]{Box-and-whisker plots for the empirical sizes (as percentages) of the permutation and bootstrap tests obtained for $\rho=0.4$. The solid, dashed, and dotted lines represent the significance level $\alpha=5\%$ and the $95\%$ and $99\%$ binomial proportion confidence intervals $[3.6\%,6.4\%]$ and $[3.2\%,6.8\%]$, respectively.}
\label{fig_sm_6}
\end{figure}
\begin{figure}[!t]
\centering
\includegraphics[width=
0.95\textwidth,height=0.85\textheight]{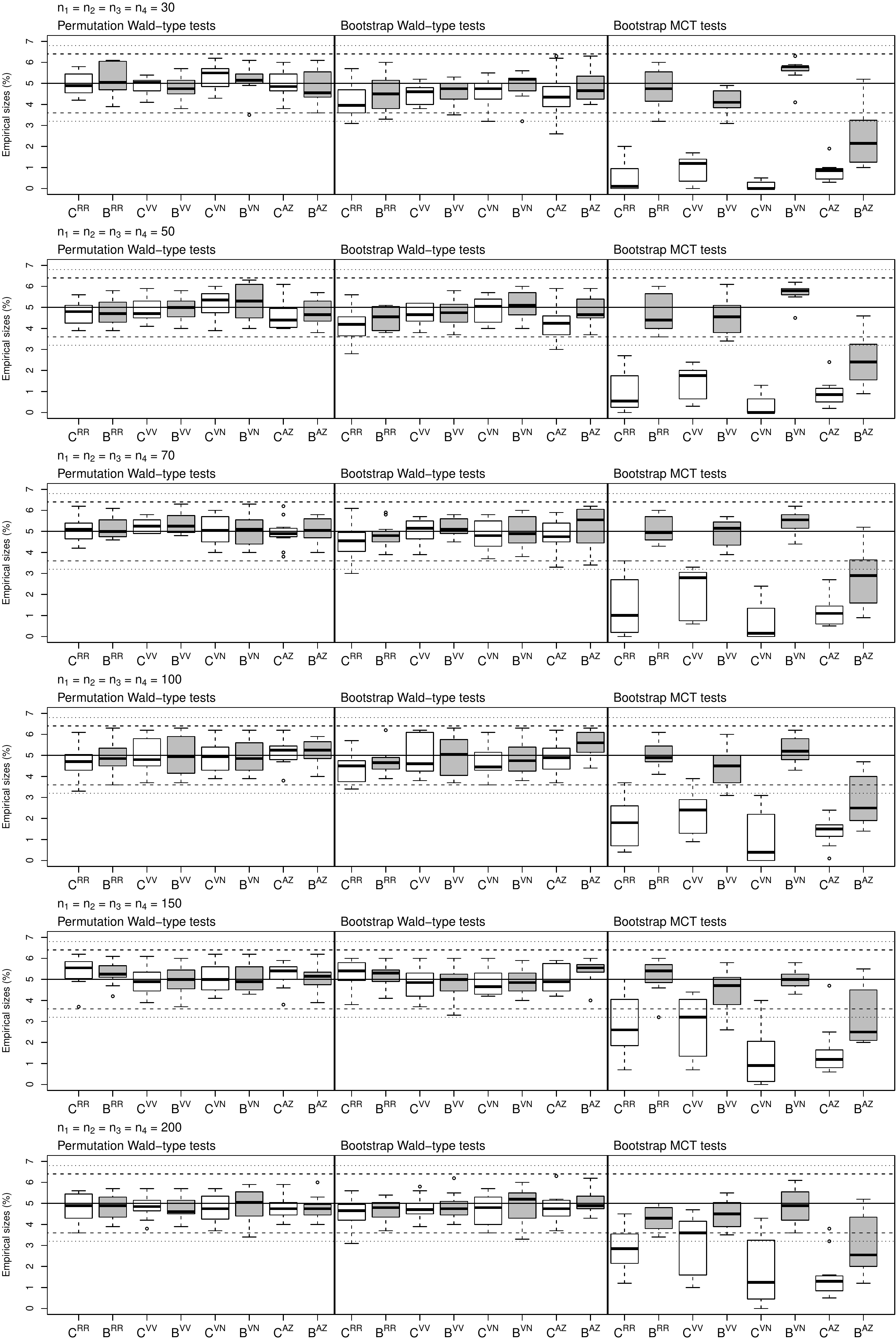}
\caption[Box-and-whisker plots for the empirical sizes of the permutation and bootstrap tests obtained for $\rho=0.7$]{Box-and-whisker plots for the empirical sizes (as percentages) of the permutation and bootstrap tests obtained for $\rho=0.7$. The solid, dashed, and dotted lines represent the significance level $\alpha=5\%$ and the $95\%$ and $99\%$ binomial proportion confidence intervals $[3.6\%,6.4\%]$ and $[3.2\%,6.8\%]$, respectively.}
\label{fig_sm_7}
\end{figure}
\begin{figure}[!t]
\centering
\includegraphics[width=
0.95\textwidth,height=0.33\textheight]{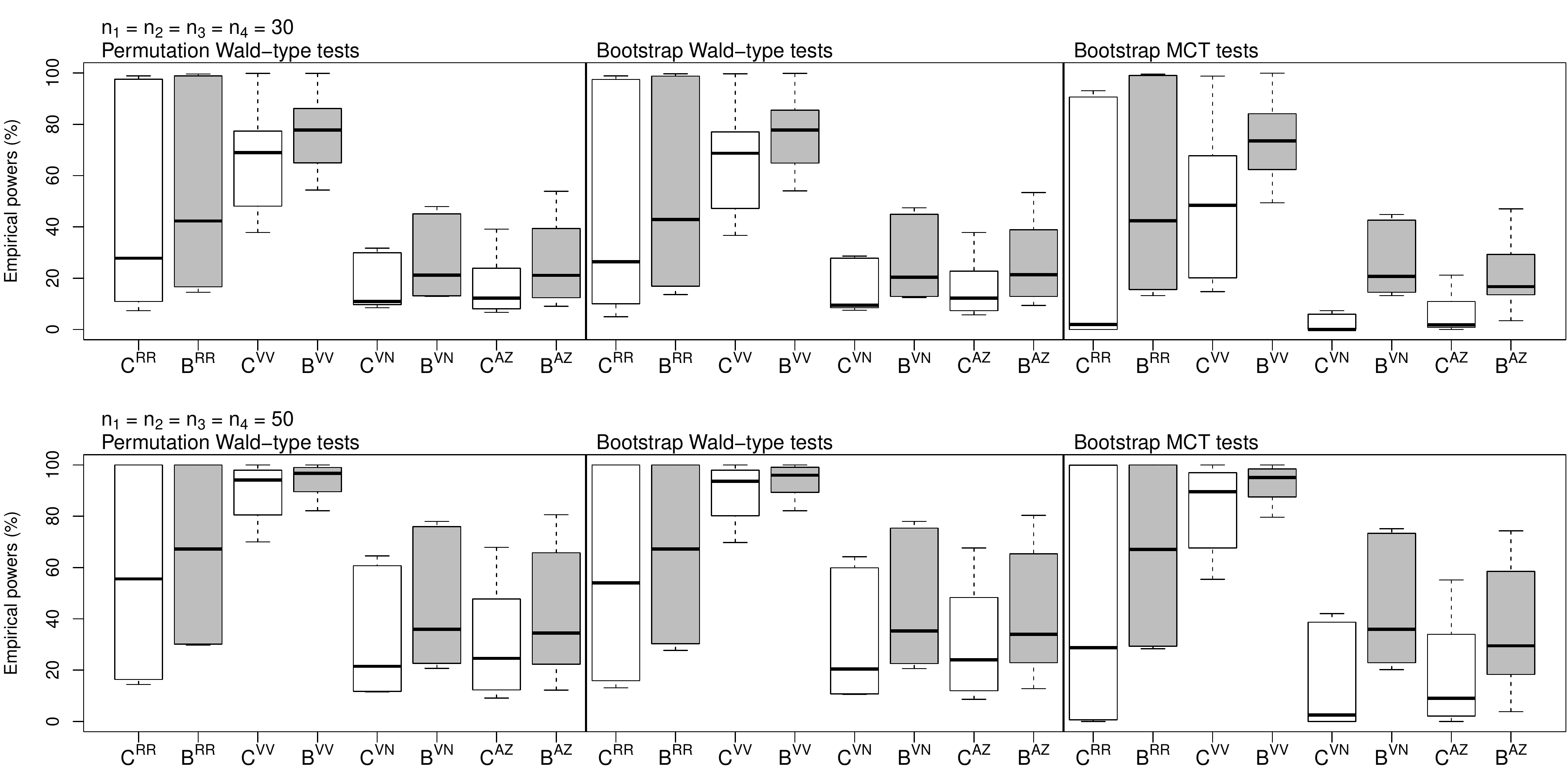}
\caption[Box-and-whisker plots for the empirical powers of the permutation and bootstrap tests obtained under normal distribution]{Box-and-whisker plots for the empirical powers (as percentages) of the permutation and bootstrap tests obtained under normal distribution.}
\label{fig_sm_8}
\end{figure}
\begin{figure}[!t]
\centering
\includegraphics[width=
0.95\textwidth,height=0.33\textheight]{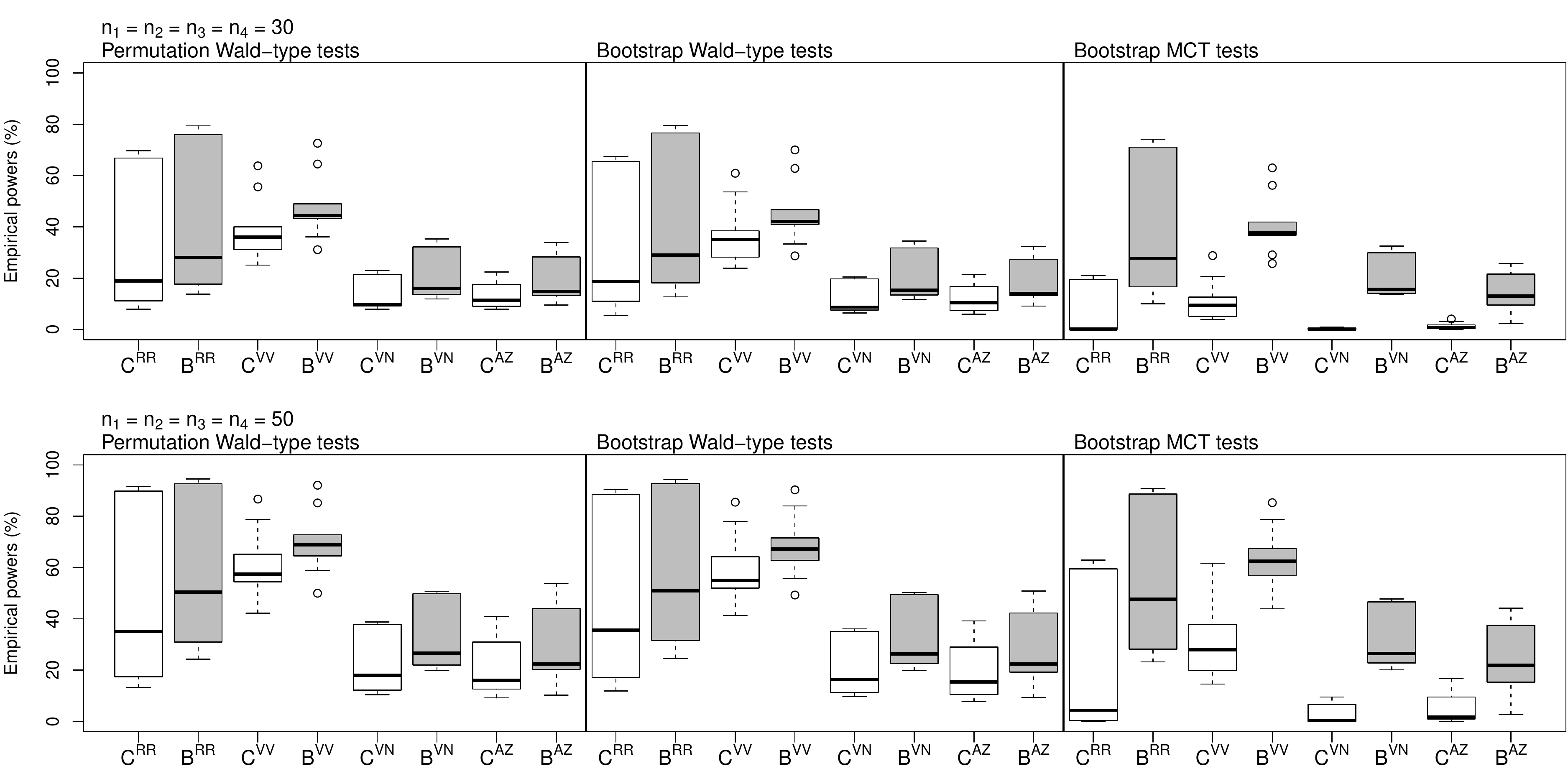}
\caption[Box-and-whisker plots for the empirical powers of the permutation and bootstrap tests obtained under $t_5$-distribution]{Box-and-whisker plots for the empirical powers (as percentages) of the permutation and bootstrap tests obtained under $t_5$-distribution.}
\label{fig_sm_9}
\end{figure}
\begin{figure}[!t]
\centering
\includegraphics[width=
0.95\textwidth,height=0.33\textheight]{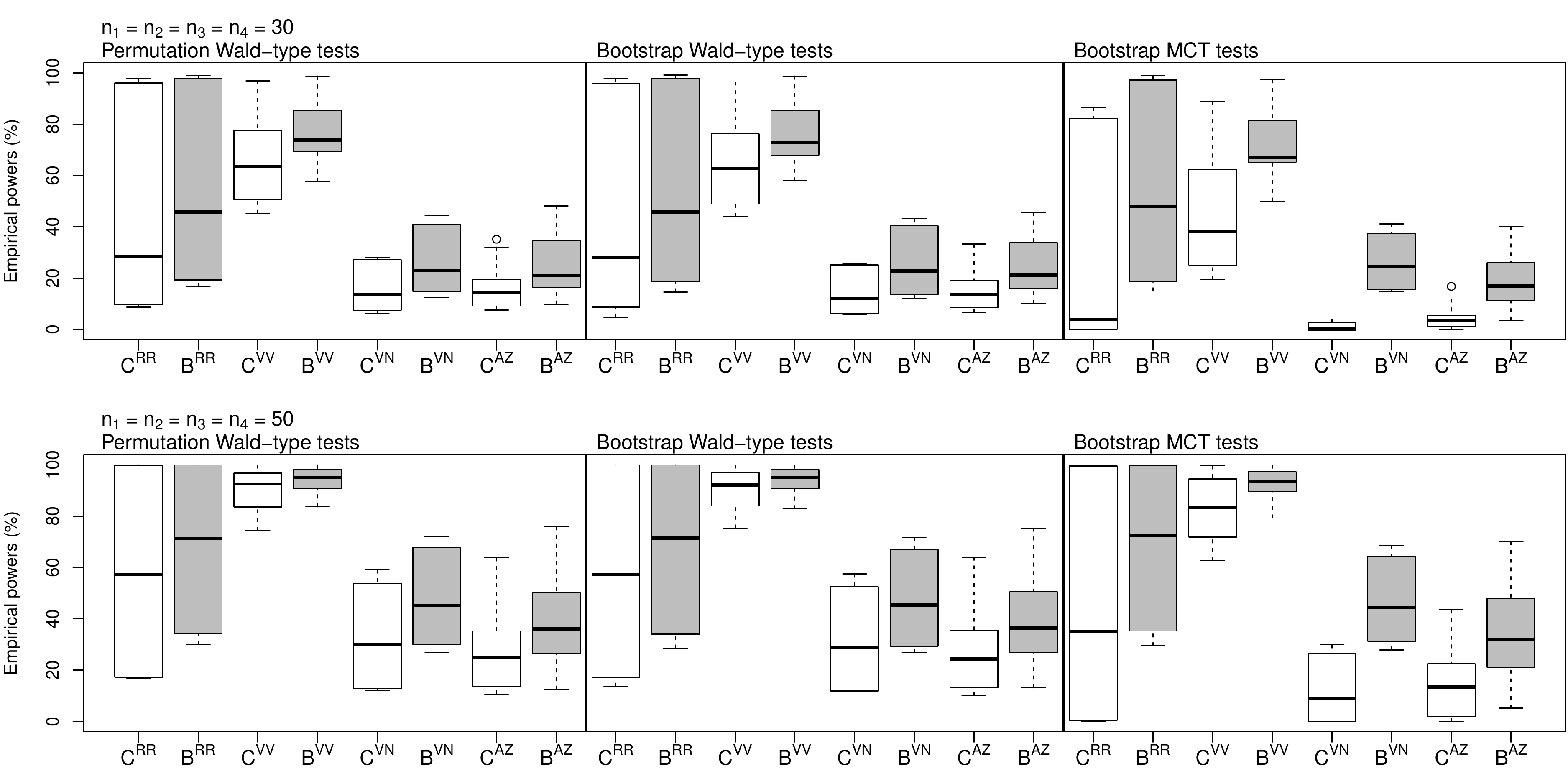}
\caption[Box-and-whisker plots for the empirical powers of the permutation and bootstrap tests obtained under $\chi_{10}^2$-distribution]{Box-and-whisker plots for the empirical powers (as percentages) of the permutation and bootstrap tests obtained under $\chi_{10}^2$-distribution.}
\label{fig_sm_10}
\end{figure}
\begin{figure}[!t]
\centering
\includegraphics[width=
0.95\textwidth,height=0.33\textheight]{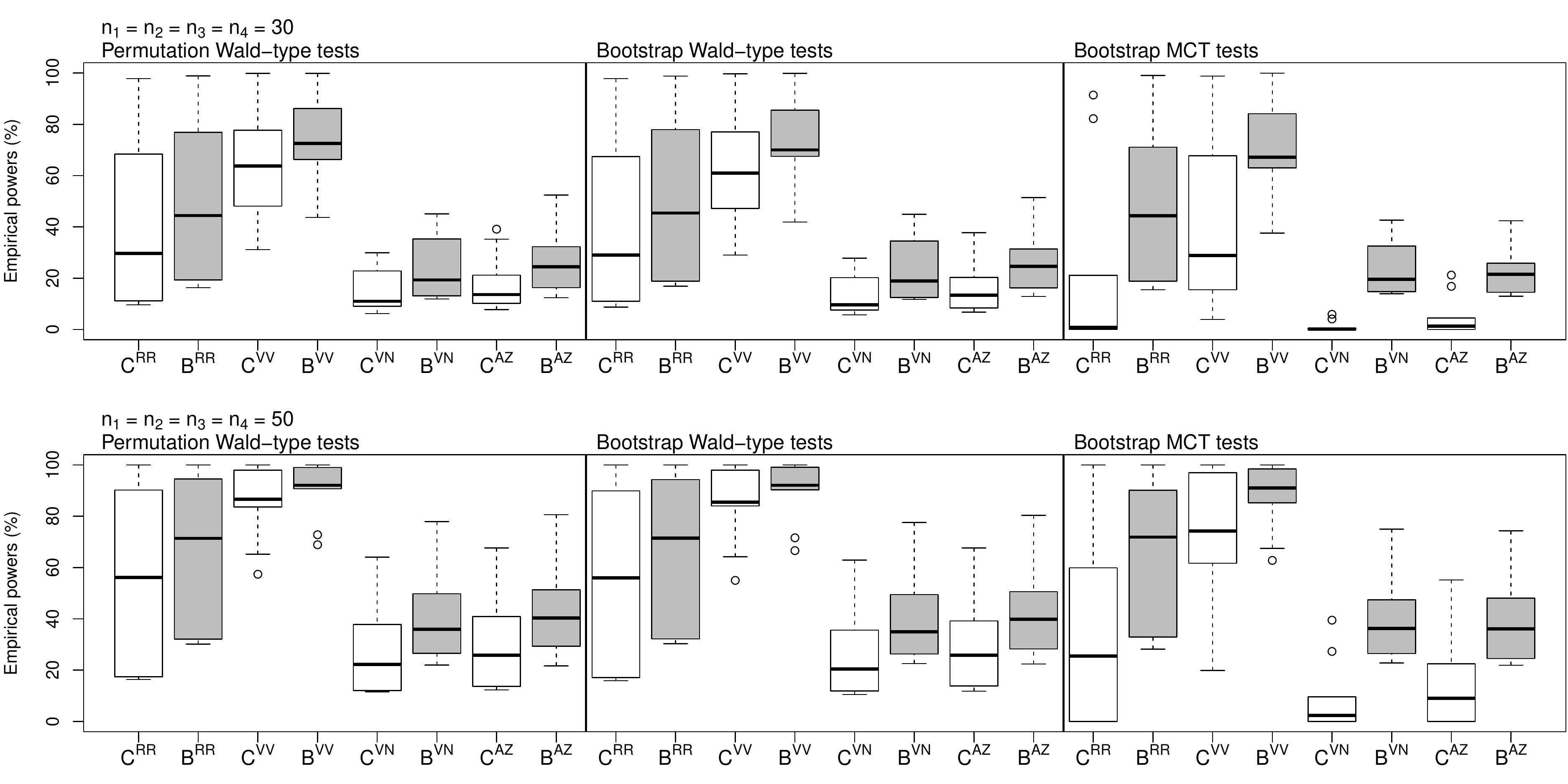}
\caption[Box-and-whisker plots for the empirical powers of the permutation and bootstrap tests obtained for $\rho=0.1$]{Box-and-whisker plots for the empirical powers (as percentages) of the permutation and bootstrap tests obtained for $\rho=0.1$.}
\label{fig_sm_11}
\end{figure}
\begin{figure}[!t]
\centering
\includegraphics[width=
0.95\textwidth,height=0.33\textheight]{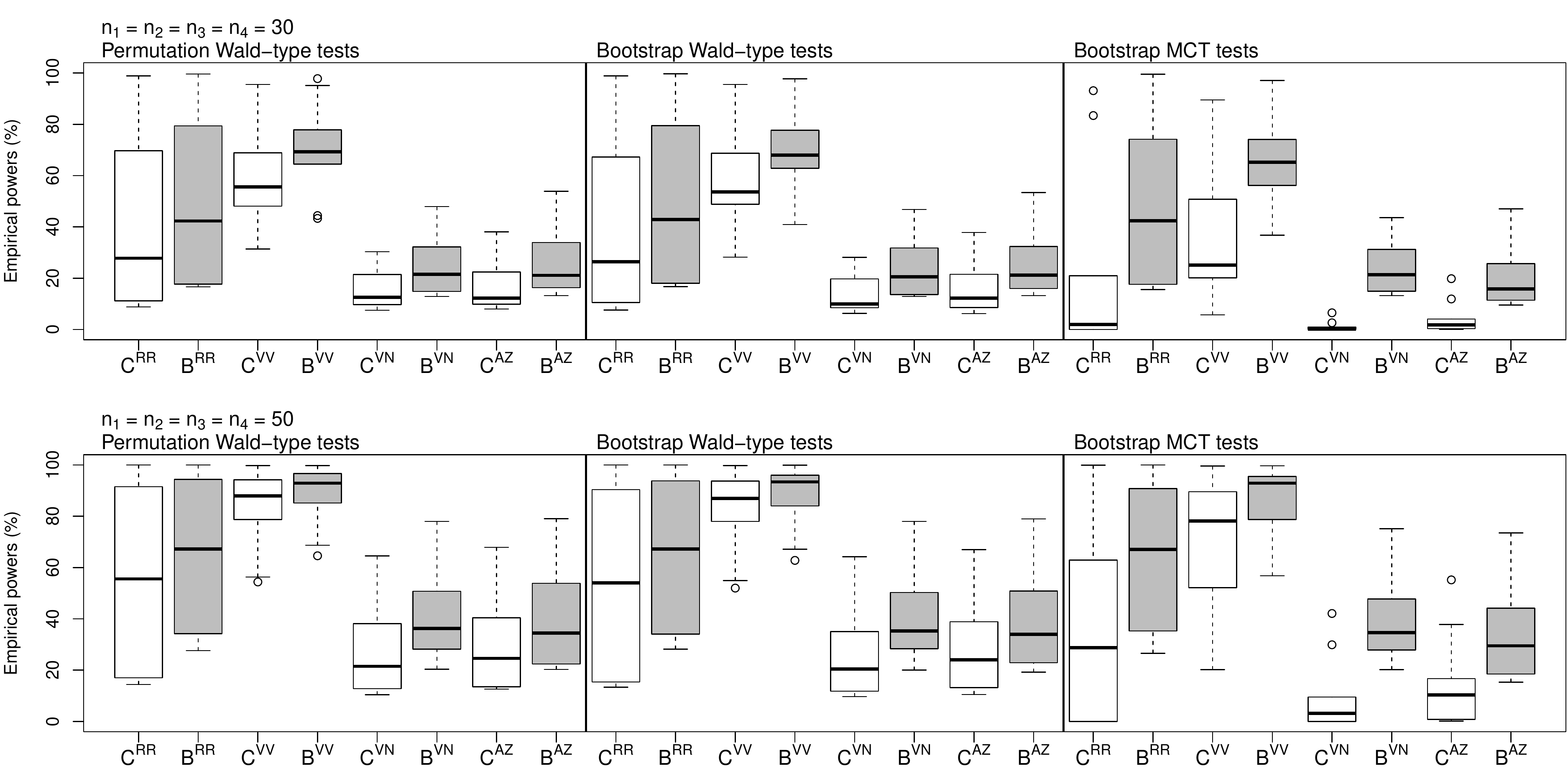}
\caption[Box-and-whisker plots for the empirical powers of the permutation and bootstrap tests obtained for $\rho=0.4$]{Box-and-whisker plots for the empirical powers (as percentages) of the permutation and bootstrap tests obtained for $\rho=0.4$.}
\label{fig_sm_12}
\end{figure}
\begin{figure}[!t]
\centering
\includegraphics[width=
0.95\textwidth,height=0.33\textheight]{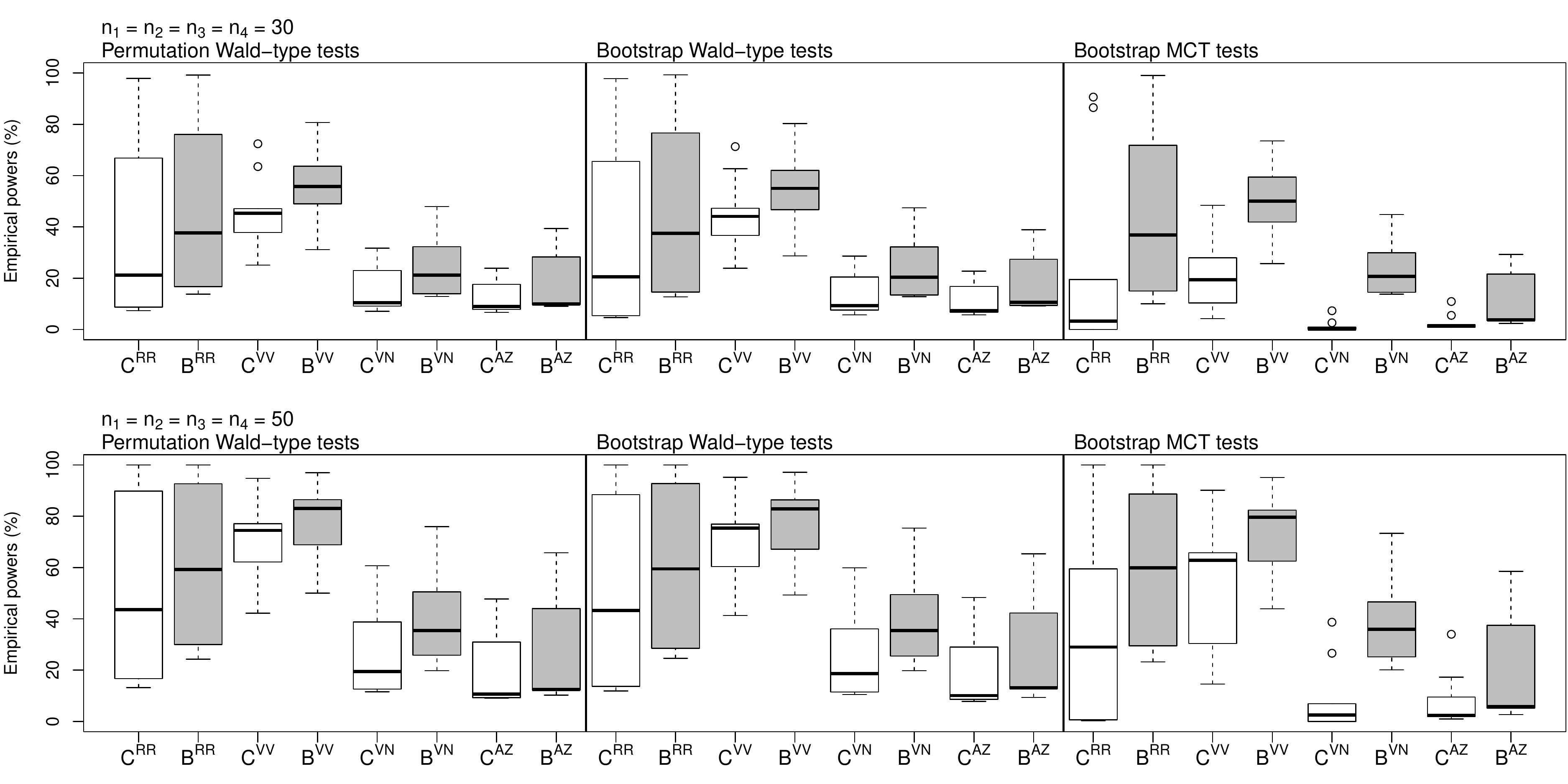}
\caption[Box-and-whisker plots for the empirical powers of the permutation and bootstrap tests obtained for $\rho=0.7$]{Box-and-whisker plots for the empirical powers (as percentages) of the permutation and bootstrap tests obtained for $\rho=0.7$.}
\label{fig_sm_13}
\end{figure}

This supplement contains all results of simulation studies of Sections 5 and 6 of the main paper. They are summarized in Figures~\ref{fig_sm_1}-\ref{fig_sm_13} and presented in Tables~\ref{tab_sm_1}-\ref{tab_sm_8}.

\listoffigures

\listoftables

\clearpage

\begingroup\fontsize{10}{12}\selectfont
\begin{longtable}[t]{lrrrrrrrrr}
\caption[Empirical sizes of all tests obtained in simulation study based on real data example]{Empirical sizes (as percentages) of all tests obtained in simulation study based on real data example of Section~6 in the main paper (Method: $a$ - asymptotic Wald-type test; $\pi$ - permutation Wald-type test; $b$ - bootstrap Wald-type test; $\max,a$ - asymptotic multiple contrast test; $\max,b$ - bootstrap multiple contrast test)}\\
\hline
Distr.&$\varphi_{C^{RR}}$ & $\varphi_{B^{RR}}$ & $\varphi_{C^{VV}}$ & $\varphi_{B^{VV}}$ & $\varphi_{C^{VN}}$ & $\varphi_{B^{VN}}$ & $\varphi_{C^{AZ}}$ & $\varphi_{B^{AZ}}$ & Method\\
\hline
\endfirsthead
\caption[]{Empirical sizes (as percentages) of all tests obtained in simulation study based on real data example of Section~6 in the main paper (Method: $a$ - asymptotic Wald-type test; $\pi$ - permutation Wald-type test; $b$ - bootstrap Wald-type test; $\max,a$ - asymptotic multiple contrast test; $\max,b$ - bootstrap multiple contrast test) \textit{(continued)}}\\
\hline
Distr.&$\varphi_{C^{RR}}$ & $\varphi_{B^{RR}}$ & $\varphi_{C^{VV}}$ & $\varphi_{B^{VV}}$ & $\varphi_{C^{VN}}$ & $\varphi_{B^{VN}}$ & $\varphi_{C^{AZ}}$ & $\varphi_{B^{AZ}}$ & Method\\
\hline
\endhead
$N$& 8.7 & 8.0 & 7.2 & 6.9 & 8.4 & 7.6 & 7.1 & 6.9 & $a$\\
& 4.7 & 4.9 & 4.9 & 4.7 & 4.8 & 4.8 & 4.2 & 4.8 & $\pi$\\
& 4.8 & 5.1 & 4.8 & 4.9 & 4.5 & 4.4 & 4.5 & 4.7 & $b$\\
& 7.8 & 7.7 & 6.3 & 6.5 & 7.4 & 7.0 & 7.0 & 6.2 & $max,a$\\
& 4.1 & 5.4 & 3.5 & 5.0 & 3.4 & 5.3 & 3.5 & 4.4 & $max,b$\\
\addlinespace
$t_5$& 14.4 & 16.7 & 9.8 & 12.3 & 16.4 & 17.2 & 9.4 & 12.9 & $a$\\
& 5.0 & 5.0 & 4.3 & 4.5 & 5.6 & 5.3 & 4.9 & 4.6 & $\pi$\\
& 4.5 & 4.9 & 4.1 & 4.8 & 5.5 & 4.8 & 4.2 & 4.2 & $b$\\
\end{longtable}
\endgroup{}

\clearpage

\begingroup\fontsize{10}{12}\selectfont

\begin{longtable}[t]{llrrrrrr}
\caption[Empirical powers of the multiple contrast tests obtained in simulation based on real data example]{Empirical powers (as percentages) of the multiple contrast tests obtained in simulation study based on real data example of Section~6 in the main paper}\\
\hline
Distr.&Test& CH-EH&CH-JH&CH-GB&EH-JH&EH-GB&JH-GB\\
\hline
\endfirsthead
\caption[]{Empirical powers (as percentages) of the multiple contrast tests obtained in simulation study based on real data example of Section~6 in the main paper \textit{(continued)}}\\
\hline
Distr.&Test& CH-EH&CH-JH&CH-GB&EH-JH&EH-GB&JH-GB\\
\hline
\endhead
$N$&$\varphi_{\max,C^{RR}}$ & 14.5 & 62.4 & 100 & 17.7 & 100 & 100.0\\
&$\varphi_{\max,B^{RR}}$ & 14.2 & 59.8 & 100 & 16.7 & 100 & 100.0\\
&$\varphi_{\max,C^{VV}}$ & 37.3 & 27.3 & 100 & 1.5 & 100 & 100.0\\
&$\varphi_{\max,B^{VV}}$ & 39.2 & 23.4 & 100 & 1.7 & 100 & 100.0\\
&$\varphi_{\max,C^{VN}}$ & 42.9 & 10.3 & 100 & 5.2 & 100 & 100.0\\
&$\varphi_{\max,B^{VN}}$ & 42.1 & 8.5 & 100 & 6.9 & 100 & 100.0\\
&$\varphi_{\max,C^{AZ}}$ & 81.7 & 28.9 & 100 & 7.3 & 100 & 100.0\\
&$\varphi_{\max,B^{AZ}}$ & 81.5 & 24.5 & 100 & 10.2 & 100 & 100.0\\
\end{longtable}
\endgroup{}

\clearpage

\begingroup\fontsize{10}{12}\selectfont

\begin{longtable}[t]{rrrrrrrrrrrr}
\caption[Empirical sizes of all tests obtained under normal distribution]{Empirical sizes (as percentages) of all tests obtained under normal distribution (Method: $a$ - asymptotic Wald-type test; $\pi$ - permutation Wald-type test; $b$ - bootstrap Wald-type test; $\max,a$ - asymptotic multiple contrast test; $\max,b$ - bootstrap multiple contrast test)}\\
\toprule
$\rho$ & $(C_1,C_2,C_3,C_4)$ & $n_i$ & $\varphi_{C^{RR}}$ & $\varphi_{B^{RR}}$ & $\varphi_{C^{VV}}$ & $\varphi_{B^{VV}}$ & $\varphi_{C^{VN}}$ & $\varphi_{B^{VN}}$ & $\varphi_{C^{AZ}}$ & $\varphi_{B^{AZ}}$ & Method\\
\midrule
\endfirsthead
\caption[]{Empirical sizes (as percentages) of all tests obtained under normal distribution (Method: $a$ - asymptotic Wald-type test; $\pi$ - permutation Wald-type test; $b$ - bootstrap Wald-type test; $\max,a$ - asymptotic multiple contrast test; $\max,b$ - bootstrap multiple contrast test) \textit{(continued)}}\\
\toprule
$\rho$ & $(C_1,C_2,C_3,C_4)$ & $n_i$ & $\varphi_{C^{RR}}$ & $\varphi_{B^{RR}}$ & $\varphi_{C^{VV}}$ & $\varphi_{B^{VV}}$ & $\varphi_{C^{VN}}$ & $\varphi_{B^{VN}}$ & $\varphi_{C^{AZ}}$ & $\varphi_{B^{AZ}}$ & Method\\
\midrule
\endhead

\endfoot
\bottomrule
\endlastfoot
0.1 & $(0.1,0.1,0.1,0.1)$ & 30 & 17.5 & 17.7 & 9.0 & 8.9 & 15.1 & 12.6 & 11.2 & 10.7 & $a$\\
&  &  & 6.1 & 5.7 & 6.0 & 5.7 & 4.5 & 4.3 & 5.3 & 5.2 & $\pi$\\
&  &  & 5.7 & 6.0 & 6.1 & 6.0 & 3.9 & 4.3 & 5.0 & 5.0 & $b$\\
&  &  & 15.9 & 16.1 & 8.4 & 8.7 & 12.2 & 11.0 & 9.1 & 9.3 & $max,a$\\
&  &  & 1.7 & 6.2 & 4.1 & 6.2 & 0.2 & 5.4 & 1.9 & 4.3 & $max,b$\\
\addlinespace
& & 50 & 9.7 & 9.4 & 7.5 & 7.9 & 10.3 & 9.3 & 7.6 & 7.1 & $a$\\
&  &  & 4.6 & 4.4 & 5.1 & 5.6 & 5.3 & 4.7 & 4.2 & 4.3 & $\pi$\\
&  &  & 4.6 & 4.5 & 4.9 & 5.3 & 4.9 & 5.2 & 4.5 & 4.3 & $b$\\
\end{longtable}
\endgroup{}

\clearpage

\begingroup\fontsize{10}{12}\selectfont

\begin{longtable}[t]{rrrrrrrrrrrr}
\caption[Empirical sizes of all tests obtained under $t_5$-distribution]{Empirical sizes (as percentages) of all tests obtained under $t_5$-distribution (Method: $a$ - asymptotic Wald-type test; $\pi$ - permutation Wald-type test; $b$ - bootstrap Wald-type test; $\max,a$ - asymptotic multiple contrast test; $\max,b$ - bootstrap multiple contrast test)}\\
\toprule
$\rho$ & $(C_1,C_2,C_3,C_4)$ & $n_i$ & $\varphi_{C^{RR}}$ & $\varphi_{B^{RR}}$ & $\varphi_{C^{VV}}$ & $\varphi_{B^{VV}}$ & $\varphi_{C^{VN}}$ & $\varphi_{B^{VN}}$ & $\varphi_{C^{AZ}}$ & $\varphi_{B^{AZ}}$ & Method\\
\midrule
\endfirsthead
\caption[]{Empirical sizes (as percentages) of all tests obtained under $t_5$-distribution (Method: $a$ - asymptotic Wald-type test; $\pi$ - permutation Wald-type test; $b$ - bootstrap Wald-type test; $\max,a$ - asymptotic multiple contrast test; $\max,b$ - bootstrap multiple contrast test) \textit{(continued)}}\\
\toprule
$\rho$ & $(C_1,C_2,C_3,C_4)$ & $n_i$ & $\varphi_{C^{RR}}$ & $\varphi_{B^{RR}}$ & $\varphi_{C^{VV}}$ & $\varphi_{B^{VV}}$ & $\varphi_{C^{VN}}$ & $\varphi_{B^{VN}}$ & $\varphi_{C^{AZ}}$ & $\varphi_{B^{AZ}}$ & Method\\
\midrule
\endhead

\endfoot
\bottomrule
\endlastfoot
0.1 & $(0.1,0.1,0.1,0.1)$ & 30 & 32.1 & 34.9 & 11.8 & 17.4 & 25.8 & 28.4 & 15.1 & 20.2 & $a$\\
&  &  & 5.7 & 5.8 & 4.7 & 4.7 & 5.3 & 4.8 & 4.9 & 4.7 & $\pi$\\
&  &  & 5.3 & 6.0 & 4.2 & 4.1 & 4.5 & 4.7 & 4.3 & 4.2 & $b$\\
&  &  & 27.1 & 32.5 & 8.3 & 15.9 & 21.3 & 25.9 & 10.3 & 19.0 & $max,a$\\
&  &  & 0.3 & 5.8 & 0.7 & 3.4 & 0.1 & 5.9 & 0.8 & 3.5 & $max,b$\\
\addlinespace
& & 50 & 19.9 & 23.0 & 10.3 & 14.9 & 18.5 & 21.9 & 10.4 & 14.0 & $a$\\
&  &  & 5.1 & 4.4 & 5.2 & 4.9 & 5.1 & 5.8 & 4.7 & 3.9 & $\pi$\\
&  &  & 4.4 & 4.6 & 5.1 & 4.6 & 5.3 & 5.4 & 4.7 & 3.2 & $b$\\
\end{longtable}
\endgroup{}

\clearpage

\begingroup\fontsize{10}{12}\selectfont

\begin{longtable}[t]{rrrrrrrrrrrr}
\caption[Empirical sizes of all tests obtained under $\chi_{10}^2$-distribution]{Empirical sizes (as percentages) of all tests obtained under $\chi_{10}^2$-distribution (Method: $a$ - asymptotic Wald-type test; $\pi$ - permutation Wald-type test; $b$ - bootstrap Wald-type test; $\max,a$ - asymptotic multiple contrast test; $\max,b$ - bootstrap multiple contrast test)}\\
\toprule
$\rho$ & $(C_1,C_2,C_3,C_4)$ & $n_i$ & $\varphi_{C^{RR}}$ & $\varphi_{B^{RR}}$ & $\varphi_{C^{VV}}$ & $\varphi_{B^{VV}}$ & $\varphi_{C^{VN}}$ & $\varphi_{B^{VN}}$ & $\varphi_{C^{AZ}}$ & $\varphi_{B^{AZ}}$ & Method\\
\midrule
\endfirsthead
\caption[]{Empirical sizes (as percentages) of all tests obtained under $\chi_{10}^2$-distribution (Method: $a$ - asymptotic Wald-type test; $\pi$ - permutation Wald-type test; $b$ - bootstrap Wald-type test; $\max,a$ - asymptotic multiple contrast test; $\max,b$ - bootstrap multiple contrast test) \textit{(continued)}}\\
\toprule
$\rho$ & $(C_1,C_2,C_3,C_4)$ & $n_i$ & $\varphi_{C^{RR}}$ & $\varphi_{B^{RR}}$ & $\varphi_{C^{VV}}$ & $\varphi_{B^{VV}}$ & $\varphi_{C^{VN}}$ & $\varphi_{B^{VN}}$ & $\varphi_{C^{AZ}}$ & $\varphi_{B^{AZ}}$ & Method\\
\midrule
\endhead

\endfoot
\bottomrule
\endlastfoot
0.1 & $(0.1,0.1,0.1,0.1)$ & 30 & 23.6 & 24.2 & 12.0 & 13.0 & 18.6 & 19.8 & 12.7 & 14.3 & $a$\\
&  &  & 5.8 & 6.0 & 4.5 & 4.6 & 4.2 & 4.1 & 5.3 & 4.8 & $\pi$\\
&  &  & 5.8 & 6.0 & 4.4 & 5.0 & 3.8 & 3.9 & 5.0 & 4.7 & $b$\\
&  &  & 21.2 & 22.5 & 9.3 & 12.5 & 14.7 & 17.2 & 10.4 & 14.0 & $max,a$\\
&  &  & 1.2 & 6.3 & 1.9 & 3.9 & 0.4 & 5.4 & 1.1 & 4.4 & $max,b$\\
\addlinespace
& & 50 & 13.7 & 14.2 & 8.0 & 8.7 & 10.2 & 10.7 & 8.1 & 9.4 & $a$\\
&  &  & 4.8 & 4.9 & 4.1 & 4.9 & 3.3 & 4.1 & 3.9 & 3.2 & $\pi$\\
&  &  & 4.9 & 4.5 & 4.4 & 4.7 & 3.2 & 4.0 & 3.1 & 3.4 & $b$\\
\end{longtable}
\endgroup{}

\clearpage

\begingroup\fontsize{10}{12}\selectfont

\begin{longtable}[!H]{rrrrrrrrrrrr}
\caption[Empirical powers of the permutation and bootstrap tests under normal distribution]{Empirical powers (as percentages) of the permutation and bootstrap tests under normal distribution (Method: $\pi$ - permutation Wald-type test; $b$ - bootstrap Wald-type test; $\max,b$ - bootstrap multiple contrast test)}\\
\toprule
$\rho$ & $(C_1,C_2,C_3,C_4)$ & $n_i$ & $\varphi_{C^{RR}}$ & $\varphi_{B^{RR}}$ & $\varphi_{C^{VV}}$ & $\varphi_{B^{VV}}$ & $\varphi_{C^{VN}}$ & $\varphi_{B^{VN}}$ & $\varphi_{C^{AZ}}$ & $\varphi_{B^{AZ}}$ & Method\\
\midrule
\endfirsthead
\caption[]{Empirical powers (as percentages) of the permutation and bootstrap tests under normal distribution (Method: $\pi$ - permutation Wald-type test; $b$ - bootstrap Wald-type test; $\max,b$ - bootstrap multiple contrast test) \textit{(continued)}}\\
\toprule
$\rho$ & $(C_1,C_2,C_3,C_4)$ & $n_i$ & $\varphi_{C^{RR}}$ & $\varphi_{B^{RR}}$ & $\varphi_{C^{VV}}$ & $\varphi_{B^{VV}}$ & $\varphi_{C^{VN}}$ & $\varphi_{B^{VN}}$ & $\varphi_{C^{AZ}}$ & $\varphi_{B^{AZ}}$ & Method\\
\midrule
\endhead

\endfoot
\bottomrule
\endlastfoot
0.1 & $(0.1,0.1,0.1,0.15)$ & 30 & 97.8 & 98.9 & 99.8 & 99.8 & 29.9 & 45.1 & 39.1 & 52.4 & $\pi$\\
 &  &  & 97.8 & 98.8 & 99.7 & 99.8 & 27.8 & 44.9 & 37.7 & 51.4 & $b$\\
 &  &  & 91.4 & 99.0 & 98.8 & 99.9 & 5.9 & 42.6 & 21.2 & 42.4 & $\max,b$\\
 \addlinespace
& & 50 & 100.0 & 100.0 & 100.0 & 100.0 & 64.1 & 77.9 & 67.6 & 80.6 & $\pi$\\
 &  &  & 100.0 & 100.0 & 100.0 & 100.0 & 62.9 & 77.6 & 67.6 & 80.4 & $b$\\
 &  &  & 100.0 & 100.0 & 100.0 & 100.0 & 39.5 & 75.0 & 55.2 & 74.3 & $\max,b$\\
 \addlinespace
& $(0.5,0.5,0.5,0.7)$ & 30 & 29.7 & 44.4 & 77.3 & 86.1 & 11.0 & 19.3 & 13.6 & 24.5 & $\pi$\\
 &  &  & 29.5 & 45.4 & 77.0 & 85.5 & 9.6 & 18.9 & 13.3 & 24.6 & $b$\\
\end{longtable}
\endgroup{}

\clearpage

\begingroup\fontsize{10}{12}\selectfont

\begin{longtable}[t]{rrrrrrrrrrrr}
\caption[Empirical powers of the permutation and bootstrap tests under $t_5$-distribution]{Empirical powers (as percentages) of the permutation and bootstrap tests under $t_5$-distribution (Method: $\pi$ - permutation Wald-type test; $b$ - bootstrap Wald-type test; $\max,b$ - bootstrap multiple contrast test)}\\
\toprule
$\rho$ & $(C_1,C_2,C_3,C_4)$ & $n_i$ & $\varphi_{C^{RR}}$ & $\varphi_{B^{RR}}$ & $\varphi_{C^{VV}}$ & $\varphi_{B^{VV}}$ & $\varphi_{C^{VN}}$ & $\varphi_{B^{VN}}$ & $\varphi_{C^{AZ}}$ & $\varphi_{B^{AZ}}$ & Method\\
\midrule
\endfirsthead
\caption[]{Empirical powers (as percentages) of the permutation and bootstrap tests under $t_5$-distribution (Method: $\pi$ - permutation Wald-type test; $b$ - bootstrap Wald-type test; $\max,b$ - bootstrap multiple contrast test) \textit{(continued)}}\\
\toprule
$\rho$ & $(C_1,C_2,C_3,C_4)$ & $n_i$ & $\varphi_{C^{RR}}$ & $\varphi_{B^{RR}}$ & $\varphi_{C^{VV}}$ & $\varphi_{B^{VV}}$ & $\varphi_{C^{VN}}$ & $\varphi_{B^{VN}}$ & $\varphi_{C^{AZ}}$ & $\varphi_{B^{AZ}}$ & Method\\
\midrule
\endhead

\endfoot
\bottomrule
\endlastfoot
0.1 & $(0.1,0.1,0.1,0.15)$ & 30 & 68.4 & 76.9 & 63.8 & 72.6 & 22.8 & 35.3 & 21.2 & 32.3 & $\pi$\\
 &  &  & 67.4 & 77.9 & 60.9 & 70.0 & 20.2 & 34.5 & 20.3 & 30.1 & $b$\\
 &  &  & 21.1 & 71.1 & 28.8 & 63.0 & 0.5 & 32.5 & 3.2 & 25.4 & $\max,b$\\
 \addlinespace
& & 50 & 90.2 & 94.5 & 86.7 & 92.1 & 37.8 & 49.8 & 40.9 & 51.3 & $\pi$\\
 &  &  & 89.9 & 94.3 & 85.5 & 90.3 & 35.6 & 49.5 & 39.2 & 48.8 & $b$\\
 &  &  & 59.9 & 90.1 & 61.7 & 85.3 & 6.7 & 47.4 & 15.5 & 42.8 & $\max,b$\\
 \addlinespace
& $(0.5,0.5,0.5,0.7)$ & 30 & 20.3 & 31.6 & 38.5 & 45.1 & 9.7 & 15.9 & 13.1 & 16.3 & $\pi$\\
 &  &  & 18.8 & 33.3 & 36.3 & 42.8 & 8.6 & 15.3 & 11.6 & 16.2 & $b$\\
\end{longtable}
\endgroup{}

\clearpage

\begingroup\fontsize{10}{12}\selectfont

\begin{longtable}[t]{rrrrrrrrrrrr}
\caption[Empirical powers of the permutation and bootstrap tests under $\chi_{10}^2$-distribution]{Empirical powers (as percentages) of the permutation and bootstrap tests under $\chi_{10}^2$-distribution (Method: $\pi$ - permutation Wald-type test; $b$ - bootstrap Wald-type test; $\max,b$ - bootstrap multiple contrast test)}\\
\toprule
$\rho$ & $(C_1,C_2,C_3,C_4)$ & $n_i$ & $\varphi_{C^{RR}}$ & $\varphi_{B^{RR}}$ & $\varphi_{C^{VV}}$ & $\varphi_{B^{VV}}$ & $\varphi_{C^{VN}}$ & $\varphi_{B^{VN}}$ & $\varphi_{C^{AZ}}$ & $\varphi_{B^{AZ}}$ & Method\\
\midrule
\endfirsthead
\caption[]{Empirical powers (as percentages) of the permutation and bootstrap tests under $\chi_{10}^2$-distribution (Method: $\pi$ - permutation Wald-type test; $b$ - bootstrap Wald-type test; $\max,b$ - bootstrap multiple contrast test) \textit{(continued)}}\\
\toprule
$\rho$ & $(C_1,C_2,C_3,C_4)$ & $n_i$ & $\varphi_{C^{RR}}$ & $\varphi_{B^{RR}}$ & $\varphi_{C^{VV}}$ & $\varphi_{B^{VV}}$ & $\varphi_{C^{VN}}$ & $\varphi_{B^{VN}}$ & $\varphi_{C^{AZ}}$ & $\varphi_{B^{AZ}}$ & Method\\
\midrule
\endhead

\endfoot
\bottomrule
\endlastfoot
0.1 & $(0.1,0.1,0.1,0.15)$ & 30 & 96.1 & 97.8 & 96.9 & 98.8 & 27.2 & 42.6 & 35.2 & 48.2 & $\pi$\\
 &  &  & 95.8 & 97.9 & 96.5 & 98.8 & 25.2 & 40.4 & 33.3 & 45.7 & $b$\\
 &  &  & 82.2 & 97.2 & 88.8 & 97.4 & 4.1 & 39.6 & 16.8 & 40.2 & $\max,b$\\
 \addlinespace
& & 50 & 99.9 & 100.0 & 100.0 & 100.0 & 59.1 & 72.0 & 63.9 & 76.0 & $\pi$\\
 &  &  & 100.0 & 100.0 & 100.0 & 100.0 & 57.5 & 71.8 & 64.1 & 75.4 & $b$\\
 &  &  & 99.6 & 100.0 & 99.7 & 100.0 & 27.3 & 67.5 & 43.5 & 70.1 & $\max,b$\\
 \addlinespace
& $(0.5,0.5,0.5,0.7)$ & 30 & 29.7 & 47.5 & 77.7 & 85.4 & 13.6 & 25.4 & 19.4 & 31.3 & $\pi$\\
 &  &  & 29.0 & 48.3 & 76.3 & 85.4 & 12.1 & 24.4 & 18.9 & 31.4 & $b$\\
\end{longtable}
\endgroup{}